\pdfoutput=1
\documentclass[usenatbib]{mn2e}
\usepackage{apjfonts}
\usepackage{amssymb}
\usepackage{amsmath}
\usepackage{ctable}
\usepackage{url}
\usepackage{fixltx2e} 
\usepackage[implicit=false,breaklinks,colorlinks,citecolor=blue]{hyperref}

\newcommand{\be}{\begin{equation}}
\newcommand{\ee}{\end{equation}}

\newcommand{\etal}{et al.}

\newcommand{\msun}{M_{\sun}}

\newcommand{\paperone}{Paper {\small I}}
\newcommand{\paperthree}{Paper {\small III}}

\newcommand{\FIREurl}{\href{http://fire.northwestern.edu}{\url{http://fire.northwestern.edu}}}
\newcommand{\gizmourl}{\href{http://www.tapir.caltech.edu/~phopkins/Site/GIZMO.html}{\url{http://www.tapir.caltech.edu/~phopkins/Site/GIZMO.html}}}
\newcommand{\movieurl}{\href{http://www.tapir.caltech.edu/~phopkins/Site/animations}{\url{http://www.tapir.caltech.edu/~phopkins/Site/animations}}}

\newcommand{\apjl}{ApJL}
\newcommand{\apjs}{ApJS}

\newcommand{\aap}{A\&A}

\newcommand\plotonesize[2]
 {\centering \leavevmode \includegraphics[width={#2\columnwidth}]{#1}}
\newcommand{\plotsidesize}[2]
 {\centering \leavevmode \includegraphics[width={#2\textwidth}]{#1}}
\newcommand{\acknowledgments}{\begin{small}\section*{Acknowledgments}\end{small}}
\newcommand\altaffilmark[1]{$^{#1}$}
\newcommand\altaffiltext[1]{$^{#1}$}
\title[Modeling Supernova Feedback]{How To Model Supernovae in Simulations of Star and Galaxy Formation
\vspace{-0.5cm}}

\vspace{-0.2cm}
\author[Hopkins \etal]{
\parbox[t]{\textwidth}{ 
Philip F.~Hopkins\thanks{E-mail:phopkins@caltech.edu}\altaffilmark{1},
Andrew Wetzel\altaffilmark{1,2,3}\thanks{Caltech-Carnegie Fellow},
Du\v{s}an Kere\v{s}\altaffilmark{4}, 
Claude-Andr{\'e} Faucher-Gigu{\`e}re\altaffilmark{5}, 
Eliot Quataert\altaffilmark{6}, 
Michael Boylan-Kolchin\altaffilmark{7}, 
Norman Murray\altaffilmark{8}, 
Christopher C.\ Hayward\altaffilmark{9,10},
Kareem El-Badry\altaffilmark{6}
} 
\vspace*{6pt} \\
\altaffiltext{1}{TAPIR, Mailcode 350-17, California Institute of Technology, Pasadena, CA 91125, USA} \\
\altaffiltext{2}{The Observatories of the Carnegie Institution for Science, Pasadena, CA 91101, USA} \\
\altaffiltext{3}{Department of Physics, University of California, Davis, CA 95616, USA} \\
\altaffiltext{4}{Department of Physics, Center for Astrophysics and Space Science, University of California at San Diego, 9500 Gilman Drive, La Jolla, CA 92093} \\ 
\altaffiltext{5}{Department of Physics and Astronomy and CIERA, Northwestern University, 2145 Sheridan Road, Evanston, IL 60208, USA} \\ 
\altaffiltext{6}{Department of Astronomy and Theoretical Astrophysics Center, University of California Berkeley, Berkeley, CA 94720}\\ \altaffiltext{7}{Department of Astronomy, The University of Texas at Austin, 2515 Speedway, Stop C1400, Austin, TX 78712, USA} \\
\altaffiltext{8}{Canadian Institute for Theoretical Astrophysics, 60 St. George Street, University of Toronto, ON M5S 3H8, Canada}\\
\altaffiltext{9}{Center for Computational Astrophysics, Flatiron Institute, 162 Fifth Avenue, New York, NY 10010, USA}\\
\altaffiltext{10}{Harvard-Smithsonian Center for Astrophysics, 60 Garden Street,
Cambridge, MA 02138, USA}
\vspace{-0.5cm}
}

\date{Submitted to MNRAS, July 2017\vspace{-0.6cm}}
\begin{document}
\maketitle
\label{firstpage}

\vspace{-0.2cm}
\begin{abstract}
\vspace{-0.2cm}

We study the implementation of mechanical feedback from supernovae (SNe) and stellar mass loss in galaxy simulations, within the Feedback In Realistic Environments (FIRE) project. We present the FIRE-2 algorithm for coupling mechanical feedback, which can be applied to any hydrodynamics method (e.g.\ fixed-grid, moving-mesh, and mesh-less methods), and black hole as well as stellar feedback. This algorithm ensures manifest conservation of mass, energy, and momentum, and avoids imprinting ``preferred directions'' on the ejecta. We show that it is critical to incorporate both momentum and thermal energy of mechanical ejecta in a self-consistent manner, accounting for SNe cooling radii when they are not resolved. Using idealized simulations of single SN explosions, we show that the FIRE-2 algorithm, independent of resolution, reproduces converged solutions in both energy and momentum. In contrast, common ``fully-thermal'' (energy-dump) or ``fully-kinetic'' (particle-kicking) schemes in the literature depend strongly on resolution: when applied at mass resolution $\gtrsim 100\,\msun$, they diverge by orders-of-magnitude from the converged solution. In galaxy-formation simulations, this divergence leads to orders-of-magnitude differences in galaxy properties, unless those models are adjusted in a resolution-dependent way. We show that {\em all models that individually time-resolve SNe converge to the FIRE-2 solution at sufficiently high resolution} ($<100\,\msun$). However, in both idealized single-SN simulations and cosmological galaxy-formation simulations, the FIRE-2 algorithm converges much faster than other sub-grid models {\em without} re-tuning parameters.  
\end{abstract}

\begin{keywords}
galaxies: formation --- galaxies: evolution --- galaxies: active --- 
stars: formation --- cosmology: theory\vspace{-0.5cm}
\end{keywords}

\vspace{-0.5cm}
\section{Introduction}
\label{sec:intro}

Stellar feedback is critical in understanding galaxy formation. Without it, gas accretes into dark matter halos and galaxies, cools rapidly on a timescale much faster than the dynamical time, collapses, fragments, and forms stars on a free fall-time \citep{bournaud:2010.grav.turbulence.lmc,hopkins:rad.pressure.sf.fb,tasker:2011.photoion.heating.gmc.evol,dobbs:2011.why.gmcs.unbound,harper-clark:2011.gmc.sims}, inevitably turning most of the baryons into stars on cosmological timescales \citep{katz:treesph,somerville99:sam,cole:durham.sam.initial,springel:lcdm.sfh,keres:fb.constraints.from.cosmo.sims}. But observations imply that, on galactic scales, only a few percent of gas turns into stars per free-fall time \citep{kennicutt98}, while individual giant molecular clouds (GMCs) disrupt after forming just a few percent of their mass in stars \citep{zuckerman:1974.gmc.constraints,williams:1997.gmc.prop,evans:1999.sf.gmc.review,evans:2009.sf.efficiencies.lifetimes}. Similarly, galaxies retain and turn into stars just a few percent of the universal baryon fraction \citep{conroy:monotonic.hod,behroozi:mgal.mhalo.uncertainties,moster:stellar.vs.halo.mass.to.z1}, and both direct observations of galactic winds \citep{martin99:outflow.vs.m,heckman:superwind.abs.kinematics,sato:2009.ulirg.outflows,steidel:2010.outflow.kinematics,coil:2011.postsb.winds} and indirect constraints on the inter-galactic and circum-galactic medium \citep[IGM/CGM;][]{aguirre:2001.igm.metal.evol.sims,pettini:2003.igm.metal.evol,songaila:2005.igm.metal.evol,oppenheimer:outflow.enrichment,martin:2010.metal.enriched.regions} require that a large fraction of the baryons have been ``processed'' in galaxies via their accretion, enrichment, and expulsion in super-galactic outflows. 

Many different feedback processes contribute to these galactic winds and ultimately the self-regulation of galactic star formation, including protostellar jets, photo-heating, stellar mass loss (O/B and AGB-star winds), radiation pressure, and supernovae (SNe) Types Ia \&\ II \citep[see][and references therein]{evans:2009.sf.efficiencies.lifetimes,lopez:2010.stellar.fb.30.dor}. Older galaxy-formation simulations could not resolve the effects of these different processes (even on relatively large scales within the galactic disk), so they used simplified prescriptions to model galactic winds. However, a new generation of high-resolution simulations has emerged with the ability to resolve multi-phase structure in the ISM and so begin to directly incorporate these distinct feedback processes \citep{hopkins:rad.pressure.sf.fb,hopkins:fb.ism.prop,tasker:2011.photoion.heating.gmc.evol,kannan:2013.early.fb.gives.good.highz.mgal.mhalo,agertz:2013.new.stellar.fb.model}. One example is the Feedback In Realistic Environments (FIRE)\footnote{\label{foot:movie}See the {\small FIRE} project website:\\
\FIREurl \\
For additional movies and images of FIRE simulations, see:\\
\movieurl} project \citep{hopkins:2013.fire}. These and similar simulations have demonstrated predictions in reasonable agreement with observations for a wide variety of galaxy properties \citep[e.g.][]{ma:2015.fire.mass.metallicity,sparre.2015:bursty.star.formation.main.sequence.fire,wetzel.2016:latte,feldmann.2016:quiescent.massive.highz.galaxies.fire}.

In a companion paper, \citet[][hereafter \paperone]{hopkins:fire2.methods}, we presented an updated version of the FIRE code.
We refer to this updated FIRE version as ``FIRE-2'' and the older FIRE implementation as ``FIRE-1''.
We explored how a wide range of numerical effects (resolution, hydrodynamic solver, details of the cooling and star formation algorithm) influence the results of galaxy-formation simulations. We compared these to the effects of feedback and concluded that mechanical feedback, particularly from Type-II SNe, has much larger effects on galaxy formation (specifically properties such as galaxy masses, star formation histories, metallicities, rotation curves, sizes and morphologies) compared to the various numerical details studied. This is consistent with a number of previous studies \citep{abadi03:disk.structure,governato04:resolution.fx,robertson:cosmological.disk.formation,stinson:2006.sne.fb.recipe,zavala:cosmo.disk.vs.fb,scannapieco:2012.aquila.cosmo.sim.compare}. However, in galaxy-formation simulations, the actual implementation of SNe feedback, and the physical assumptions associated with it, often differ significantly between different codes. This can have significant effects on the predictions for galaxy formation \citep[see][]{scannapieco:2012.aquila.cosmo.sim.compare,rosdahl:2016.sne.method.isolated.gal.sims,kim:agora.isolated.disk.test}.

In this paper, we present a detailed study of the algorithmic implementation of SNe feedback and its effects, in the context of the FIRE-2 simulations. We emphasize that there are two {\em separate} aspects of mechanical feedback that must be explored.

First, the {\em numerical} aspects of the algorithmic coupling. Given some feedback ``products'' (mass, metals, energy, momentum) from a star, these must be deposited in the surrounding gas. {\em Any} good algorithm should respect certain basic considerations: conservation (of mass, energy, and momentum), statistical isotropy\footnote{Throughout the text, we use the term ``statistical isotropy'' to refer to a specific, desirable property of the numerical feedback-coupling algorithm. Namely, that the algorithm does not un-physically systematically bias the ejecta into certain directions (or otherwise ``imprint'' preferred directions) for numerical reasons. Of course, ejecta may be intrinsically anisotropic in the SN frame, and there can be global anisotropies sourced by e.g.\ pressure gradients and galaxy morphology, but these can only be captured properly if the ejecta-coupling {\em algorithm} is statistically isotropic.} (avoiding imprinting preferred directions that either depend on the numerical grid axes or the arbitrary gas configuration around the feedback source), and convergence. We will show that accomplishing these is non-trivial, and that many algorithms in common use (including the older algorithm that we used in FIRE-1\footnote{To be specific (this will be discussed below): the FIRE-1 algorithm used the ``non-conservative method'' defined in \S~\ref{sec:feedback:mechanical:vector}, with a less-accurate SPH approximation of the solid angle subtended by neighbors ($\omega_{b}$ defined in \S~\ref{sec:feedback:mechanical:weighting} set $\propto m_{b}/\rho_{b}$), and only coupled to the nearest neighbors for each SN instead of using the bi-directional search defined in \S~\ref{sec:feedback:mechanical:neighbor.finding} and needed to ensure statistical isotropy.}) do not respect all of them. 

\begin{footnotesize}
\ctable[
  caption={{\normalsize FIRE-2 simulations run to $z=0$ used for our case studies}\label{tbl:sims}},center,star
  ]{lcccccccr}{
\tnote[ ]{Parameters describing the FIRE-2 simulations from \citet{hopkins:fire2.methods} that we use for our case studies. Halo and stellar properties listed refer only to the original ``target'' halo around which the high-resolution region is centered. All properties listed refer to our highest-resolution simulation using the standard, default FIRE-2 physics and numerical methods. All units are physical. 
{\bf (1)} Simulation Name: Designation used throughout this paper. 
{\bf (2)} $M_{\rm halo}^{\rm vir}$: Virial mass \citep[following][]{bryan.norman:1998.mvir.definition} of the ``target'' halo at $z=0$.
{\bf (3)} $R_{\rm vir}$: Virial radius at $z=0$. 
{\bf (4)} $M_{\ast}$: Stellar mass of the central galaxy at $z=0$. 
{\bf (5)} $R_{1/2}$: Half-mass radius of the stars in the central $M_{\ast}$ at $z=0$. 
{\bf (6)} $m_{i,\,1000}$: Mass resolution: the baryonic (gas or star) particle/element mass, in units of $1000\,\msun$. The DM particle mass is always larger a factor $\approx 5$ (the universal ratio). 
{\bf (7)} $\epsilon_{\rm gas}^{\rm MIN}$: Minimum gravitational force softening reached by the gas in the simulation (gas softenings are adaptive so always exactly match the hydrodynamic resolution or inter-particle spacing); the Plummer-equivalent softening is $\approx 0.7\,\epsilon_{\rm gas}$.
{\bf (8)} $r_{\rm DM}^{\rm conv}$: Radius of convergence in the dark matter (DM) properties, in DM-only simulations. This is based on the \citet{power:2003.nfw.models.convergence} criterion using the best estimate from \citet{hopkins:fire2.methods} as to where the DM density profile is converged to within $<10\%$. The DM force softening is much less important and has no appreciable effects on any results shown here, so is simply fixed to $40\,$pc for all runs here.
}
}{
\hline\hline
Simulation & $M_{\rm halo}^{\rm vir}$ & $R_{\rm vir}$ & $M_{\ast}$ & $R_{1/2}$ & $m_{i,\,1000}$ & $\epsilon_{\rm gas}^{\rm MIN}$ & $r_{\rm DM}^{\rm conv}$ & Notes \\
Name \, & $[\msun]$ & $[{\rm kpc}]$ & $[\msun]$  & $[{\rm kpc}]$ & $[1000\,\msun]$ & $[{\rm pc}]$ & $[{\rm pc}]$ & \, \\ 
\hline 
{\bf m10q} & 8.0e9 & 52.4 & 1.8e6 & 0.63 & 0.25 & 0.52 & 73 & Isolated dwarf in an early-forming halo. Forms a dSph with a bursty SFH.  \\
{\bf m12i} & 1.2e12 & 275 & 6.5e10 & 2.9 & 7.0 & 0.38 & 150 & ``Latte'' primary halo from \citet{wetzel.2016:latte}. Thin disk with a flat SFH.  \\  
\hline\hline
}
\end{footnotesize}

Second, the {\em physics} of the coupling must be explored. At any finite resolution, there is a ``sub-grid scale'' -- the space or mass between a star particle and the center of the nearest gas resolution element, for example. An ideal implementation of the feedback coupling should exactly reproduce the converged solution, if we were to populate that space with infinite resolution -- in other words, our coupling should be equivalent to ``down grading'' the resolution of a high-resolution case, given the {\em same} physical assumptions used in the larger-scale simulation. We use a suite of simulations of isolated SNe (with otherwise identical physics to our galaxy-scale simulations) to show that a well-posed algorithm of this nature must account for {\em both} thermal {\em and} kinetic energy of the ejecta as they couple in a specific manner. This forms the basis for the default treatment of SNe in the FIRE simulations (introduced in \citealt{hopkins:2013.fire}), and similar to subsequent implementations in simulations by e.g.\ \citealt{kimm.cen:escape.fraction,rosdahl:2016.sne.method.isolated.gal.sims}). In contrast, we show that coupling only thermal or kinetic energy leads to strongly resolution-dependent errors, which in turn can produce order-of-magnitude too-large or too-small galaxy masses. To predict reasonable masses, such models must be modified (a.k.a.\ ``re-tuned'') at each resolution level. This is even more severe in ``delayed cooling'' or ``target temperature'' models which are explicitly intended for low-resolution applications, and are not designed to converge to the exact solution at high resolution. This explains many seemingly contradictory conclusions in the literature regarding the implementation of feedback. In contrast, we will show that the mechanical feedback models proposed here reproduce the high-resolution solution in idealized problems at {\em all} resolution levels that we explore, converge much more rapidly in cosmological galaxy-formation simulations, and (perhaps most importantly) represent the solution towards which other less-accurate ``sub-grid'' SNe treatments (at least those which do not artificially modify the cooling physics) converge at very high resolution.

Our study here is relevant for simulations of the ISM and galaxy formation with mass resolution in the range $\sim 10-10^{6}\,\msun$; we will show that at resolution higher than this, the numerical details have weak effects because early SN blastwave evolution is explicitly well-resolved. Conversely, at lower resolution than this, treating individual SN events becomes meaningless (necessitating a different sort of ``sub-grid'' approach).

In \S~\ref{sec:methods} we provide a summary of the FIRE-2 simulations (\S~\ref{sec:methods:all}), a detailed description of the numerical algorithm for mechanical feedback coupling (\S~\ref{sec:feedback:mechanical}), and a detailed motivation and description of the physical breakdown between kinetic and thermal energy (\S~\ref{sec:feedback:mechanical:sedov}). We note that \paperone\ includes complete details of all aspects of the simulations here, necessary to fully reproduce our results. In \S~\ref{sec:feedback:mechanical:ideal.tests} we validate the numerical coupling algorithm (conservation, statistical isotropy, and convergence) and explore the effects of alternative coupling schemes on full galaxy formation simulations. In \S~\ref{sec:feedback:mechanical:tests} we validate the physical breakdown of coupled kinetic/thermal energy, compare this to simulations of individual SN explosions at extremely high resolution, and explore how different choices which neglect these physics alter the predictions of full galaxy formation simulations. We briefly discuss non-convergent alternative models (e.g.\ ``delayed cooling'' and ``target temperature'' models) but provide more detailed tests of these in the Appendices. In \S~\ref{sec:discussion} we summarize our conclusions. Additional tests are discussed in the Appendices.

\vspace{-0.5cm}
\section{Methods \&\ Physical Motivation}
\label{sec:methods}

\subsection{Overview \&\ Methods other than Mechanical Feedback}
\label{sec:methods:all}

The simulations in this paper were run as part of the Feedback in Realistic Environments (FIRE) project, using the FIRE-2 version of the code detailed in \paperone. Our default simulations are exactly those in \paperone; we will vary the SNe algorithm to explore how this alters galaxy formation, but all other simulation properties, physics, and numerical choices are held fixed. For detailed exploration of how those numerical details alter galaxy formation, we refer to \paperone.  The simulations were run using {\small GIZMO}\footnote{A public version of {\small GIZMO} is available at \gizmourl} \citep{hopkins:gizmo}, in its meshless finite-mass MFM mode. This is a mesh-free, finite-volume Lagrangian Godunov method which provides adaptive spatial resolution together with conservation of mass, energy, momentum, and angular momentum, and the ability to accurately capture shocks and fluid mixing instabilities (combining advantages of both grid-based and smoothed-particle hydrodynamics methods). For extensive test problems see \citet{hopkins:gizmo,hopkins:mhd.gizmo,hopkins:cg.mhd.gizmo,hopkins:gizmo.diffusion}; for tests of the methods specific to these simulations see \paperone.

These simulations are cosmological ``zoom-in'' runs that follow the Lagrangian region that surrounds a galaxy at $z=0$ (out to several virial radii) from seed perturbations at $z=100$. Gravity is solved for collisional (gas) and collisionless (stars and dark matter) species with adaptive gravitational softening so hydrodynamic and force softening are always matched. Gas cooling is followed self-consistently from $T=10-10^{10}\,$K including free-free, Compton, metal-line, molecular, fine-structure, dust collisional, and cosmic ray processes, photo-electric and photo-ionization heating by both local sources and a uniform but redshift-dependent meta-galactic background, and self-shielding. Gas is turned into stars using a sink-particle prescription (gas which is locally self-gravitating at the resolution scale following \citealt{hopkins:virial.sf}, self-shielding/molecular following \citealt{krumholz:2011.molecular.prescription}, Jeans unstable, and denser than $n_{\rm crit}>1000\,{\rm cm^{-3}}$ is converted into star particles on a free-fall time). Star particles are then treated as single-age stellar populations with all IMF-averaged feedback properties calculated from {\small STARBURST99} \citep{starburst99} assuming a \citet{kroupa:2001.imf.var} IMF. We then explicitly treat feedback from SNe (both Types Ia and II), stellar mass loss (O/B and AGB winds), and radiation (photo-ionization and photo-electric heating and UV/optical/IR radiation pressure), with implementations at the resolution-scale described in \paperone\ and here.

\paperone\ provides a complete description of all aspects of the numerical methods. In this paper, we study the mechanical feedback algorithm, used for SNe and stellar mass loss. In a companion paper (henceforth \paperthree), we study the radiation feedback algorithm. 

For simplicity, we focus our study here on two example galaxies: {\bf m10q} is a dwarf galaxy and {\bf m12i} is a Milky Way (MW)-mass galaxy. Table~\ref{tbl:sims} lists their properties. Both were studied extensively in \paperone. The star formation history, stellar mass, and mean stellar-mass weighted metallicity of each galaxy as a function of cosmic time, as well as the $z=0$ baryonic and dark matter mass profiles and rotation curves, will be discussed below. We have explicitly verified that the conclusions drawn here regarding mechanical feedback from our {\bf m10q} and {\bf m12i} simulations are robust across simulations of several different galaxies/halos at dwarf and MW mass scales, respectively.

\vspace{-0.5cm}
\subsection{Mechanical Feedback Coupling Algorithm}
\label{sec:feedback:mechanical}

\begin{figure*}
\plotsidesize{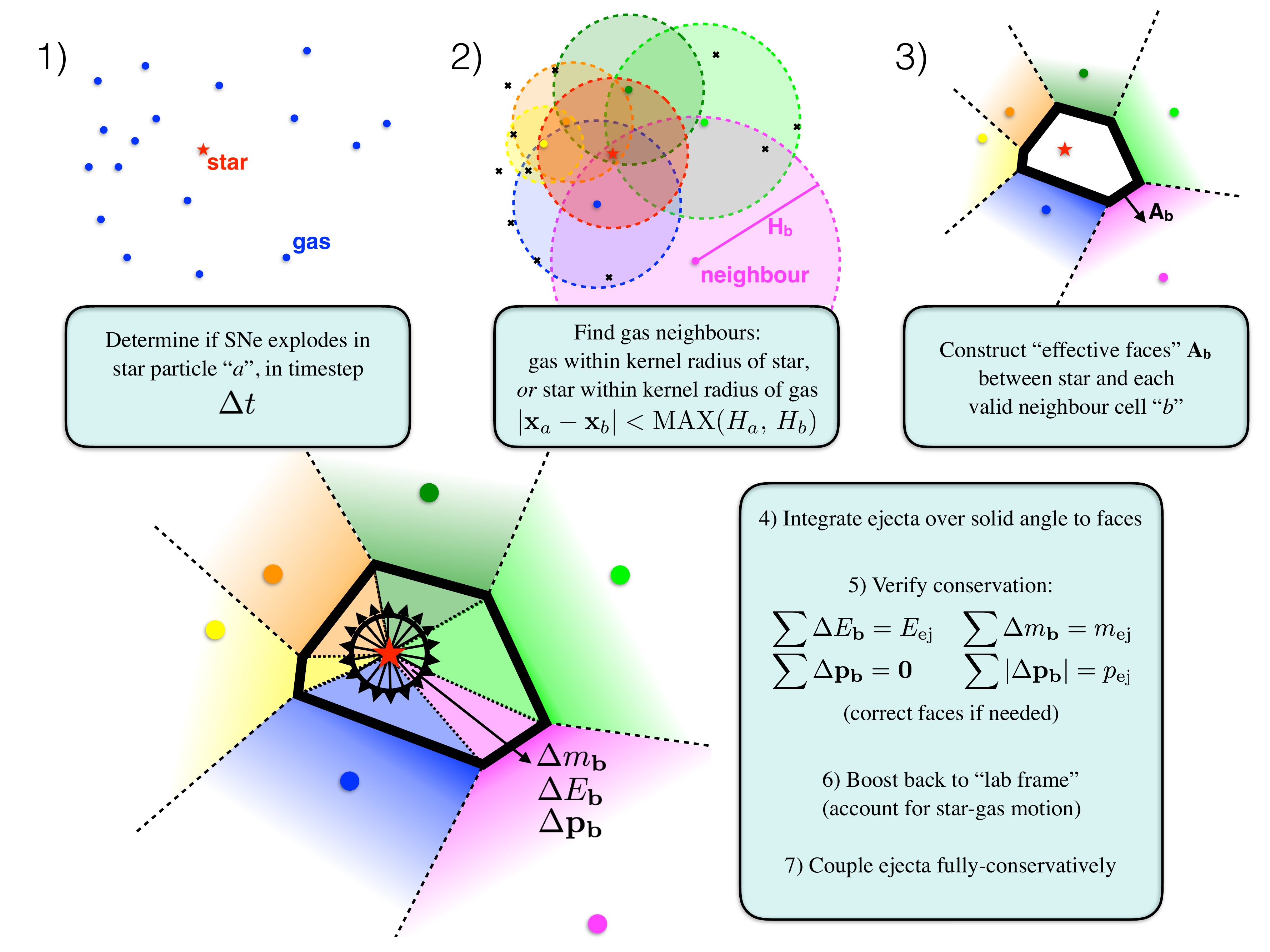}{0.99}
    \vspace{-0.25cm}
    \caption{Cartoon illustrating the numerical algorithm for coupling mechanical feedback, described in detail in \S~\ref{sec:feedback:mechanical}. 
    {\bf (1)} We determine from stellar evolution tracks (\paperone, Appendix~A) whether a star particle is a source of mechanical feedback (SNe, stellar mass loss, short-range radiation pressure) at a given timestep.
    {\bf (2)} We identify valid interacting neighbors for the star (\S~\ref{sec:feedback:mechanical:neighbor.finding}). The search includes not only gas within the nearest neighbor search radius, $H_{a}$, of the star, but {\em also} gas for which the star falls within the gas element's search radius, $H_{b}$. This typically corresponds to lower-density gas, and in this example, the two neighbours (magenta+lime) to the right of the star would not be included if only gas within $H_{a}$ were used. This would artificially prevent us from coupling momentum in that direction, violating statistical isotropy.
    {\bf (3)} Construct the ``effective faces'' of the interacting gas elements, as seen by the star (\S~\ref{sec:feedback:mechanical:weighting}). This example uses a Voronoi tesselation, which is similar to the result from our default MFM method.
    {\bf (4)} Integrate the feedback quantities (mass, metals, energy, momentum), assumed to be isotropically emitted from the star (in its rest frame), over the solid angle subtended by the effective face of each gas element, to determine the fluxes into each cell. Integrate through to the face to account for the $PdV$ work done between star and face (\S~\ref{sec:feedback:mechanical:sedov}). 
    {\bf (5)} Verify that the fluxes maintain machine-accurate mass/energy/momentum conservation: if not, re-normalize the faces to correct them such that this is satisfied (\S~\ref{sec:feedback:mechanical:vector}).
    {\bf (6)} Boost to the lab frame, to account for any relative star-gas motion (\S~\ref{sec:feedback:mechanical:assignment}). 
    {\bf (7)} Finally, couple the fluxes, maintaining exact conservation, giving the updated mass, metallicity, thermal and kinetic energy, and momentum to the gas element.
        \label{fig:mechanical.fb.cartoon}}
\end{figure*}

\subsubsection{Determining When Events Occur}
\label{sec:feedback:mechanical:event.detection}

Once a star particle forms, the SNe rate is taken from stellar evolution models, assuming the particle represents an IMF-averaged population of a given age (since it formed) and abundances (inherited from its progenitor gas element). Given the particle masses and timesteps ($\Delta t \sim 100-1000\,$yr) for young star particles, the expected number of SNe per particle per timestep is always $\ll 1$. To determine if an event occurs, we therefore draw from a binomial distribution at each timestep given the expected rate $\langle N \rangle = (dN/dM_{\ast}\,dt)\,m_{i}\,\Delta t$, where $(dN/dM_{\ast}\,dt)$ is the IMF-averaged SNe rate per unit mass for a single stellar population of the age and metallicity of the star particle and $m_{i}$ is the star particle mass. For continuous mass-loss processes such as O/B or AGB winds, an ``event'' occurs every timestep, with mass loss $\Delta M_{\ast} = \Delta t\,\dot{M}_{\ast}$ and the associated kinetic luminosity. See \paperone\ for details and tabulations of the relevant rates.

Consider a time $t_{a}$ (timestep $\Delta t$), during which a mechanical feedback ``event'' occurs sourced at some location ${\bf x}_{a}$ (for example, the location of a star particle ``$a$'' in which a SN explodes). Our focus in this paper is how to treat this event. Fig.~\ref{fig:mechanical.fb.cartoon} provides an illustration of our algorithm.
We first define a set of conserved quantities: mass $m_{\rm ej}$, metals $m_{Z,\,{\rm ej}}$, momentum $p_{\rm ej}=m_{\rm ej}\,v_{\rm ej}$, and energy $E_{\rm ej}$, which must be ``injected'' into the neighboring gas via some numerical fluxes.

\vspace{-0.5cm}
\subsubsection{Finding Neighbors to Couple}
\label{sec:feedback:mechanical:neighbor.finding}

We define an effective neighbor number $N_{\ast}$ the same as for the hydrodynamics, $N_{\ast} = (4\pi/3)\,H_{a}^{3}\,\bar{n}_{a}(H_{a})$ where, $\bar{n}_{a}=\sum W({\bf x}_{ba}\equiv {\bf x}_{b}-{\bf x}_{a},\,H_{a})$, $W$ is the kernel function, and $H_{a}$ is the search radius around the star (set by $N_{\ast}$, which is the ``fixed'' parameter).\footnote{In this paper we will use a cubic spline for $W$, but other choices have weak effects on our conclusions (because $W$ will be re-normalized anyways in the assignment of ``weights'' for feedback). We adopt $N_{\ast}=64$ for reasons discussed below. The equation for $N_{\ast}(H_{a})$ is non-linear, so it is solved iteratively in the neighbor search; see \citet{springel:entropy}.}  Thus we obtain all gas elements $b$ within a radius $|{\bf x}_{ba}| < H_{a}$.

However, severe pathologies can occur if feedback is coupled {\em only} to the nearest neighboring gas to the star. For example, in an infinitely thin, dense disk of gas surrounding the star particle, with a tenuous atmosphere in the vertical direction above/below the disk, the closest $N_{\ast}$ elements to ${\bf x}_{a}$ likely will be in the disk -- so searching only within $H_{a}$ will fail to ``see'' the vertical directions, thus coupling all feedback within the disk, despite the fact that the disk subtends a vanishingly small portion of the sky as seen from the star. Our solution to this is to use the same approach used in the hydrodynamic solver (in all mesh-free methods; SPH and MFM/MFV): we include {\em both} elements with $|{\bf x}_{ba}| < H_{a}$ and $|{\bf x}_{ba}| < H_{b}$. That is, we additionally include any gas elements whose kernel encompasses the star. In the disk example, the closest ``atmosphere elements'' above/below the disk necessarily have their own kernel radii, $H_{b}$, that overlap the disk, so this guarantees ``covering'' by elements in the vertical direction. This is demonstrated in Fig.~\ref{fig:mechanical.fb.cartoon}. The importance of including these elements is validated in our tests below, where we show that failure to include these neighbors artificially biases the feedback deposition.

We impose a maximum cutoff radius, $r_{\rm max}$, on the search, to prevent pathological situations for which there is no nearby gas so feedback would be deposited at unphysically large distances. Specifically, we impose $r_{\rm max}=2\,$kpc.
This corresponds to where the ram pressure of free-expanding ejecta falls below the thermal pressure in even low-density circum-galactic conditions ($T \sim 10^{4}$\,K at $n\gtrsim 0.001\,{\rm cm^{-3}}$). However, our results are not sensitive to this choice, because it affects a vanishingly small number of events.

\vspace{-0.5cm}
\subsubsection{Weighting the Deposition: The Correct ``Effective Area''}
\label{sec:feedback:mechanical:weighting}

Having identified interacting neighbors, $b$, we must deposit the injected quantities according to some weighting scheme. Each neighbor resolution element gets a weight $\tilde{\omega}_{b}$ that determines the fraction of the injected quantity it receives. Of course, this must be normalized to properly conserve quantities, so we first calculate an un-corrected weight, $\omega_{b}$, and then assign
\begin{align}
\label{eqn:weight.renorm.scalar} \tilde{\omega}_{b} \equiv \frac{\omega_{b}}{\sum_{c}\,\omega_{c}}
\end{align}
so that $\sum_{b} \tilde{\omega}_{b}=1$, exactly. 

Naively, a simple weight scheme might use $\omega_{b}=1$, or $\omega_{b}=W({\bf x}_{ba},\,H_{a})$. However, for quasi-Lagrangian schemes for which the different gas elements have approximately equal masses ($m_{b}\sim$\,constant), this is effectively mass-weighting the feedback deposition, which is not physical. In the example of the infinitely thin disk, because most of the neighbor elements lie within the disk, the disk-centered elements would again receive most of the feedback, despite the fact that they cover a vanishingly small portion of the sky from the source.

If the feedback is emitted statistically isotropically from the source ${\bf x}_{a}$, the correct solution is to integrate the injection into each solid angle and determine the total solid angle $\Delta \Omega_{b}$ subtended by a given gas resolution element, i.e.\ adopt $\omega_{b} = \Delta \Omega_{b}/4\pi$. This is shown in Fig.~\ref{fig:mechanical.fb.cartoon}. Given a source at ${\bf x}_{a}$ and neighbors at ${\bf x}_{b}$, we can construct a set of faces that enclose ${\bf x}_{a}$ with some convex hull. Each face has a vector oriented area ${\bf A}_{b}$; if the face is symmetric it subtends a solid angle on the sky as seen by ${\bf x}_{a}$ of 
\begin{align}
\label{eqn:solidangle}\omega_{b} & \equiv \frac{1}{2}\,\left(1-\frac{1}{\sqrt{1+({\bf A}_{b}\cdot \hat{\bf x}_{ba})/(\pi\,|{\bf x}_{ba}|^{2})}}\right) \approx \frac{\Delta\Omega_{b}}{4\pi}
\end{align}
(This simply interpolates between $\sim A_{b}/4\pi\,r_{b}^{2}$ for $r_{b}^{2} = |{\bf x}_{ba}|^{2} \gg A_{b} \equiv |{\bf A}_{b}\cdot \hat{\bf x}_{ba}|$, and $1/2$ for $r_{b}^{2} \ll A_{b}$.)\footnote{Eq.~\ref{eqn:solidangle} is exact for a face ${\bf A}_{b}$ which is rotationally symmetric about the axis $\hat{\bf x}_{ba}$; for asymmetric ${\bf A}_{b}$, evaluating $\Delta\Omega_{b}$ exactly requires an expensive numerical quadrature. If this is done exactly, Eq.~\ref{eqn:weight.renorm.scalar} is unnecessary: $\sum_{b}\omega_{b}=1$ is guaranteed. We have experimented with an exact numerical quadrature; but it is extremely expensive and has no measurable effect on our results compared to simply using Eq.~\ref{eqn:weight.renorm.scalar} \&\ \ref{eqn:solidangle} for all ${\bf A}_{b}$ (Eq.~\ref{eqn:solidangle} is usually accurate to $<1\%$, and the most severe discrepancies do not exceed $\sim 10\%$, and these are normalized out by Eq.~\ref{eqn:weight.renorm.scalar}).}

No unique convex hull exists. One solution, for example, would be to construct a Voronoi tesselation around ${\bf x}_{a}$, with both the star particle ${\bf x}_{a}$ and the locations of all neighbors ${\bf x}_{b}$ as mesh-generating points. However, we already have an internally consistent value of ${\bf A}_{b}$, namely, the definition ${\bf A}_{b}^{\rm hydro}$ of the ``effective faces'' used in the hydrodynamic equations (the faces that appear in the discretized Euler equations: e.g.\ $d{\bf U}_{a}/dt = - \sum_{b}\,{\bf F}_{ab}({\bf U}) \cdot {\bf A}_{b}$, where ${\bf U}$ is a conserved quantity and ${\bf F}$ is its flux). For a Voronoi moving-mesh code (e.g.\ {\small AREPO}), this is the Voronoi tesselation. For SPH as implemented in {\small GIZMO}, this is ${\bf A}_{b}^{\rm hydro} = [\bar{n}_{a}^{-2}\,\partial W(r_{b},\,H_{a})/\partial r_{b} + \bar{n}_{b}^{-2}\,\partial W(r_{b},\,H_{b})/\partial r_{b}]\,\hat{\bf x}_{ba}$. For MFM/MFV the expression is more complicated but is given in Eq.~18 in \citet{hopkins:gizmo}.\footnote{In MFM/MFV, the effective face ${\bf A}_{b}$ is given by:
\begin{align}
\label{eqn:mfm.face.area.def} {\bf A}_{b} &\equiv  \bar{n}_{a}^{-1}\,\bar{\bf q}_{b}({\bf x}_{a}) + \bar{n}_{b}^{-1}\,\bar{\bf q}_{a}({\bf x}_{b})\\ 
\label{eqn:mfm.face.area.def.sub1} \bar{\bf q}_{b}({\bf x}_{a}) &\equiv {\bf E}_{a}^{-1} \cdot {\bf x}_{ba}\, W({\bf x}_{ba},\,H_{a}) \\
\label{eqn:mfm.face.area.def.sub2} {\bf E}_{a} &\equiv \sum_{c}\,({\bf x}_{ca} \otimes {\bf x}_{ca}) \,W({\bf x}_{ca},\,H_{a}) 
\end{align}
where ``$\cdot$'' and ``$\otimes$'' denote the inner and outer product, respectively.} We therefore adopt ${\bf A}_{b} = {\bf A}_{b}^{\rm hydro}$ -- the ``effective face area'' that the neighbor gas elements would share with ${\bf x}_{a}$ in the hydrodynamic equations if the source (star particle) were a gas element. Fig.~\ref{fig:sne.fb.coupling.tests} demonstrates that this is sufficient to ensure the coupling into each solid angle is statistically isotropic in the frame of the SN.

While we find that weighting by solid angle is important, at the level of accuracy here, the exact values of ${\bf A}_{b}^{\rm hydro}$ given by SPH, MFM, or Voronoi formalisms differ negligibly, and we can use them interchangeably with no detectable effects on our results. This is not surprising: \citet{hopkins:gizmo} showed that the Voronoi tesselation is simply the limit for a sharply-peaked kernel of the MFM faces.

\vspace{-0.5cm}
\subsubsection{Dealing With Vector Fluxes (Momentum Deposition)}
\label{sec:feedback:mechanical:vector}

If we were only considering sources of scalar conserved quantities (e.g.\ mass $m_{\rm ej}$ or metals $m_{Z,\,{\rm ej}}$), we would be done. We simply define a numerical flux $\Delta m_{b} = \tilde{\omega}_{b}\,m_{\rm ej}$ into each neighbor element (subtracting the same from our ``source'' star particle), and we are guaranteed both machine-accurate conservation ($\sum_{b}\,\Delta m_{b} = m_{\rm ej}$) and the correct spatial distribution of ejecta. 

However, the situation is more complex for a vector flux, specifically here, momentum deposition. 
If the ejecta have some uniform radial velocity, ${\bf v}_{\rm ej} = v_{\rm ej}\,\hat{r}$, away from the source, ${\bf x}_{a}$, then one might naively define the corresponding momentum flux $\Delta {\bf p}_{b} = \tilde{\omega}_{b}\,m_{\rm ej}\,v_{\rm ej}\,\hat{\bf x}_{ba} = p_{\rm ej}\,\tilde{\omega}_{b}\,\hat{\bf x}_{ba} $. However, then $\sum_{b} \Delta {\bf p}_{b} = p_{\rm ej}\,\sum_{b}\,\tilde{\omega}_{b}\,\hat{\bf x}_{ba}$. But this is {\em not} guaranteed to vanish: the deposition can violate linear momentum conservation, if $\boldsymbol{\psi}_{a} \equiv \sum_{b}\,\tilde{\omega}_{b}\,\hat{\bf x}_{ba} \ne {\bf 0}$. The correct $\boldsymbol{\psi}_{a}={\bf 0}$ is only guaranteed if (1) the coupled momentum $\Delta {\bf p}_{b}$ is the {\em exact} solution of the integral of $p_{\rm ej}\,(4\pi\,|{\bf r}|^{2})^{-1}\hat{\bf r}\cdot d{\bf A}_{b}(\theta,\,\phi)$ (where ${\bf r}$ is the vector from ${\bf x}_{a}$ to a location ${\bf x}$ on the surface ${\bf A}_{b}$), and (2) the faces of the convex hull close exactly ($\sum_{b}\,{\bf A}_{b}={\bf 0}$). Even in a Cartesian grid (which trivially satisfies (2)), condition (1) can only be easily evaluated if we assume (incorrectly) that the feedback event occurs exactly at the center or corner of a cell; in Voronoi meshes and mesh-free methods (1) is only possible to satisfy with an expensive numerical quadrature, and (2) is only satisfied up to some integration accuracy.

In practice, $\Delta {\bf p}_{b} = p_{\rm ej}\, \tilde{\omega}_{b}\,\hat{\bf x}_{ba}$ is a good approximation to the integral in condition (1), and is again exact for faces symmetric about $\hat{\bf x}_{ba}$, and (2) is satisfied up to second-order integration errors in our MFM/MFV methods, so the dimensionless $|\boldsymbol{\psi}_{a}|\ll 1$ is small. However, we wish to ensure machine-accurate conservation, so we must impose a {\em tensor} re-normalization condition, not simply the scalar re-normalization in Eq.~\ref{eqn:weight.renorm.scalar}: we therefore define the six-dimensional vector weights $\hat{\bf x}_{ba}^{\pm}$: 
\begin{align}
\label{eqn:vector.weight.def} \hat{\bf x}_{ba} &\equiv \frac{{\bf x}_{ba}}{|{\bf x}_{ba}|} = \sum_{+,\,-}\,\hat{\bf x}_{ba}^{\pm} \\ 
\label{eqn:vector.weight.def.sub1} (\hat{\bf x}^{+}_{ba})^{\alpha} &\equiv {|{\bf x}_{ba}|^{-1}}\,{\rm MAX}({\bf x}_{ba}^{\alpha},\,0)\,{\Bigr|}_{\alpha=x,\,y,\,z}\\
\label{eqn:vector.weight.def.sub2} (\hat{\bf x}^{-}_{ba})^{\alpha} &\equiv {|{\bf x}_{ba}|^{-1}}\,{\rm MIN}({\bf x}_{ba}^{\alpha},\,0)\,{\Bigr|}_{\alpha=x,\,y,\,z}
\end{align} 
i.e.\ the unit vector component in the plus (or minus) $x,\,y,\,z$ directions ($\alpha$ refers to these components), for each neighbor. 
We can then define a vector weight $\tilde{\bf w}_{b}$: 
\begin{align}
\label{eqn:vector.weight.normalized} \bar{\bf w}_{b} &\equiv \frac{{\bf w}_{b}}{\sum_{c}\,|{\bf w}_{c}|} \\ 
\label{eqn:vector.weight.normalized.sub1} {\bf w}_{b} &\equiv \omega_{b}\, \sum_{+,\,-}\,\sum_{\alpha}\,(\hat{\bf x}_{ba}^{\pm})^{\alpha}\,\left( f_{\pm}^{\alpha} \right)_{a} \\ 
\label{eqn:vectornorm} \left( f_{\pm}^{\alpha} \right)_{a} &\equiv \left\{ \frac{1}{2}\,\left[1 +  \left( \frac{\sum_{c}\,\omega_{c}\,|\hat{\bf x}_{ca}^{\mp}|^{\alpha}}{\sum_{c}\,\omega_{c}\,|\hat{{\bf x}}_{ca}^{\pm}|^{\alpha}} \right)^{2}\right]\right\}^{1/2} 
\end{align}
This is evaluated in two passes over the neighbor list.\footnote{The function $(f_{\pm})$ in Eq.~\ref{eqn:vectornorm} is derived by requiring ${\bf 0} = \sum \Delta {\bf p}_{b}$. Component-wise, this becomes $0 = \sum (\Delta {\bf p}_{b})^{\alpha} = p_{\rm ej}/(\sum_{c}\,|{\bf w}_{c}|)\,\left[\left( f_{+}^{\alpha}\,\sum_{b} \omega_{b}\,(\hat{\bf x}_{ba}^{+})^{\alpha} + 
f_{-}^{\alpha}\,\sum_{b} \omega_{b}\,(\hat{\bf x}_{ba}^{-})^{\alpha} \right) \right]$. Since $p_{\rm ej}$ and $\sum_{c}\,|{\bf w}_{c}|$ are positive-definite, the term in brackets must vanish ($f_{+}^{\alpha}\,\boldsymbol{\psi}_{+}^{\alpha}=f_{-}^{\alpha}\,\boldsymbol{\psi}_{-}^{\alpha}$, if we define $\boldsymbol{\psi}_{\pm}^{\alpha} \equiv \sum_{b}\,\omega_{b}\,|\hat{\bf x}_{ba}^{\pm}|^{\alpha}$). But we also wish to minimize the effect of the correction factor $f_{\pm}$ on the total momentum coupled (ensuring $f_{\pm}\approx 1$), so we minimize the least-squares penalty function $\Delta^{2}_{\boldsymbol{\psi}} =\| [(f_{+}^{\alpha}\boldsymbol{\psi}_{+}^{\alpha})^{2} + (f_{-}^{\alpha}\boldsymbol{\psi}_{-}^{\alpha})^{2} ] - [ (\boldsymbol{\psi}_{+}^{\alpha})^{2} + (\boldsymbol{\psi}_{-}^{\alpha})^{2} ] \|$. The $f_{\pm}$ in Eq.~\ref{eqn:vectornorm} is the unique function which simultaneously guarantees ${\bf 0} = \sum \Delta {\bf p}_{b}$ (i.e.\ $f_{+}^{\alpha}\,\boldsymbol{\psi}_{+}^{\alpha}=f_{-}^{\alpha}\,\boldsymbol{\psi}_{-}^{\alpha}$) and $\Delta^{2}_{\boldsymbol{\psi}}=0$. It is easy to see that $f_{\pm}\rightarrow 1$, as it should, if $\boldsymbol{\psi}_{+}=\boldsymbol{\psi}_{-}$, i.e.\ when $\sum \Delta {\bf p}_{b} = {\bf 0}$ without the need for an additional correction.}

It is straightforward to verify (and we show explicitly in tests below) that the approach above guarantees momentum conservation to machine accuracy. Ignoring these correction terms can (if the neighbors are ``badly ordered,'' e.g.\ all lie the same direction), lead to order-unity errors in momentum conservation, and the fractional error $|\sum_{b} \Delta {\bf p}_{b}| / p_{\rm ej} = |\boldsymbol{\psi}_{a}|$ depends only on the spatial distribution of neighbors in the kernel, not on the resolution.

Physically, we should think of the vector weights $\bar{\bf w}$ as accounting for asymmetries about the vector $\hat{\bf x}_{ab}$ in the faces ${\bf A}_{b}$. If the faces were all exactly symmetric (e.g.\ the neighbor elements were perfectly isotropically distributed), then the net momentum integrated into each face would indeed point exactly along $\hat{\bf x}_{ab}$. But, typically, they are not, so we must account for this in order to properly retain momentum conservation.

\vspace{-0.5cm}
\subsubsection{Assigning Fluxes and Including Gas-Star Motion}
\label{sec:feedback:mechanical:assignment}

Finally, we can assign fluxes: 
\begin{align}
\label{eqn:flux.m} \Delta m_{b} &= |\bar{\bf w}_{b}|\,m_{\rm ej} \\ 
\label{eqn:flux.z} \Delta m_{Z,\,b} &= |\bar{\bf w}_{b}|\,m_{Z,\,{\rm ej}} \\ 
\label{eqn:flux.e} \Delta E_{b} &= |\bar{\bf w}_{b}|\,E_{\rm ej} \\ 
\label{eqn:flux.p} \Delta {\bf p}_{b} &= \bar{\bf w}_{b}\,p_{\rm ej}
\end{align}
which the definitions above guarantee will {\em exactly} satisfy: 
\begin{align} 
\label{eqn:flux.m.conservation} \sum\,\Delta m_{b} &= m_{\rm ej} \\ 
\label{eqn:flux.z.conservation} \sum\,\Delta m_{Z,\,b} &= m_{Z,\,{\rm ej}} \\ 
\label{eqn:flux.e.conservation} \sum\,\Delta E_{b} &= E_{\rm ej} \\ 
\label{eqn:flux.p.conservation1} \sum\,|\Delta {\bf p}_{b}| &= p_{\rm ej} \\ 
\label{eqn:flux.p.conservation2} \sum\,\Delta {\bf p}_{b} &= {\bf 0} 
\end{align}
Our definitions also ensure that the fraction of ejecta entering a gas element is as close as possible (as much as allowed by the strict conservation conditions above) to the fraction of solid angle subtended by the element, as would be calculated self-consistently by the hydrodynamic method in the code, i.e.\ 
\begin{align}
\label{eqn:weight.area.equivalence} |\bar{\bf w}_{b}| \approx \frac{\Delta {\mathbf \Omega}_{b}^{\rm hydro}}{4\pi}
\end{align}
Moreover, in the limit where Eq.~\ref{eqn:solidangle} is exact (the faces ${\bf A}_{b}$ are symmetric about $\hat{\bf x}_{ba}$), and they close exactly ($\sum_{b}{\bf A}_{b}={\bf 0}$; i.e.\ good element order), then $(f_{\pm})=1$ and $\sum_{c}\,|{\bf w}_{c}|=1$, i.e.\ $\bar{\bf w}_{b}\rightarrow \omega_{b}\,\hat{\bf x}_{ba}$ and our naive estimate is both exact and conservative, and no normalization of the weights is necessary. In practice, as noted above, we find that the deviations (in the sum) from this perfectly-ordered case are usually small (percents-level), but there are always pathological element configurations where they can be large, and maintaining good conservation requires the corrected terms above.

Implicitly, we have been working in the frame moving with the feedback ``source'' (${\bf x}_{a}={\bf 0}$, ${\bf v}_{a} \equiv d{\bf x}_{a}/dt = {\bf 0}$), in which the source is statistically isotropic. However, in coupling the fluxes to surrounding gas elements, we also must account for the frame motion. Boosting back to the lab/simulation frame, the total ejecta velocity entering an element is of course $\Delta m_{b}^{-1}\,\Delta {\bf p}_{b} + {\bf v}_{a}$. This change of frame has no effect on the mass fluxes, but it does modify the momentum and energy fluxes: to be properly conservative, we must take:
\begin{align}
\label{eqn:flux.mz.framecorr} \Delta m_{b}^{\prime} &\equiv \Delta m_{b}\ \ \ , \ \ \  \Delta m_{Z,\,b}^{\prime} \equiv \Delta m_{Z,\,b} \\ 
\label{eqn:flux.p.framecorr} \Delta {\bf p}_{b}^{\prime} &\equiv \Delta {\bf p}_{b} + \Delta m_{b}\,{\bf v}_{a} \\ 
\label{eqn:flux.e.framecorr} \Delta E_{b}^{\prime} &\equiv \Delta {E}_{b} + \frac{1}{2\,\Delta m_{b}}\,\left( |\Delta {\bf p}_{b}^{\prime}|^{2} - |\Delta {\bf p}_{b}|^{2} \right)
\end{align}
where the prime (e.g.\ ``$ \Delta m_{b}^{\prime}$'') notation denotes the lab frame.
Note that the extra momentum added to the neighbors ($\sum_{b}\,\Delta m_{b}\,{\bf v}_{a} = m_{\rm ej}\,{\bf v}_{a}$) is exactly the momentum lost by the feedback source $a$, by virtue of its losing $m_{\rm ej}$ in mass.\footnote{The de-boosted energy equation, Eq.~\ref{eqn:flux.e.framecorr}, assumes that the gas surrounding the star has initial gas-star relative velocities small compared to the ejecta velocity. A more general expression is presented in Appendix~\ref{sec:energy.cons.w.motion}.}

These fluxes are simply added to each neighbor in a fully-conservative manner: 
\begin{align}
\label{eqn:flux.m.coupling} m_{b}^{\rm new} &= m_{b} + \Delta m_{b}^{\prime} \\ 
\label{eqn:flux.z.coupling} (Z\,m_{b})^{\rm new} &= Z^{\rm new}\,m_{b}^{\rm new} = (Z\,m_{b}) + \Delta m_{Z,\,b}^{\prime} \\ 
\label{eqn:flux.p.coupling} {\bf p}_{b}^{\rm new} &= m_{b}^{\rm new}\,{\bf v}_{b}^{\rm new} = {\bf p}_{b} + \Delta {\bf p}_{b}^{\prime} \\ 
\label{eqn:flux.e.coupling} E_{b}^{\rm new} &= E_{\rm kinetic}^{\rm new} + U_{\rm internal}^{\rm new} = E_{b} + \Delta E_{b}^{\prime}
\end{align}
So the updated vector velocity ${\bf v}$ of the element follows from its updated momentum and mass (and its metallicity follows from its updated metal mass and total mass); the energy $E$ here is a {\em total} energy, so the updated internal energy $U$ of the element follows from its updated total energy ($E$), kinetic energy (from ${\bf v}$), and mass (this is the usual procedure in finite-volume updates with conservative hydrodynamic schemes).

The terms accounting for the relative gas-star motion are necessary to ensure exact conservation. For SNe, they have essentially no effect. However, for slow stellar winds (e.g.\ AGB winds with $v_{\rm wind}\sim 10\,{\rm km\,s^{-1}}$), the relative star-gas velocity can be much larger than the wind velocity ($|{\bf v}_{b} - {\bf v}_{a}| \gg v_{\rm wind}$), which means the shock energy and post-shock temperature of the winds colliding with the ISM is much higher than would be calculated ignoring these terms, which may significantly change their role as a feedback agent \citep{conroy:2014.agb.heating.quenching}.

\vspace{-0.5cm}
\subsection{Sub-Grid Physics: Unresolved Sedov-Taylor Phases}
\label{sec:feedback:mechanical:sedov}

A potential concern if naively applying the above prescription for SNe is that low-resolution simulations are unable to resolve the Sedov-Taylor (S-T) phase, during which the expanding shocked bubble is energy-conserving (the cooling time is long compared to the expansion time) and does $P\,dV$ work on the gas, converting energy into momentum, until it reaches some terminal radius where the residual thermal energy has been lost and the blastwave becomes a cold, momentum-conserving shell. This would, if properly resolved, modify the input momentum ($\Delta p_{b}$) and energy ($\Delta E_{b}$) felt by the gas element $b$.

\vspace{-0.5cm}
\subsubsection{Motivation: Individual SN Remnant Evolution}
\label{sec:feedback:mechanical:sedov:background}

Idealized, high-resolution simulations (with element mass $m_{i}\ll \msun$) have shown that there is a robust radial terminal momentum, $p_{\rm t}$, of the swept-up gas in the momentum-conserving phase, from a single explosion, given by: 
\begin{align}
\label{eqn:terminal.p}\frac{p_{\rm t}}{\msun\,{\rm km\,s^{-1}}} &\approx 4.8\times10^{5}
\left(\frac{E_{\rm ej}}{10^{51}\,{\rm erg}}\right)^{\frac{13}{14}}
\left(\frac{n}{{\rm cm^{-3}}}\right)^{-\frac{1}{7}}
f(Z)^{\frac{3}{2}}\\
\label{eqn:terminal.p.zdep} f(Z) &\approx 
\begin{cases}
	{\displaystyle 2 \, \ \ \ \ \ \ \ \ \ \ \ \ \ \ \ \ \ \ \ \ \ \ \ \hfill { (Z/Z_{\sun}<0.01)}} \\
	{\displaystyle (Z/Z_{\sun})^{-0.14}\ \ \ \ \ \hfill { (0.01 \le Z/Z_{\sun})}} 
\end{cases}
\end{align}
where $n$ and $Z$ are the gas number density and metallicity surrounding the explosion.
The expression above is from \citet{cioffi:1988.sne.remnant.evolution} (where we restrict $f(Z)$ to the minimum metallicity they consider), but similar expressions have been found in a wide range of other studies \citep[for discussion see][]{draine:1991.snr.with.xrays,slavin:snr.expansion,thornton98,martizzi:sne.momentum.sims,walch.naab:sne.momentum,kim.ostriker:sne.momentum.injection.sims,haid:snr.in.clumpy.ism,iffrig:sne.momentum.magnetic.no.effects,hu:photoelectric.heating,li:multi.sne.sims,gentry:clustered.sne.momentum.enhancement}, with variations up to a factor $\sim 2$, which we explore below.\footnote{We adopt the specific expression from \citet{cioffi:1988.sne.remnant.evolution}, as opposed to that from more recent work, for consistency with the previous FIRE-1 simulations.} 

We validate this expression in simulations below. But physically, this follows from simple cooling physics: taking $E_{0}\sim E_{51}\,10^{51}\,{\rm erg}$ and converting an order-unity fraction to thermal energy within a swept-up mass $M$ gives a temperature $T\sim 10^{6}\,K\,(M/3000\,E_{51}\,M_{\sun})^{-1}$, so when $M \gtrsim M_{\rm cool} \sim 3000\,E_{51}\,\msun$, $T$ drops to $< 10^{6}$\,K and the gas moves into the peak of the cooling curve where radiative losses are efficient \citep{rees:1977.tcool.tdyn.vs.mhalo}. While energy-conserving, the shell momentum scales as $p\sim M\,v \sim \sqrt{2\,M\,E_{0}}$, so the terminal momentum is $p_{\rm t} \sim \sqrt{2\,M_{\rm cool}\,E_{0}} \sim 5\times10^{5}\,E_{51}\,\msun\,{\rm km\,s^{-1}}$.

One important caveat: these scalings (and our implementations below) are developed for single ``events'' (e.g.\ explosions), as opposed to continuous events (e.g.\ approximately constant rates of stellar mass-loss over long time periods). ``Continuous'' feedback can, in principle, produce different scalings \citep[see e.g.\ the discussion in][]{weaver:1977.wind.bubble.expansion,mckee:bubble.expansion,freyer:2006.massive.star.wind.egy,cafg:2012.egy.cons.bal.winds,gentry:clustered.sne.momentum.enhancement}. It is still the case that winds must either expand in some energy-conserving fashion (doing $P\,dV$ work) or cool, and so a scaling qualitatively like those here must still apply -- however, details of when cooling occurs (which set the exact ``terminal momentum''), in continuous cases, are much less robust to the environment, density profile, ability of the surrounding medium to confine the wind, and temperature range of the reverse shock (see references above and e.g.\ \citealt{harper.clark:stellar.bubbles.energy.missing,rosen:2014.xray.energy.wind.clusters}). Moreover, there is growing evidence that stellar mass loss is highly ``bursty'' or ``clumpy'' with most of the kinetic luminosity associated with smaller time-or-spatial scale ejection events and/or clumps \citep{2003ApJ...582L..39F,young:2003.clumpy.wind.models,repolust:clumpy.stellar.winds,2007Natur.447.1094Z,2010ApJ...724L.133A,2012A&A...537A..35C}. In those cases, treating each ``event'' with the scalings above is appropriate. Because the kinetic luminosity in stellar mass-loss is an order-of-magnitude lower than that associated with SNe, even relatively large changes in our treatment of stellar mass-loss (e.g.\ assuming the ejecta are entirely radiative, so the terminal momentum is the initial momentum) have little effect on galaxy scales (if SNe are also present). We therefore, for simplicity, apply the same scalings to all mechanical feedback. But this certainly merits more detailed study in future work.

\vspace{-0.5cm}
\subsubsection{Numerical Treatment}
\label{sec:subgrid.terminal.momentum}

To account for potentially unresolved energy-conserving phases, we first calculate the momentum that {\em would} be coupled to the gas element, assuming the blastwave were energy conserving throughout that {\em single} element, which is simply $\Delta p_{b}^{\prime} \rightarrow \Delta p_{b}^{\prime} (1 + m_{b}/\Delta m_{b})^{1/2}$. We then compare this to the terminal momentum $p_{\rm t}$ (assume each neighbor $b$ sees the appropriate ``share'' of the terminal momentum according to its share of the ejecta mass), and assign the actual coupled momentum to be the smaller of the two.\footnote{\citet{kimm.cen:escape.fraction} introduce a smooth interpolation function rather than a simple threshold in Eq.~\ref{eqn:dp.subgrid.sub1}; we have experimented with variations of this and find no detectable effects.} In other words: 
\begin{align}
\label{eqn:dp.subgrid.sub1} \Delta {\bf p}_{b}^{\rm new} &\equiv {\rm MIN}\left[ \Delta {\bf p}_{b}^{\rm energy-conserving}\ ,\ 
 \Delta {\bf p}_{b}^{\rm terminal} \right] \\ 
\label{eqn:dp.subgrid} &= \Delta {\bf p}_{b}\ {\rm MIN}\left[ \sqrt{1 + \frac{m_{b}}{\Delta m_{b}} }\  , 
\ \frac{p_{\rm t}}{p_{\rm ej}} \right]
\end{align}
(where recall $p_{\rm ej} = \sqrt{2\,m_{\rm ej}\,E_{\rm ej}}$). 
Because the coupled $\Delta E$ is the {\em total} energy and is not changed, this remains manifestly energy-conserving (the energy that implicitly goes into the $PdV$ work increasing ${\bf p}$ is automatically moved from thermal to kinetic energy). This is done in the rest frame (before boosting back to the lab frame).

Consider the two limits: (1) when $p_{t}/p_{\rm ej} < (1 + m_{b}/\Delta m_{b})^{1/2}$, the physical statement is that the cooling radius is un-resolved. Because $\Delta {\bf p}_{b} = p_{\rm ej}\,\bar{\bf w}_{b}$, multiplying by $p_{\rm t}/p_{\rm ej}$ simply replaces the ``at explosion'' initial $p_{\rm ej}$ with the terminal $p_{\rm t}$ -- in other words, exactly the momentum that the element $b$ {\em should} see, if we had properly resolved the S-T phase between ${\bf x}_{a}$ and ${\bf x}_{b}$. On the other hand: (2) when $p_{t}/p_{\rm ej} > (1 + m_{b}/\Delta m_{b})^{1/2}$, the cooling radius is resolved; so we simply assume the blastwave is energy-conserving at the location of coupling. Because, by definition, the coupled momentum will be less then $p_{\rm t}$, the actual momentum coupling is, in this limit, largely irrelevant -- we essentially couple thermal energy and rely on the hydrodynamic code to actually {\em solve} for the correct $PdV$ work as the blastwave expands.\footnote{Note that we do not need to make any distinction between the free-expansion radius, post-shock (reverse shock) radius, etc, in our formalism, because the fully-conservative coupling -- which {\em exactly} solves the elastic two-body gas collision between ejecta and gas resolution element -- automatically assigns the correct values in either limit. For example, if $m_{b} \ll m_{\rm ej}$, our coupling will automatically determine that element $b$ should simply be ``swept up'' with velocity ${\bf v}_{b} \approx {\bf v}_{\rm ej}$ (free-expansion); if $m_{b} \gg m_{\rm ej}$, the gas is automatically assigned the appropriate post-shock temperature.} 

{Strictly speaking, the expressions in Eq.~\ref{eqn:dp.subgrid.sub1}-\ref{eqn:dp.subgrid} are expected if the relative gas-star velocities (${\bf v}_{b}-{\bf v}_{a}$) surrounding the explosion are either (a) small or (b) uniform. In Appendix~\ref{sec:energy.cons.w.motion} we present the more exact scalings, as well as the appropriate boost/de-boost corrections for momentum and energy, accounting for arbitrary gas-star motions.}

We show in \S~\ref{sec:feedback:mechanical:tests}, Figs.~\ref{fig:sne.convergence.momentum}-\ref{fig:sne.convergence.energy} that this algorithm reproduces the exact results of much higher-resolution, converged simulations of SN blastwaves even when the coupling is applied in lower-resolution simulations -- just as intended. 

To be fully consistent, we also need to account for the loss of thermal energy (via radiation) in limit (1), when the cooling radius is un-resolved. The effective cooling radius $R_{\rm cool}$ is exactly determined by the expression for $p_{\rm t}$, because at the end of the energy-conserving phase ($R_{\rm shock} = R_{\rm cool}$), $(1/2)\,(m_{\rm ej} + m_{\rm swept}[R_{\rm cool}])\,v_{f}^{2} = (1/2)\,m_{\rm ej}\,v_{\rm ej}^{2}$ and $p_{\rm t} = m_{\rm swept}[R_{\rm cool}]\,v_{f}$, giving $R_{\rm cool} \approx 28.4\,{\rm pc}\,(n/{\rm cm^{-3}})^{-3/7}\,(E_{\rm ej}/10^{51}\,{\rm erg})^{2/7}\,f(Z)$ for $p_{\rm t}$ in Eq.~\ref{eqn:terminal.p}. Following \citet{thornton98}, the post-shock thermal energy outside $R_{\rm cool}$ decays $\propto (r/R_{\rm cool})^{-6.5}$, so we first calculate the post-shock thermal energy of element $b$ that would be added by the ejecta, $\Delta U_{b} = E_{\rm ej} - \Delta {\rm KE}$ (where $\Delta {\rm KE}$ is the change in kinetic energy, i.e.\ based on the coupled energy and momentum) in our usual fully-conservative manner, then if $r_{b} \equiv |{\bf x}_{ba}| > R_{\rm cool}$ we reduce this accordingly: $\Delta U_{b} \rightarrow \Delta U_{b}\,(|{\bf x}_{ba}|/R_{\rm cool})^{-6.5}$. In practice, because {\em by definition} this correction to $\Delta U_{b}$ only appears outside the cooling radius (where the post-shock cooling time is short compared to the expansion time), we find that the inclusion/exclusion of this correction term has no detectable effects on our simulations (see \S~\ref{sec:feedback:mechanical:tests:effects}); if we do not include it, the thermal energy is simply radiated away in the next timestep, as it should be. Still, we include the term for consistency. 

We can (and do, for the sake of consistency) apply the full treatment described above to continuous stellar mass loss as well as SNe, using the differential $E_{\rm ej}$ (and enforcing $p_{\rm t} \ge p_{\rm ej}$), but the ``multiplier'' is small because the winds are injected continuously so the $E_{\rm ej}$ in a single timestep is small.

Finally, the calculations of $p_{t}$ and Eqs.~\ref{eqn:dp.subgrid.sub1}-\ref{eqn:dp.subgrid} are done independently for each neighbor $b$. In effect, we are considering each solid angle face $\Delta\Omega_{b}$ to be an independent ``cone'' with its own density and metallicity, in which an independent energy-conserving solution is considered. \citet{haid:snr.in.clumpy.ism} have performed a detailed simulation study of SNe in inhomogeneous environments and showed explicitly that almost all of the (already weak) effect of different inhomogeneous initial conditions (in e.g.\ turbulent, clumpy, multi-phase media) in their study and others is properly captured by considering each element surrounding the SN as an independent cone, which is assigned its own density-dependent solution according to the single homogeneous scaling above. In fact, once the density and metallicity dependence are accounted for as we do, residual systematic uncertainties in Eq.~\ref{eqn:terminal.p} are remarkably small ($\sim 10-50\%$) -- much smaller than uncertainties in the SNe rate itself!

\begin{figure}
\begin{tabular}{c}
    \vspace{-0.15cm}
\includegraphics[width=0.98\columnwidth]{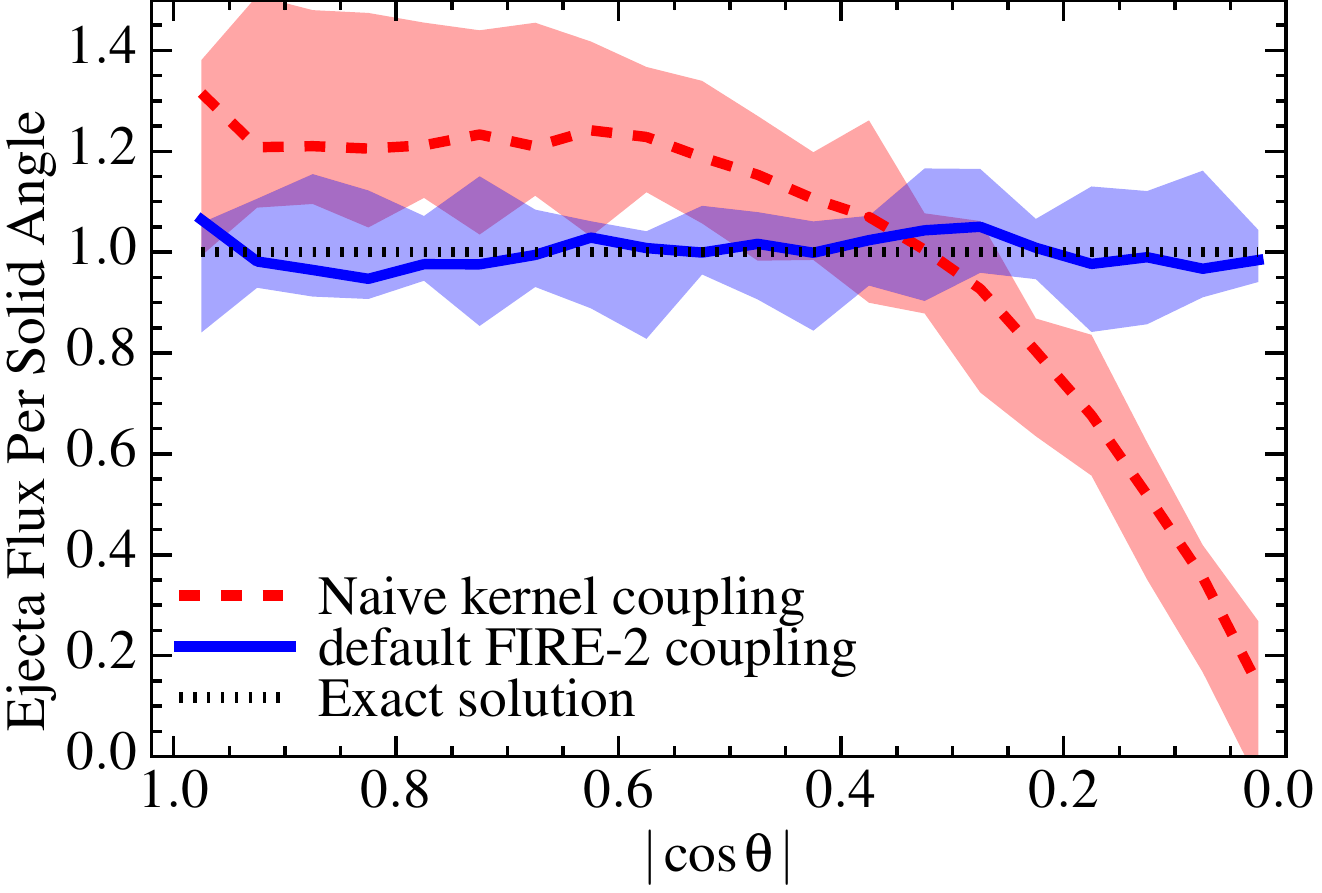} \\
    \hspace{-0.15cm}
\includegraphics[width=0.99\columnwidth]{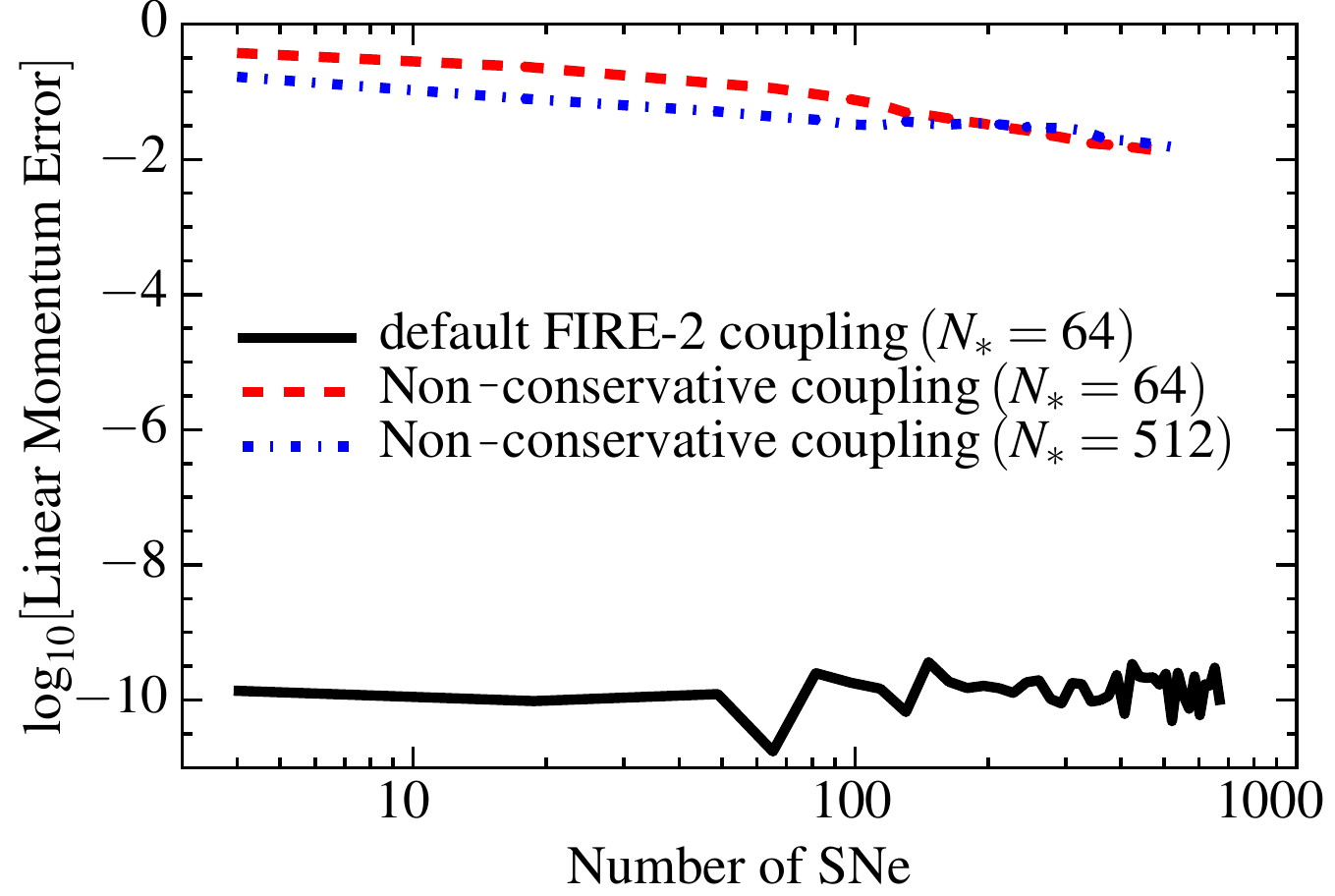}
\end{tabular}
    \vspace{-0.25cm}
    \caption{Numerical tests of the mechanical feedback coupling algorithm in \S~\ref{sec:feedback:mechanical:tests}.
    {\em Top:} Statistical Isotropy Test: We detonate a single SN in the center of a thin disk, generated by randomly sampling a vertical Gaussian profile with gas particles, and we measure the resulting momentum/mass/metal flux deposited to neighbor gas elements as a function of polar angle $\theta$ ($\cos{\theta}=\pm1$ is midplane, $\cos{\theta}=0$ polar). We repeat $100$ times and show the median ({\em lines}) and $95\%$ interval ({\em shaded}). The result {\em should} be statistically isotropic (``Exact Solution''). Our ``Default FIRE-2 Coupling'' method (Fig.~\ref{fig:mechanical.fb.cartoon}) recovers this with noise owing to the finite number of particles coupled. The ``Naive Kernel Coupling'' model only includes neighbors within the search radius $H_{\ast}$ of the SN (not those where $H_{\ast} < |{\bf x}_{\ast}-{\bf x}_{\rm gas}| < H_{\rm gas}$; \S~\ref{sec:feedback:mechanical:neighbor.finding}) and weights deposition by a simple kernel function (effectively mass-weighting) instead of the solid angle subtended by the element (\S~\ref{sec:feedback:mechanical:weighting}). This naive coupling biases ejecta to couple into the midplane and suppresses coupling in the polar direction.
    {\em Bottom:} Momentum Conservation Test: We detonate SNe at random locations in the same system and measure the total fractional error in the linear momentum of the box (error $L_{1}(t) \equiv | \sum_{a} m_{a}\,{\bf v}_{a}(t) | / \sum p_{\rm coupled}$, where $p_{\rm coupled} = N(t)\,p_{\rm ej}$ is the total magnitude of the momentum injected by all events). This is the net deviation from exact momentum conservation, relative to the total coupled. Our ``Default FIRE-2 Coupling'' uses a tensor re-normalization scheme to keep these errors at machine accuracy (see \S~\ref{sec:feedback:mechanical:vector}). The ``Non-Conservative Coupling'' scheme removes this re-normalization (but is otherwise identical); fractional conservation errors for a single event can then be order-unity! The fractional error declines with SNe number as $\sim N_{\rm SNe}^{-1/2}$ because of cancellations; increasing the coupled neighbor number $N_{\ast}$ reduces the errors but inefficiently.
    \vspace{-0.4cm}
    \label{fig:sne.fb.coupling.tests}}
\end{figure}

\begin{figure}
\plotonesize{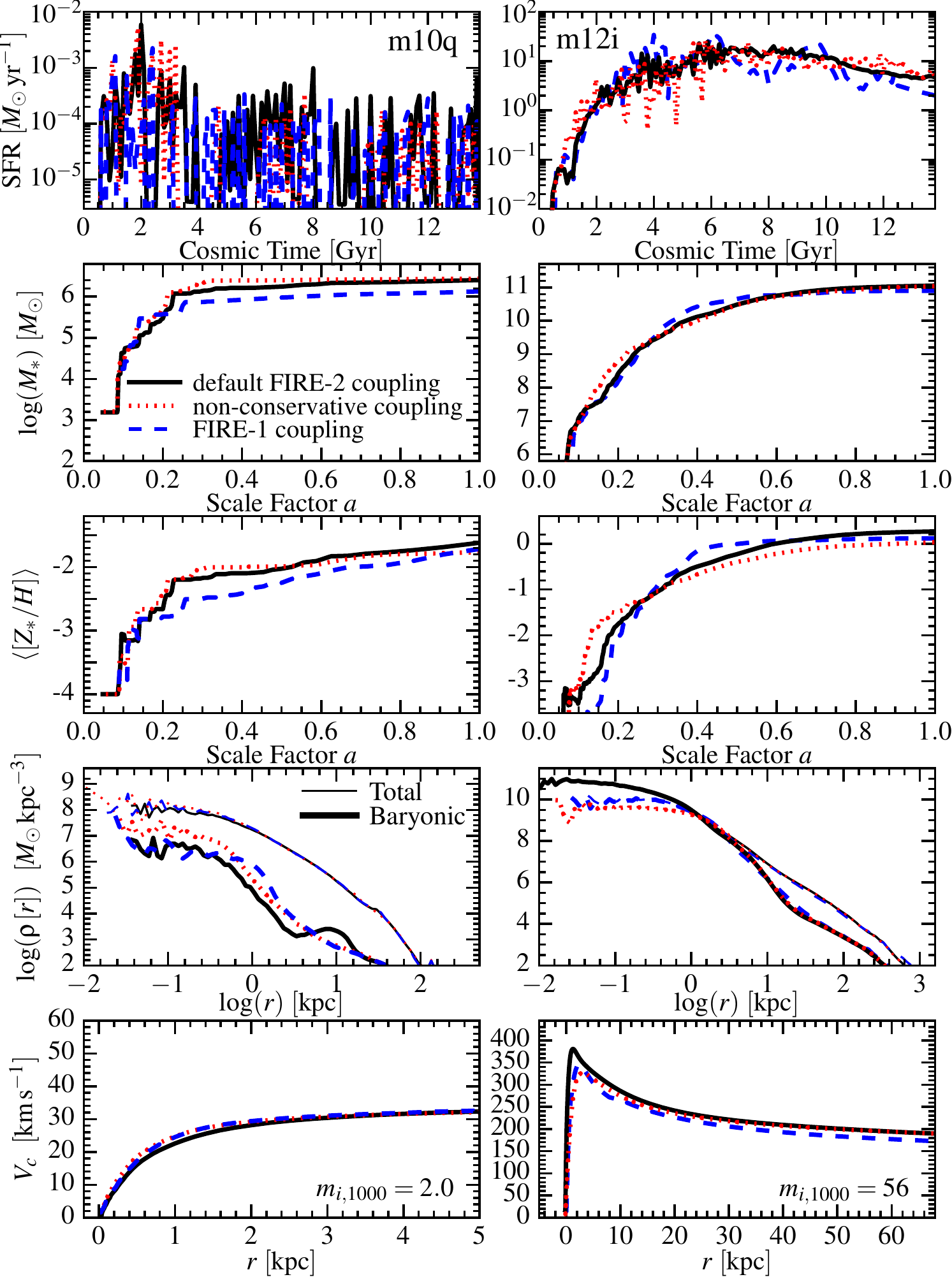}{0.99}
    \vspace{-0.25cm}
    \caption{Comparison of different algorithms for coupling the same mechanical feedback to gas, in zoom-in cosmological simulations. {\em Top:} Star formation history (averaged in $100\,$Myr intervals) of the primary $z=0$ galaxy. 
     {\em Second:} Total stellar mass in box (dominated by primary galaxy) vs.\ scale factor $a=1/(1+z)$. 
     {\em Middle:} Stellar mass-weighted average metallicity vs.\ scale factor.
     {\em Third:} Baryonic ({\em thick}) and total ({\em thin}) mass density profiles (averaged in spherical shells) as a function of radius around the primary galaxy at $z=0$. 
     {\em Bottom:} Rotation curves (circular velocity $V_{c}$ versus radius) in the primary galaxy. 
     In each panel we compare three models: 
    {\bf (1)} {\em Default:} Our most accurate, fully-conservative algorithm (\S~\ref{sec:feedback:mechanical}). Ejecta are deposited in neighboring gas elements weighted by the solid angle subtended ``as seen by'' the SN, guaranteeing mass, energy, momentum conservation, and statistical isotropy.
    {\bf (2)} {\em Non-conservative coupling:} ``Naive'' coupling with momentum deposited along the vector connecting the center-of-mass from star to neighbor gas element, without tensor renormalization (\S~\ref{sec:feedback:mechanical:vector}). While simple, this algorithm can violate linear momentum conservation (imparting {\em net} momentum to the gas). 
    {\bf (3)} {\em FIRE-1 coupling:} SNe algorithm from FIRE-1: it used the non-conservative method (\S~\ref{sec:feedback:mechanical:vector}), a less-accurate SPH approximation of the solid angle (essentially a volume-weighting, $\omega_{b}\propto m_{b}/\rho_{b}$; \S~\ref{sec:feedback:mechanical:weighting}), and only used the nearest neighbors for each SN instead of the bi-directional search needed to ensure statistical isotropy (\S~\ref{sec:feedback:mechanical:neighbor.finding}). 
    Despite these algorithmic differences, results at this resolution for dwarfs and massive galaxies are similar.
    \vspace{-0.4cm}
    \label{fig:sf.history.sne.algorithm}}
\end{figure}
%
%

\begin{figure*}
\begin{tabular}{cccc}
\vspace{-0.1cm}
\hspace{-0.20cm}
\includegraphics[width=0.49\columnwidth]{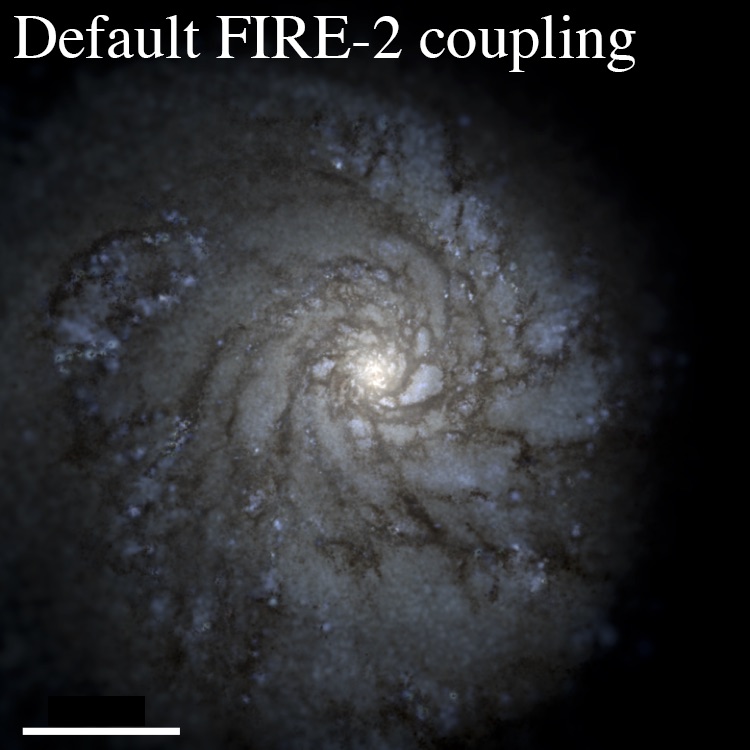} &
\hspace{-0.60cm}
\includegraphics[width=0.49\columnwidth]{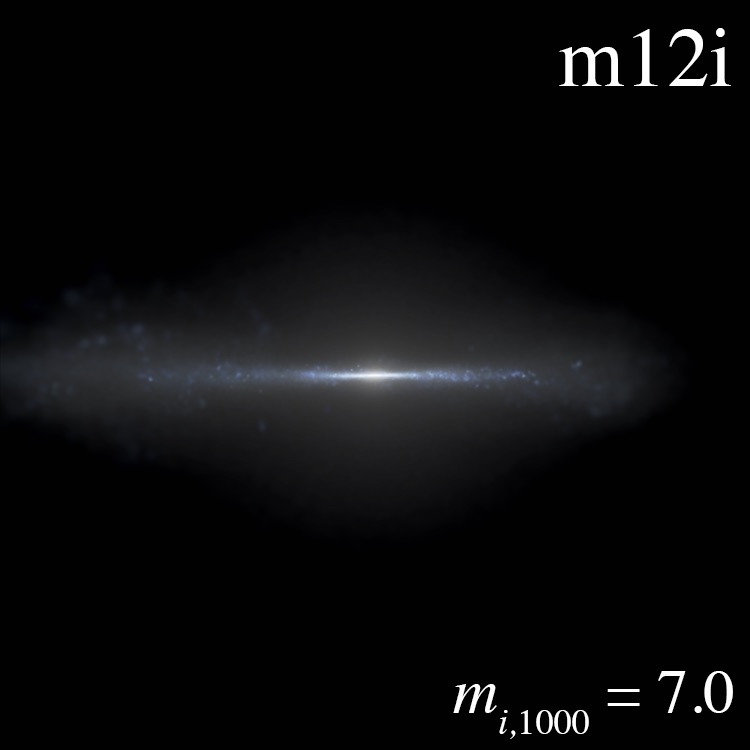} &
\hspace{-0.20cm}
\includegraphics[width=0.49\columnwidth]{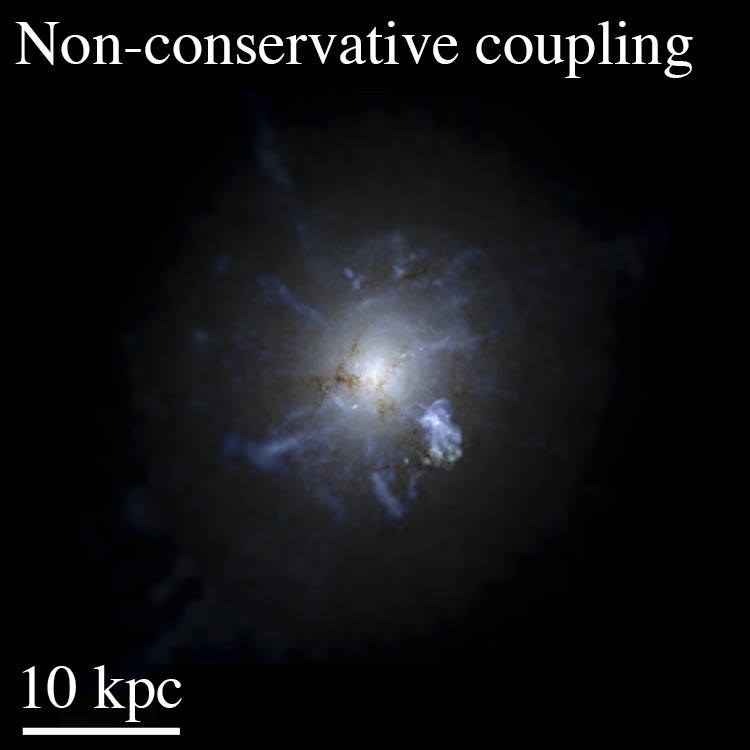} &
\hspace{-0.50cm}
\includegraphics[width=0.49\columnwidth]{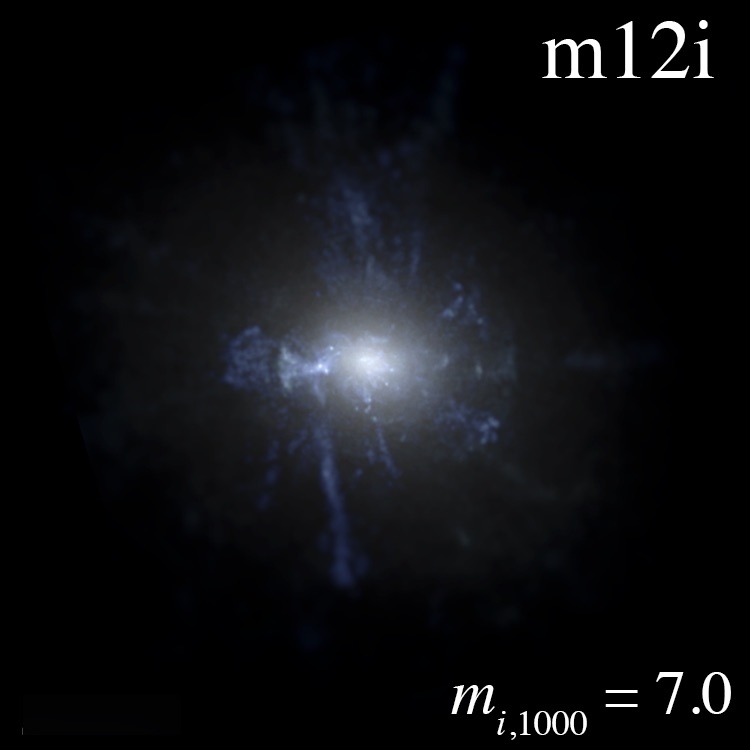} \\
\vspace{-0.09cm}
\hspace{-0.20cm}
\includegraphics[width=0.49\columnwidth]{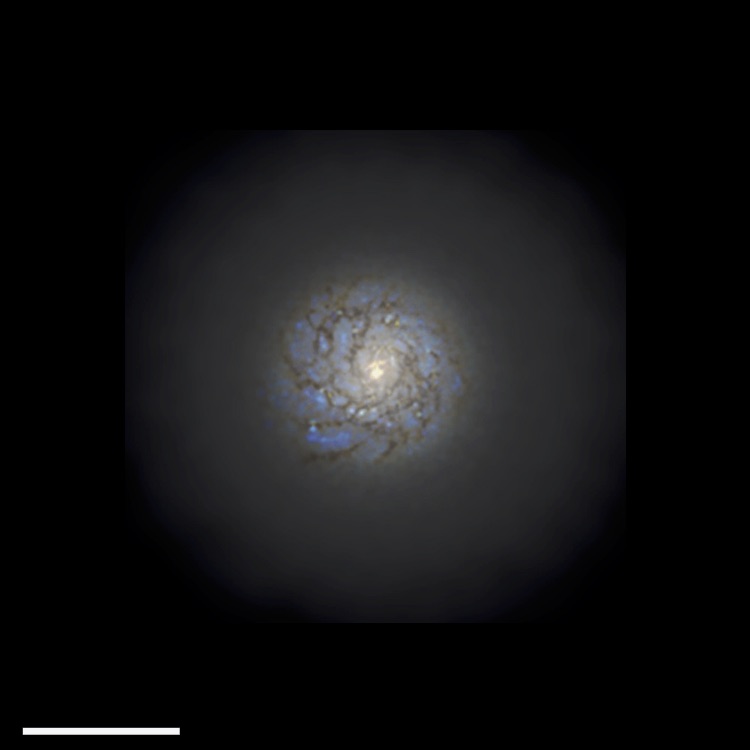} &
\hspace{-0.60cm}
\includegraphics[width=0.49\columnwidth]{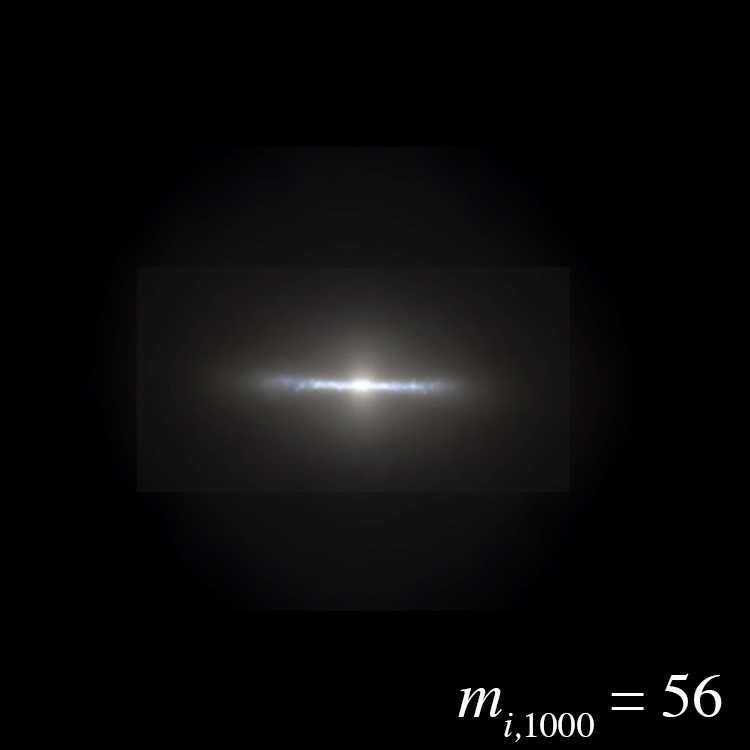} &
\hspace{-0.20cm}
\includegraphics[width=0.49\columnwidth]{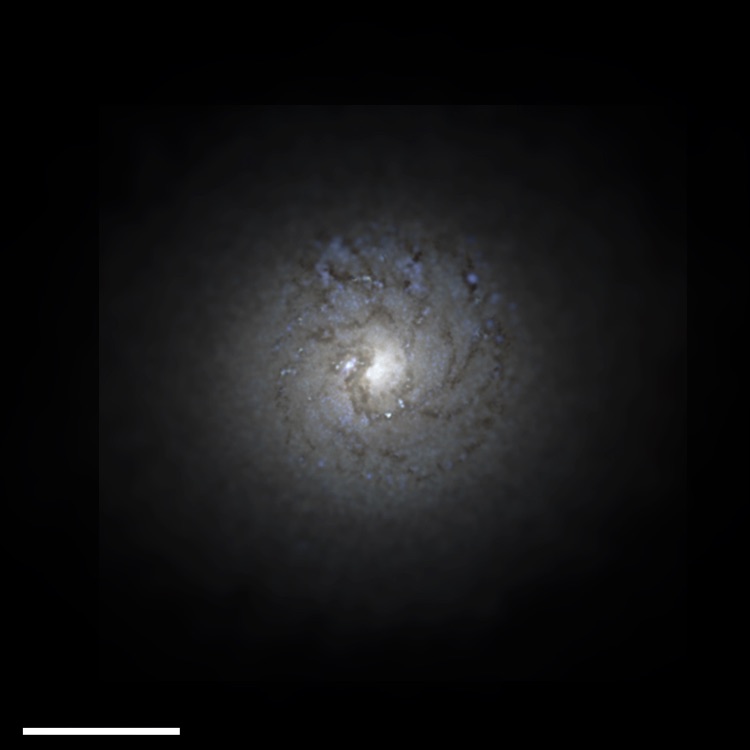} &
\hspace{-0.50cm}
\includegraphics[width=0.49\columnwidth]{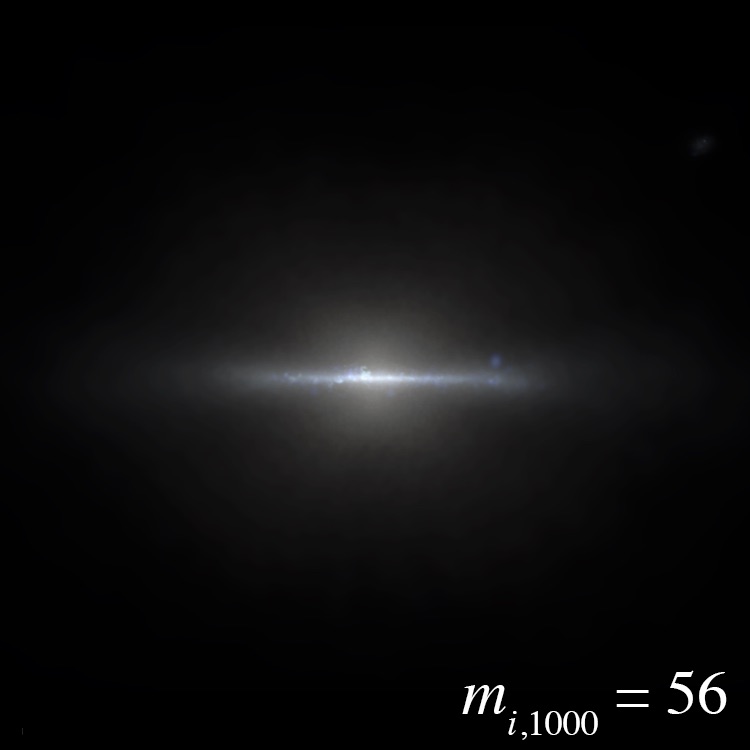} \\
\end{tabular}
    \vspace{-0.1cm}
    \caption{Mock images in HST bands of our {\bf m12i} run at $z=0$ as a function of mass resolution but for the ``default'' ({\em left}) and ``non-conservative'' ({\em right}) SNe coupling algorithm from Fig.~\ref{fig:sf.history.sne.algorithm}. In the non-conservative algorithm, the error in {\em momentum}  is independent of resolution, so the (worst-case) error in {\em velocity} scales as $|\Delta {\bf v}_{\rm err}|\sim 100\,{\rm km\,s^{-1}}\,(7000/m_{i,\,1000})$. At low resolution ($m_{i,\,1000}=56$ shown, or $m_{i,\,1000}=450$ in Fig.~\ref{fig:sf.history.sne.algorithm}) the maximum velocity error is $\sim 10\,{\rm km\,s^{-1}}$ (comparable to random motions in the ISM), so the galaxy is well-behaved. However, at high resolution $|\Delta {\bf v}_{\rm err}|\sim 100\,{\rm km\,s^{-1}}$, so the errors blow up the galaxy! Specifically, some young star-forming regions accumulate a coherent error in momentum conservation and are launched out of the galaxy, generating the visible streams and preventing thin-disk formation. In our default coupling these errors vanish, so the galaxy is well behaved: at higher resolution the disk is even thinner and more extended.
    \vspace{-0.5cm}
    \label{fig:images.resolution.nonsymmetric}}
\end{figure*}

\vspace{-0.5cm}
\subsubsection{Implied Resolution Requirements}
\label{sec:feedback:mechanical:sedov:resolution}

Eq.~\ref{eqn:dp.subgrid} demonstrates that with sufficiently small element mass ($m_{b}$ below some critical $m_{\rm crit}$), the cooling radius is resolved -- i.e.\ we are in limit (2) above: $(1 + m_{b}/\Delta m_{b})^{1/2} < p_{t}/p_{\rm ej}$. This corresponds to $m_{b} 
\ll m_{\rm crit} = 2500\,\msun\,|\bar{\bf w}|\,E_{51}^{6/7}\,(n/{\rm cm^{-3}})^{-2/7}\,f(Z)^{3}$. Because the kernel function is strongly peaked, most of ejecta energy/momentum/mass is deposited in the nearest few elements, so $|\bar{\bf w}|\sim 1/$few; so $m_{\rm crit} \sim 500\,\msun\,(n/{\rm cm^{-3}})^{-2/7}$.
This is a {\em mass} resolution criterion: as noted above, the cooling {\em radius} depends on density, $R_{\rm cool} \sim n^{-1/3}$, such that an almost invariant mass $M_{\rm cool} \equiv m_{\rm swept}(R_{\rm cool}) \sim ({\rm a\ few})\,m_{\rm crit}$ is enclosed inside $R_{\rm cool}$.

Similar results are found in \citet{hu:photoelectric.heating} (their Appendix~B): they show, for example, that with $\sim 100\,\msun$ resolution, the blastwave momentum is almost perfectly recovered (within $<10\%$ of simulations with element/particle masses $\sim 0.01\,\msun$). Even higher-order effects such as the blastwave mass-loading, velocity structure, shell position, etc, are recovered almost perfectly once the shell has propagated into the momentum-conserving phase.

In our cosmological simulations of isolated dwarf galaxies, we begin to satisfy $m_{b} < m_{\rm crit}$. However, in MW-mass simulations, this remains unattainable for now. Therefore, ignoring the correction for an unresolved S-T phase in massive galaxies can significantly under-estimate the effects of feedback. We consider explicit resolution tests below which validate these approximate scalings.

\vspace{-0.5cm}
\section{Numerical Tests: The Coupling Algorithm}
\label{sec:feedback:mechanical:ideal.tests}

We now consider detailed numerical tests of the SNe coupling scheme. Specifically, we first consider tests of the pure algorithm used to deposit feedback from \S~\ref{sec:feedback:mechanical:neighbor.finding}-\ref{sec:feedback:mechanical:assignment}, independent of the feedback physics (energy, momentum, rates, etc.).

\vspace{-0.5cm}
\subsection{Validation: Ensuring Correct Coupling Isotropy, Weights, and Exact Conservation}
\label{sec:feedback:mechanical:ideal.tests:validation}

Fig.~\ref{fig:sne.fb.coupling.tests} considers two simple validation tests (for conservation and statistical isotropy) of our algorithm in a pure hydrodynamic test problem.
We initialize a periodic box of arbitrarily large size centered on ${\bf x}=\mathbf{0}$, filled with particles of equal mass, $m$, meant to represent a patch of a vertically-stratified disk. There is no gravity and the gas is forced to obey an exactly isothermal equation of state with vanishingly small pressure. The particles are laid down randomly with a uniform probability distribution in the $x$ and $y$ dimensions and probability $dp$ along the $z$ dimension such that $dp \propto  m^{-1}\langle\rho(z)\,\rangle \,dx\,dy\,dz$, where $\langle \rho(z) \rangle = \exp{(-z^{2}/2h^{2})}$. Initial velocities are zero. We define $m$ and code units such that $h$ is equal to the mean inter-particle separation in the midplane. The desired density distribution is therefore obeyed on average but with a noisy particle distribution, as in a real simulation.

The top panel of Fig.~\ref{fig:sne.fb.coupling.tests} shows the results after a single SN detonated at the center of the box, using the standard FIRE-2 coupling scheme to deposit its ejecta. Because of the enforced equation-of-state, the coupled thermal energy is instantly dissipated -- all that is retained is momentum, mass, and metals. We measure the amount deposited in each direction -- each unit solid angle ``as seen by'' the SN. By construction, our algorithm is {\em supposed} to couple the ejecta statistically isotropically.
But because the ejecta must be deposited discretely in a finite number of neighbors, in any single explosion the deposition is noisy: it occurs only along the directions where there are neighbors. We therefore re-generate the box and repeat $100$ times, and plot the resulting mean distribution and scatter. We confirm that our default algorithm correctly deposits ejecta statistically isotropically, on average. However, if we instead consider a simpler algorithm where the search for neighbors to couple the SN (\S~\ref{sec:feedback:mechanical:neighbor.finding}) is done only using particles within a nearest-neighbor radius $H_{a}$ of the SN (excluding particles outside $H_{a}$ but for which the SN is inside {\em their} nearest-neighbor radius $H_{b}$), {\em or} if we weight the deposition ``per neighbor'' by a simple kernel weight (\S~\ref{sec:feedback:mechanical:weighting}), in this case the cubic spline kernel ($\omega_{b} = W({\bf x}_{ba},\,H_{a})$); then we obtain a biased ejecta distribution. The bias is as expected: most of the ejecta go into the disk midplane direction, because on average there are more particles in this direction, and they are closer, as opposed to in the vertical direction, where the density decreases. In a real simulation, this is a serious concern: momentum and energy would be preferentially coupled in the plane of the galaxy disk, rather than ``venting'' in the vertical direction as they should, simply because more particles are in the disk!

In the bottom panel of Fig.~\ref{fig:sne.fb.coupling.tests}, we repeat our setup, but now we repeatedly detonate SNe throughout the box at fixed time intervals, each in a random position. After each SN we measure the total momentum of all gas elements, $|{\bf p}| \equiv | \sum m_{a}\,{\bf v}_{a} |$, and define the dimensionless, fractional linear momentum error as the ratio of this to the total ejecta momentum that has been injected, $L_{1}=|{\bf p}| / \sum p_{\rm ej}$. Linear momentum conservation demands ${\bf p} = {\bf 0}$. In our standard FIRE-2 algorithm, we confirm momentum is conserved to machine accuracy. However, re-running without the tensor renormalization in \S~\ref{sec:feedback:mechanical:vector}, we see quite large errors, with $L_{1}\sim 0.1-1$ for a single SN, decreasing slowly with the number of SNe in the box only because the errors add incoherently (so $L_{1}$ gradually decreases $\propto N_{\rm SNe}^{-1/2}$ as a Poisson process). We can decrease $L_{1}$ in the non-conservative algorithm by increasing the number of gas neighbors used for the SN deposition, but this is inefficient and reduces the spatial resolution.

\vspace{-0.5cm}
\subsection{Tests in FIRE Simulations: Effects of Algorithmic SNe Coupling}
\label{sec:feedback:mechanical:ideal.tests:firesims}

In Figs.~\ref{fig:sf.history.sne.algorithm}-\ref{fig:images.resolution.nonsymmetric},\footnote{Mock images in Fig.~\ref{fig:images.resolution.nonsymmetric} are computed as $ugr$ composites, ray-tracing from each star after using its age and metallicity to determine the intrinsic spectrum from \citet{starburst99} and accounting for line-of-sight dust extinction with a MW-like extinction curve and dust-to-metals ratio following \citet{hopkins:lifetimes.letter}.} we examine how the algorithmic choices discussed above alter the formation history of galaxies in cosmological simulations. We compare:

\begin{enumerate}

\item{{\em Default}: Our default FIRE-2 coupling. This manifestly conserves mass, energy, and momentum; correctly deposits the ejecta in an unbiased (statistically isotropic) manner; and accounts for the Lagrangian distribution of particles in all directions.}

\item{{\em Non-conservative:} Coupling that neglects the tensor correction from \S~\ref{sec:feedback:mechanical:vector}, which Fig.~\ref{fig:sne.fb.coupling.tests} showed was necessary to maintain exact momentum conservation. We stress that the scalar mass and energy from SNe are still manifestly conserved here; only vector momentum is imperfectly added.}

\item{{\em FIRE-1 Coupling:} Our older scheme from FIRE-1, which used the non-conservative formulation, conducted the SNe neighbor search only ``one-directionally'' (ignoring neighbors with at distances $>H_{a}$), as defined in \S~\ref{sec:feedback:mechanical:neighbor.finding}, and scaled the deposition ``weights'' $\omega_{b}$ defined in \S~\ref{sec:feedback:mechanical:weighting} with volume ($\omega_{b} \propto m_{b}/\rho_{b}$; the ``SPH-like'' weighting; see \citealt{price:2012.sph.review}), as opposed to solid angle. Fig.~\ref{fig:sne.fb.coupling.tests} shows this leads to unphysically anisotropic momentum deposition.}

\end{enumerate}

\begin{figure*}
\plotsidesize{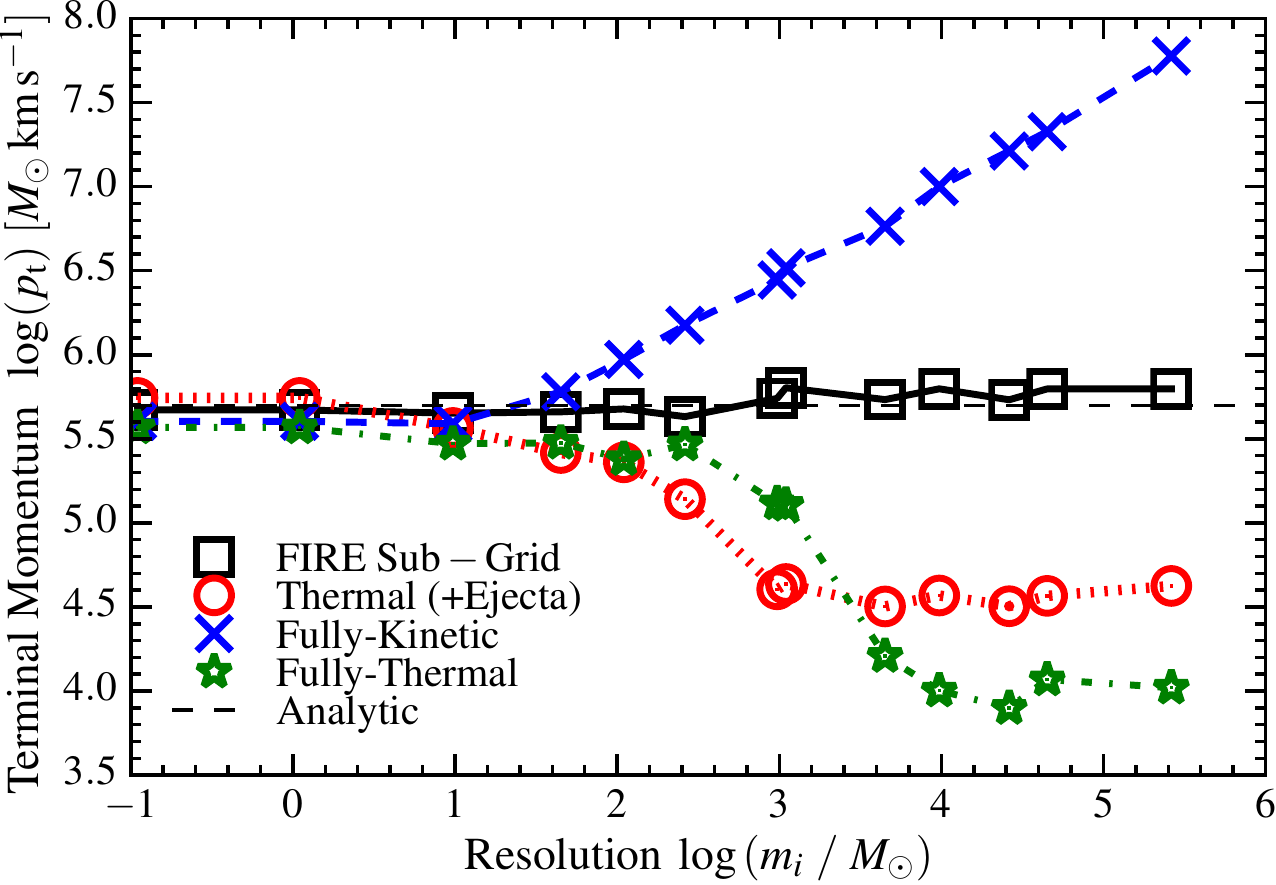}{0.8}
    \vspace{-0.25cm}
    \caption{Convergence of ``sub-grid'' implementations of SNe feedback (\S~\ref{sec:feedback:mechanical:sedov}). We explode a single SN (ejecta mass $=10.4\,\msun$, kinetic energy $=10^{51}\,{\rm erg}$, yields in \paperone\ Appendix~A) in a box of uniform density and metallicity ($n=1\,{\rm cm^{-3}}$, $Z=Z_{\sun}$), with the same cooling physics as our standard FIRE-2 simulations, using varying gas particle mass resolution, across $m_{i} = 0.1 - 10^{6}\,\msun$, as shown on the x-axis. Once the blastwave is well into the momentum-conserving stage, we measure the terminal gas momentum $p_{t}$. We compare: 
    {\bf (1)} {\em Analytic:} the analytic result for $p_{t}$ (Eq.~\ref{eqn:terminal.p}). 
    {\bf (2)} {\em FIRE Sub-Grid:} Our default implementation. This couples SN mass, metals, energy, and momentum, in a {\em manifestly} conservative, statistically isotropic manner, with the coupled momentum following Eq.~\ref{eqn:dp.subgrid}. 
    {\bf (3)} {\em Thermal (+Ejecta):} The coupling algorithm is the same, but the coupled momentum is {\em only} the original ejecta momentum (no $PdV$ work is accounted for) and the energy is always the initial $10^{51}\,{\rm erg}$ at coupling (no un-resolved radiation assumed). At low resolution this means the energy coupled is almost entirely thermal.
    {\bf (4)} {\em Fully-Kinetic:} We couple $100\%$ of the $10^{51}\,{\rm erg}$ as kinetic energy (all in momentum), regardless of resolution. 
    {\bf (5)} {\em Thermal Only:} We couple $100\%$ of the $10^{51}\,{\rm erg}$ as thermal energy (none in momentum). 
    At $m_{i}\ll 10\,\msun$, the ejecta free-expansion phase is resolved and all methods produce an identical, well-resolved Sedov Taylor phase and terminal momentum in excellent agreement with the analytic value. At $m_{i} \gtrsim 100\,\msun$, the cooling radius becomes un-resolved. At this low resolution, ``Thermal (+Ejecta)'' (``Fully-Thermal'') under-estimates the terminal momentum by a factor $\sim 15$ ($\sim 60$), because the $PdV$ work done in the energy-conserving phase is missed (``Thermal (+Ejecta)'' simply returns the original ejecta momentum; ``Fully-Thermal'' produces a small residual). At low resolution ``Fully-Kinetic'' over-estimates the terminal momentum by a factor $\sim 100\,(m_{i} / 10^{5}\,\msun)^{1/2}$, because it assumes that $PdV$ work continues well after the remnant should cool. Our FIRE sub-grid model, by construction, agrees within $\sim 10\%$ with the exact/high-resolution solution, {\em independent of resolution}.
    \vspace{-0.4cm}
    \label{fig:sne.convergence.momentum}}
\end{figure*}

Fig.~\ref{fig:sf.history.sne.algorithm} ({\em left}) shows that the detailed choice of coupling algorithm has essentially no effect in dwarf galaxies, because of their stochastic, bursty star formation and outflows and irregular/spheroidal morphologies.
That is, a ``galaxy wide explosion'' remains such regardless of exactly how individual SN are deposited.
Indeed, we find that this independence from the coupling algorithm persists at any resolution that we test.
We do not show visual morphologies of dwarf galaxies in Fig.~\ref{fig:images.resolution.nonsymmetric}, because they are essentially the same in all cases (see also \paperone). For MW-mass halos, we find only a weak dependence of galaxy properties in Fig.~\ref{fig:sf.history.sne.algorithm} on the SNe algorithm (see Appendix~\ref{sec:resolution} for demonstration of this at various resolution levels). The non-conservative implementations generally show a lower central stellar density at $<1\,$kpc, owing to burstier intermediate-redshift star formation, because the momentum conservation errors allow more ``kicking out'' of material in dense regions, as discussed further below.

At low and intermediate resolution, the MW-mass simulations all exhibit ``normal'' disky visual morphologies, without strong dependence on the SNe algorithm.
However, at high resolution the ``non-conservative'' run essentially destroys its disk! This is in striking contrast to the ``default'' run, where the disk continues to become thinner and more extended at higher resolution (a trend seen in several MW-mass halos studied in \paperone).
Note that the formation history and mass profile are not dramatically different in the two runs, so what has ``gone wrong'' in the non-conservative case? The problem is, as noted in \S~\ref{sec:feedback:mechanical:vector}, the momentum conservation error in the non-conservative algorithm is zeroth-order -- it depends only on the spatial distribution of and number of neighbor gas elements within the kernel, not on the absolute mass/spatial scale of that kernel. Because we keep the number of neighbors seen by the SN fixed with changing mass resolution, this means that the fractional errors (i.e.\ the net linear momentum error deposited per SN) does not converge away. Meanwhile, the individual gas element masses get smaller at high resolution -- so the net linear velocity ``kick'' becomes larger. The ``worst-case'' error for a single SN would be an order-unity fractional violation of momentum conservation, implying a kick $|\Delta {\bf v}_{\rm err}| \sim p_{t}/m_{a} \sim 100\,{\rm km\,s^{-1}}\,(7000/m_{i,\,1000})$; at low and intermediate resolution even this worst-case gives $|\Delta {\bf v}_{\rm err}|\lesssim 10\,{\rm km\,s^{-1}}$ (comparable to the thin-disk velocity dispersion) so this is not a serious issue. But at our highest resolution, the non-conservative ``worst case scenario'' occurs where in some star-forming regions, net momentum is coherently deposited all in one direction owing to a pathological local particle distribution: the cloud then coherently ``self-ejects'' or ``bootstraps'' itself out of the disk. The thin disk is destroyed in the process, and the most extreme examples of this are visibly evident as ``streaks'' of stars from self-ejected clumps flying out of the galaxy center!

We also re-ran a ``non-conservative'' simulation of {\bf m12i} at high resolution ($m_{i,\,1000}=7.0$) with a crude ``cap'' or upper limit arbitrarily imposed for the fraction of the momentum allowed to couple to any one particle, and to the maximum velocity change per event (of $50\,{\rm km\,s^{-1}}$). This is presented in Appendix~\ref{sec:non.con.extreme}. In that case, the system does indeed form a thin, extended disk, similar to our default coupling. This confirms that the ``self-destruction'' of the disk is driven by rare cases with large momentum errors, rather than small errors in ``typical'' cases. 

As noted above, our older FIRE-1 algorithm used the ``non-conservative'' formulation. The MW-mass simulations published with that algorithm were all lower-resolution, where $|\Delta {\bf v}_{\rm err}|\lesssim 10\,{\rm km\,s^{-1}}$, so these errors were not obvious (at dwarf masses, the lower metallicities and densities meant the cooling radii of blastwaves were explicitly resolved, so as Fig.~\ref{fig:sf.history.sne.algorithm} shows, the effects were even smaller, and their irregular morphologies meant perturbations to thin disks were not possible). However, running that algorithm in MW-mass halos at higher resolution led to similar errors as shown in Fig.~\ref{fig:images.resolution.nonsymmetric}. This, in fact, motivated the development of the new FIRE-2 algorithm. 

We have confirmed that all of the conclusions above are not unique to the two halos above: we have re-run halos {\bf m09} ($\sim 10^{9}\,\msun$), {\bf m10v} ($\sim 10^{10}\,\msun$), {\bf m11q}, {\bf m11v} ($\sim 10^{11}\,\msun$), {\bf m12f} and {\bf m12m} ($\sim 10^{12}\,\msun$) from \paperone\ with ``Default'' and ``Non-conservative'' implementations. All halos $\sim 10^{9}-10^{11}\,\msun$ show the same {\em lack} of effect from the coupling scheme as our {\bf m10q} run here; the $\sim 10^{12}\,\msun$ halos all show the same systematic dependencies as our {\bf m12i} run. 

In Appendix~\ref{sec:grid} we briefly discuss algorithms that ensure manifest momentum conservation by simply coupling a pre-determined momentum in the Cartesian $\pm x$, $\pm y$, $\pm z$ directions (independent of the local mesh or particle geometry). We do not adopt such a method because (a) it ignores the physically correct geometry of the mesh in irregular-mesh or mesh-free methods, and (b) it imprints preferred directions onto the simulation, which forces disks to align with the simulation coordinate axes, introducing spurious numerical torques that can significantly reduce disk angular momentum (as often seen in grid-based codes).

\begin{figure}
\plotonesize{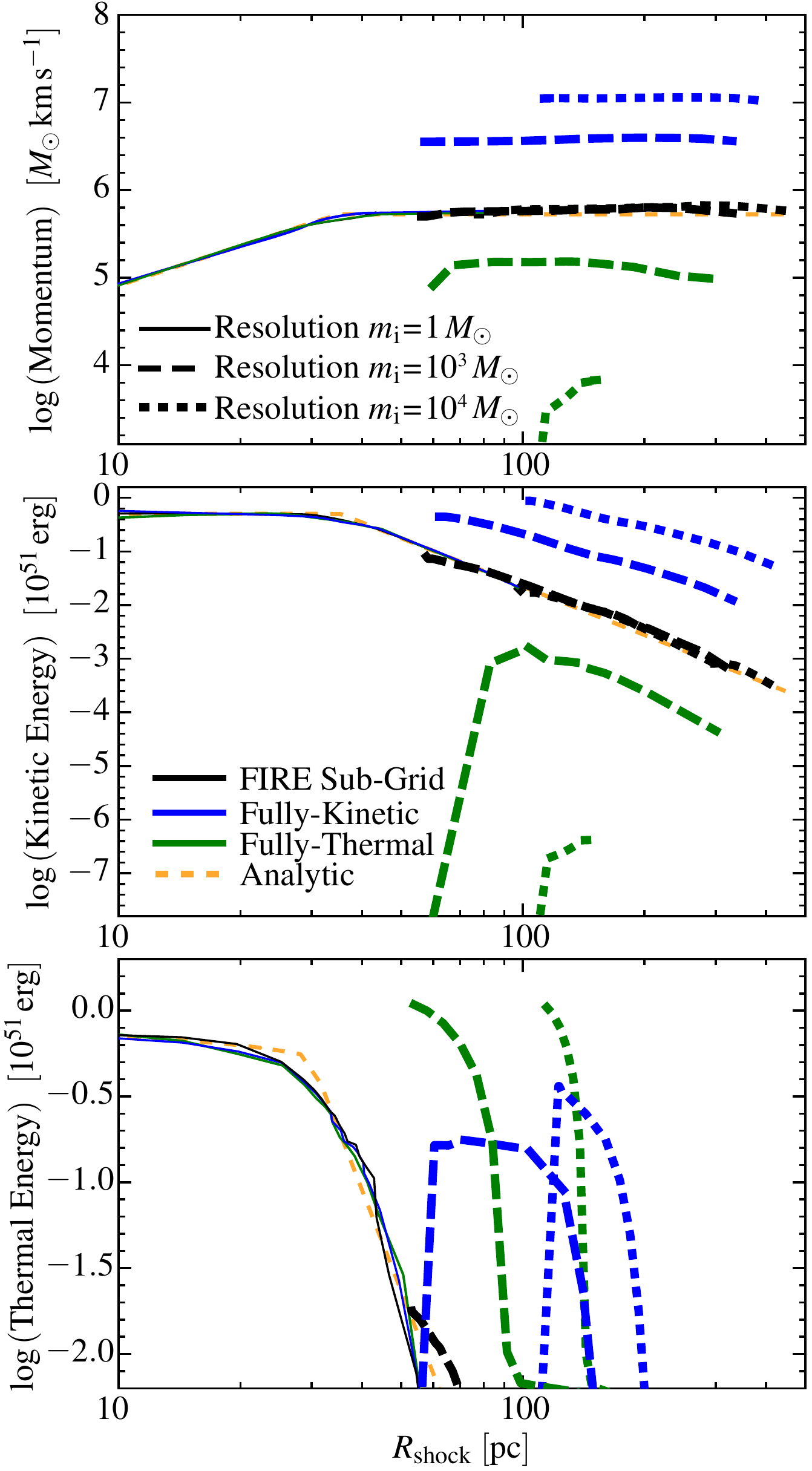}{0.99}
    \vspace{-0.25cm}
    \caption{Time evolution of blastwave momentum ({\em top}), kinetic energy ({\em middle}), and thermal energy ({\em bottom}), 
    as a function of shock position $R_{\rm shock}(t)$, in the test simulations of SNe treatments from Fig.~\ref{fig:sne.convergence.momentum}. For clarity we show just three resolution levels. ``Analytic'' assumes the explosion is energy-conserving until the terminal momentum from Eq.~\ref{eqn:terminal.p} is reached, and subsequently momentum-conserving. At high resolution, all methods resolve the blast and agree with the analytic curve. 
Because density is fixed, poor mass resolution means the smallest $R_{\rm shock}$ (earlier stages) are not resolved; but the FIRE method reproduces the high-resolution solution {\em at the same radius} at all resolution levels. ``Fully-Thermal'' or ``Fully-Kinetic'' models systematically under or over-estimate the momentum and kinetic energy, in a strongly resolution-dependent manner. (The excess thermal energy at low resolution in ``Fully-Kinetic'' models is caused by the shock from the blastwave moving faster than it should, physically). 
   \vspace{-0.4cm}
        \label{fig:sne.convergence.energy}}
\end{figure}

\begin{figure}
\plotonesize{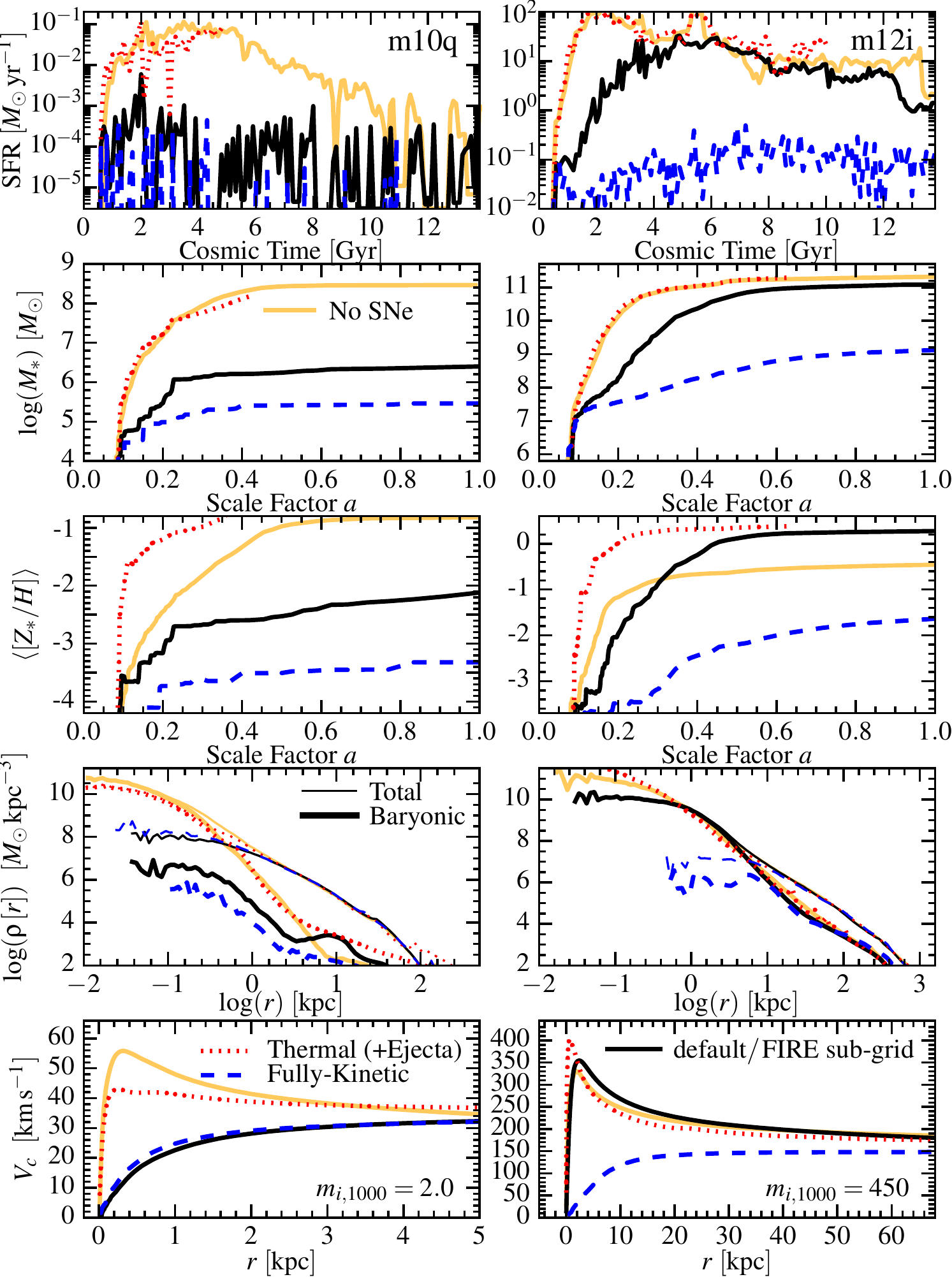}{0.99}
    \vspace{-0.25cm}
    \caption{Comparison of the alternative sub-grid assumptions (\S~\ref{sec:feedback:mechanical:sedov}) from Figs.~\ref{fig:sne.convergence.momentum}-\ref{fig:sne.convergence.energy}, in zoom-in cosmological simulations.  
    {\bf (1)} {\em Default:} the ``FIRE Sub-Grid'' model, which matches the correct momentum and energy from high-resolution solutions of individual blastwave explosions at all resolution levels.
    {\bf (2)} {\em ``Thermal (+Ejecta)'':} This couples just the initial ejecta momentum (ignoring any un-resolved $PdV$ work done), with the remaining energy in pure thermal form. From our high-resolution tests, this {\em under}-estimates the final momentum (kinetic energy) of SNe by a factor $\sim 16$ ($\sim 250$) at the low resolution of the tests here. Therefore our predictions resemble a ``no SNe'' case, and stars form extremely rapidly in the early Universe.
   {\bf (3)}  {\em ``Fully-Kinetic'':} This couples the $100\%$ of the SNe ejecta energy as kinetic energy: our high-resolution tests show this {\em over}-estimates the momentum (kinetic energy) by a factor $\sim 10$ ($\sim 40$) for {\bf m10q} (with $m_{i,\,1000}=2$) and $\sim 160$ ($\sim 2500$) for {\bf m12i} (with $m_{i,\,1000}=450$). Not surprisingly, star formation is overwhelmingly suppressed (a MW-mass halo forms a $<10^{9}\,\msun$ dwarf). 
   {\bf (4)} {\em No SNe:} No supernovae included.
   \vspace{-0.4cm}
    \label{fig:sf.history.sne.subgrid}}
\end{figure}

\begin{figure}
\plotonesize{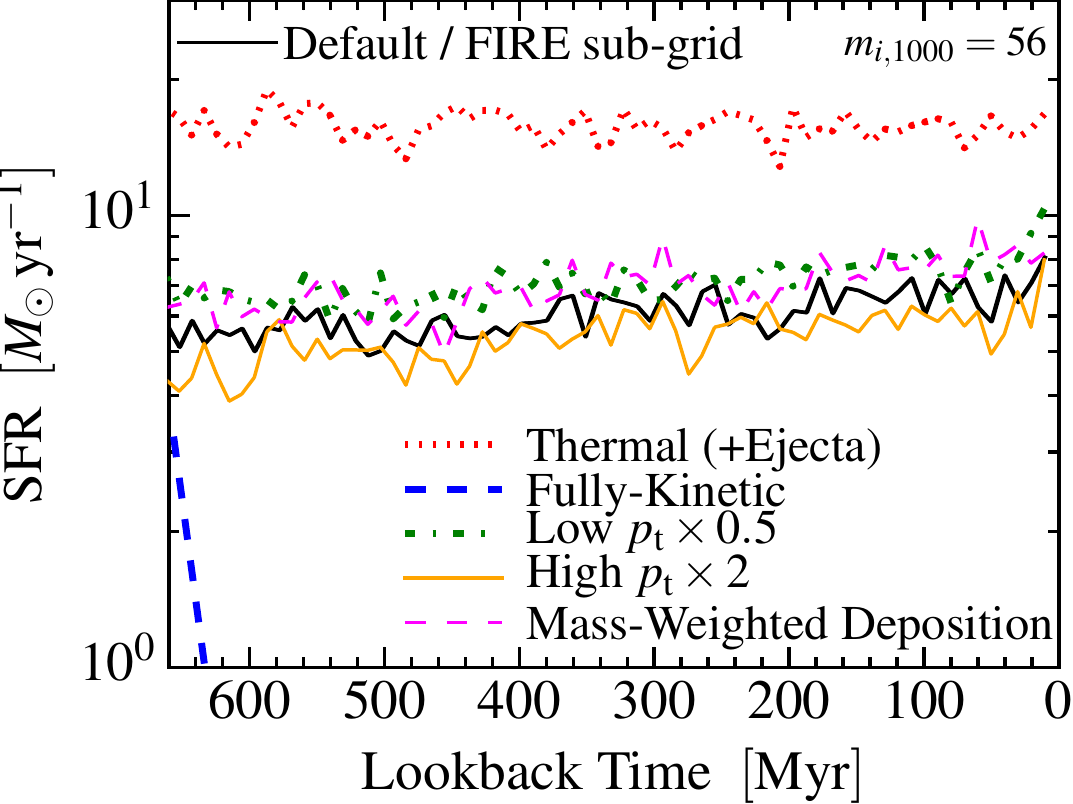}{0.99}
    \vspace{-0.25cm}
    \caption{Comparison of the sub-grid SNe treatments from Figs.~\ref{fig:sne.convergence.momentum}-\ref{fig:sf.history.sne.subgrid}. In each case, we restart our ``default'' {\bf m12i} simulation (resolution $m_{i,\,1000}=56$) at $z=0.07$ and run to $z=0$ ($\sim 700\,$Myr) using different feedback implementations. This allows us to examine how different physics change the predicted SFR within the galaxy on smaller-than-cosmological timescales given an {\em identical} initial galaxy (removing the non-linear effects over cosmological time present in Fig.~\ref{fig:sf.history.sne.subgrid}).
As in Fig.~\ref{fig:sf.history.sne.subgrid}, we compare our default model, which correctly captures the momentum, thermal, and kinetic energy of SNe at this resolution level ($m_{i,\,1000}=56$), to the ``Thermal (+Ejecta)'' model which under-estimates the momentum (kinetic energy) of SNe by a factor $\sim 16$ ($\sim 250$) at this resolution, and the ``Fully-Kinetic'' model which over-estimates the momentum (kinetic energy) by a factor $\sim 50$ ($\sim 400$) at this resolution. As expected, the ``Thermal (+Ejecta)'' model produces much higher SFRs (the SFR cannot become much higher than $\sim 20\,\msun\,{\rm yr^{-1}}$, here, because it is limited by the free-fall time), while the ``Fully-Kinetic'' model overwhelmingly suppresses the SFR (the gaseous disk essentially ``explodes'' in the first dynamical time). We then re-run our ``default FIRE sub-grid'' model, but systematically increase (decrease) the terminal momentum $p_{t}$ in Eq.~\ref{eqn:terminal.p} by a factor of $2$ (much larger than physical uncertainties; see \S~\ref{sec:feedback:mechanical:sedov}). Higher $p_{t}$ (i.e.\ larger assumed cooling radii) produce slightly more-suppressed star formation (as expected) but with only a $\sim 50\%$ change in SFR for a factor $\sim 4$ change in $p_{t}$. 
   \vspace{-0.4cm}
    \label{fig:sf.z0.sne.subgrid}}
\end{figure}

\begin{figure*}
\begin{tabular}{cc}
\includegraphics[width=0.33\textwidth]{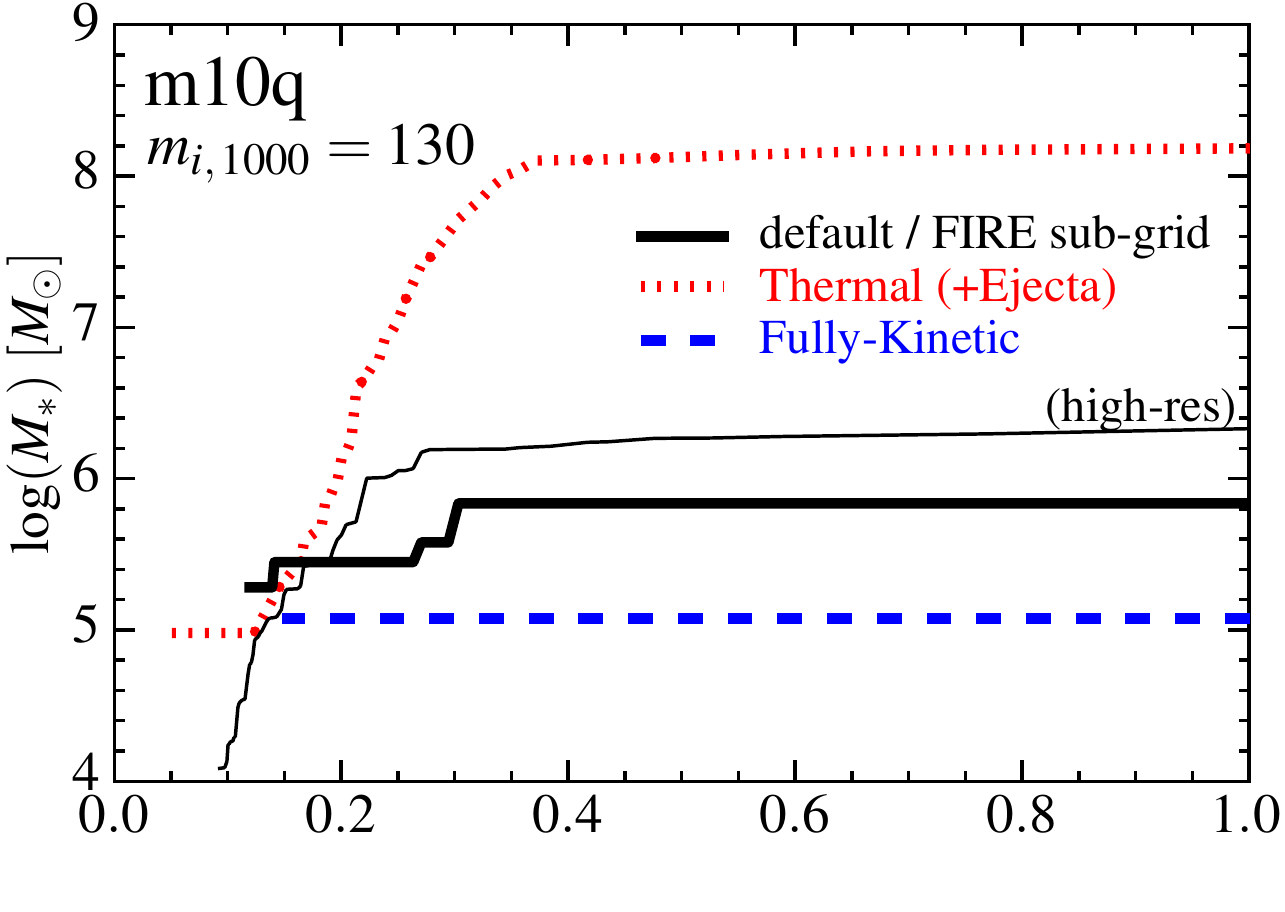} &
\hspace{-0.5cm}
\includegraphics[width=0.33\textwidth]{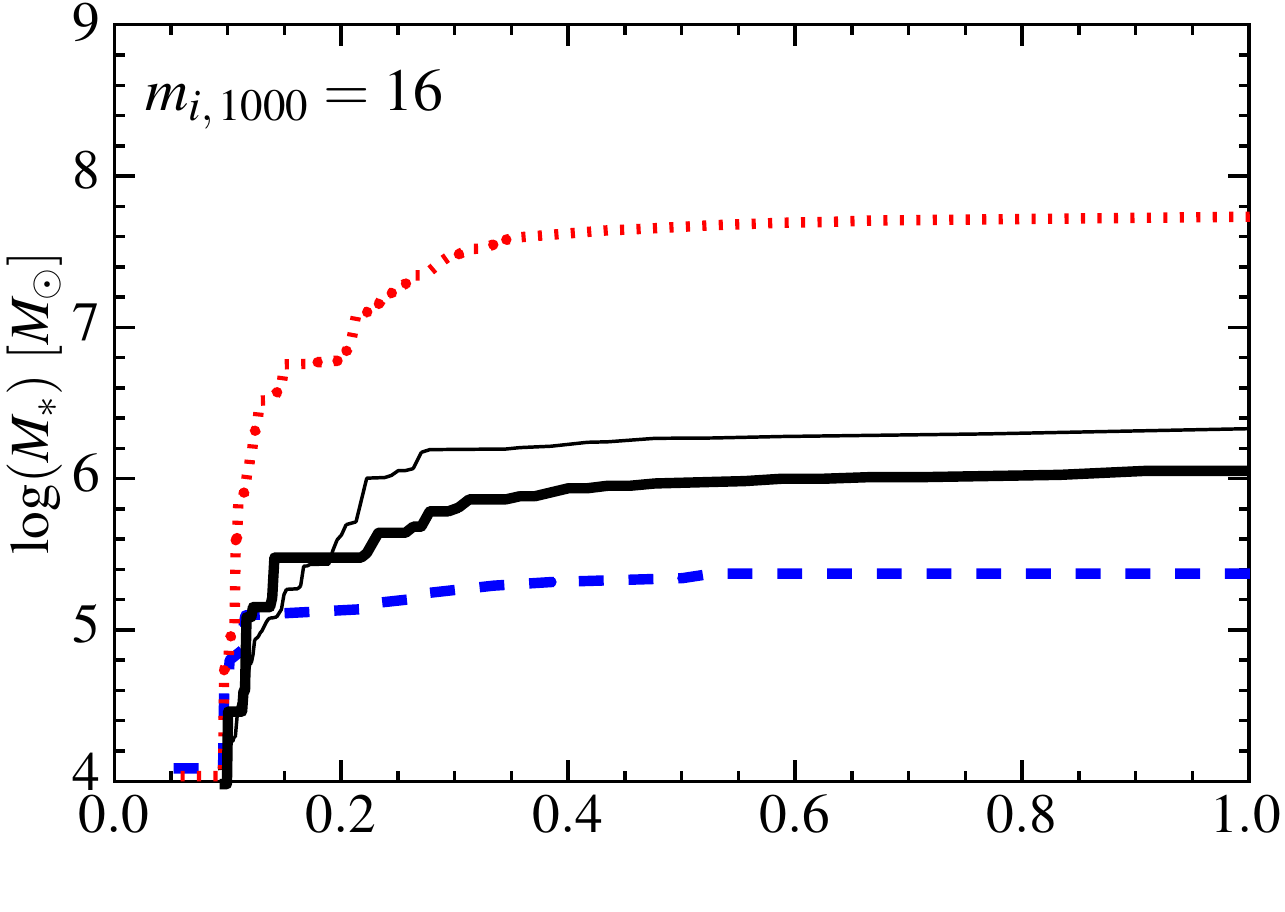} \\
\end{tabular}
\begin{tabular}{ccc}
\includegraphics[width=0.33\textwidth]{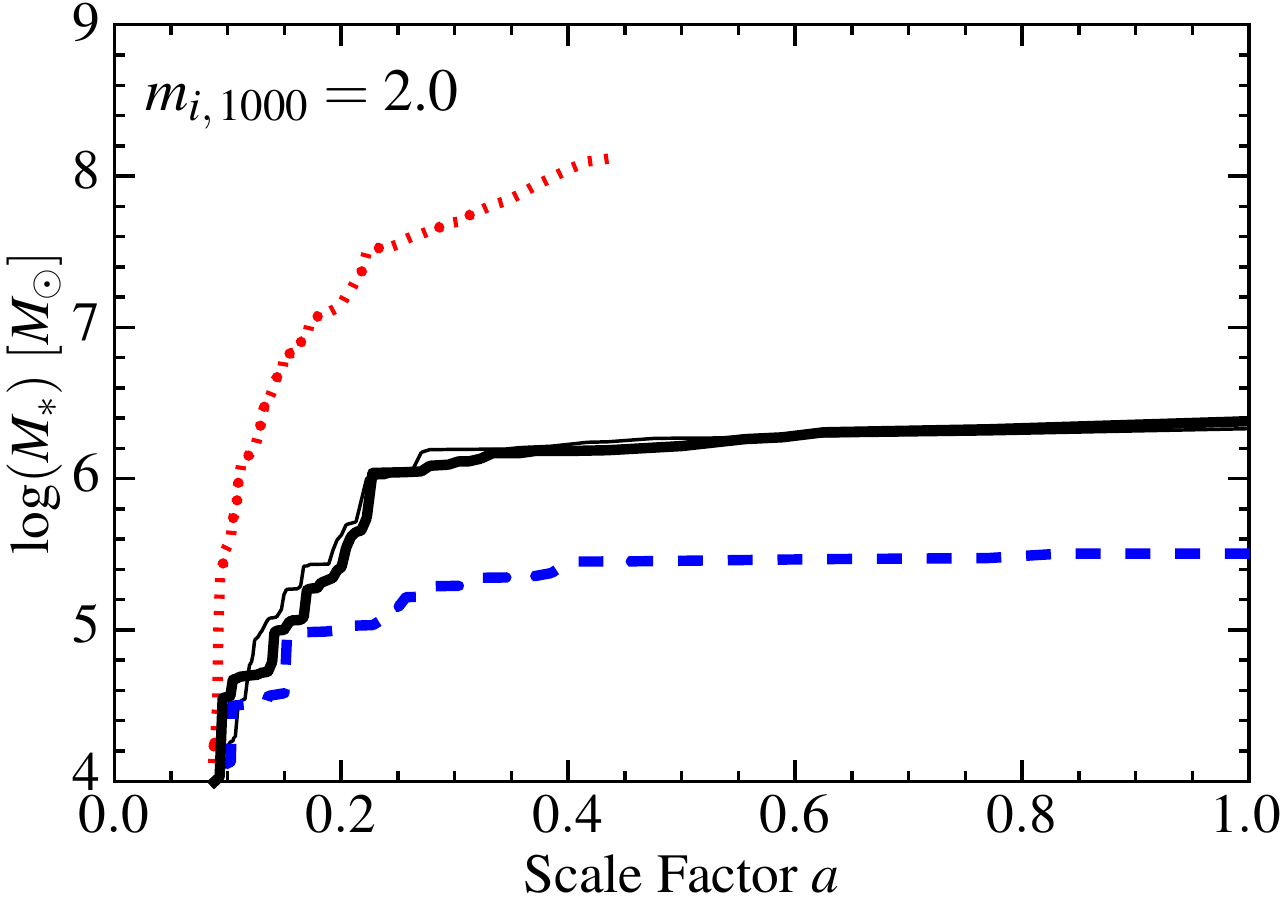} &
\hspace{-0.5cm}
\includegraphics[width=0.33\textwidth]{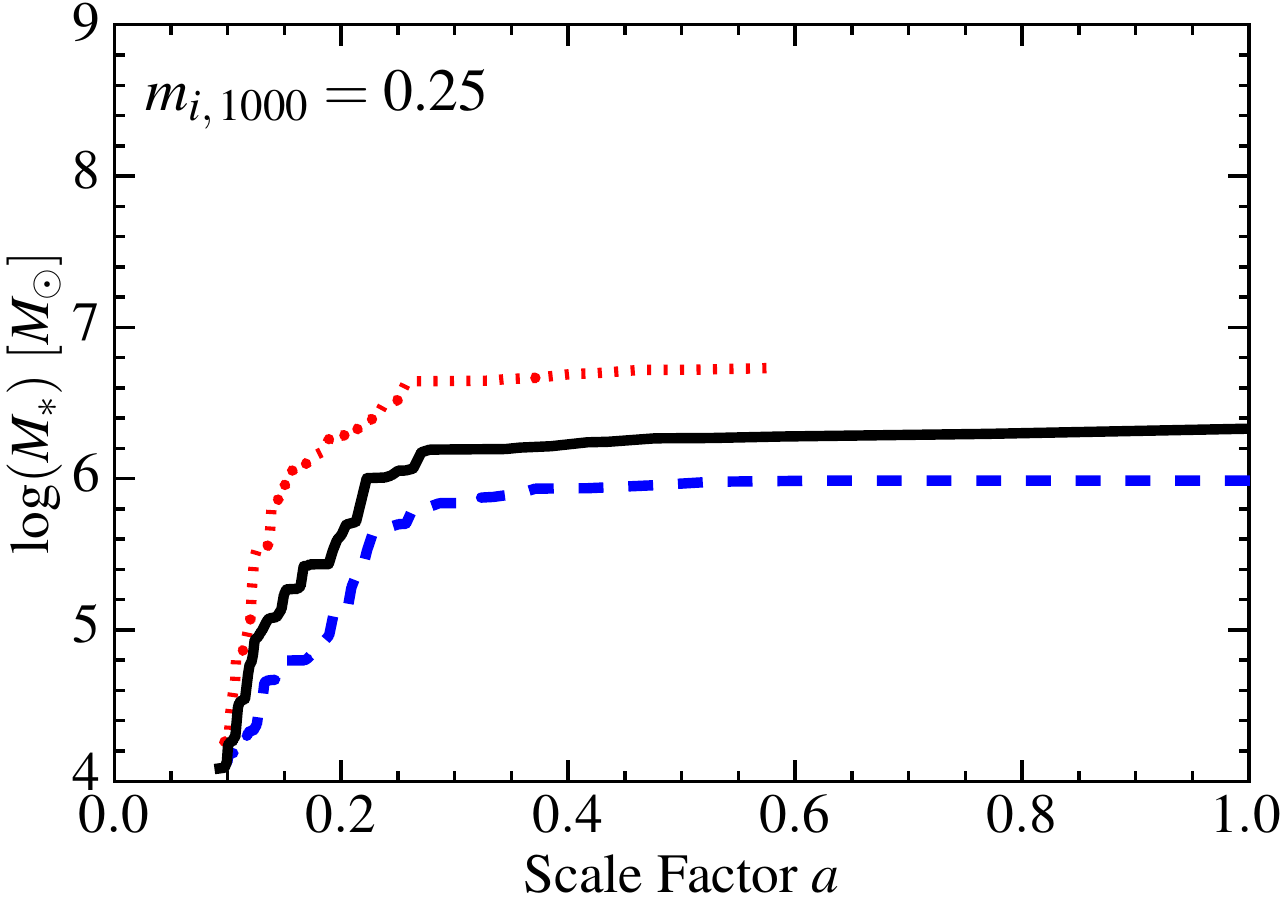} &
\hspace{-0.5cm}
\includegraphics[width=0.33\textwidth]{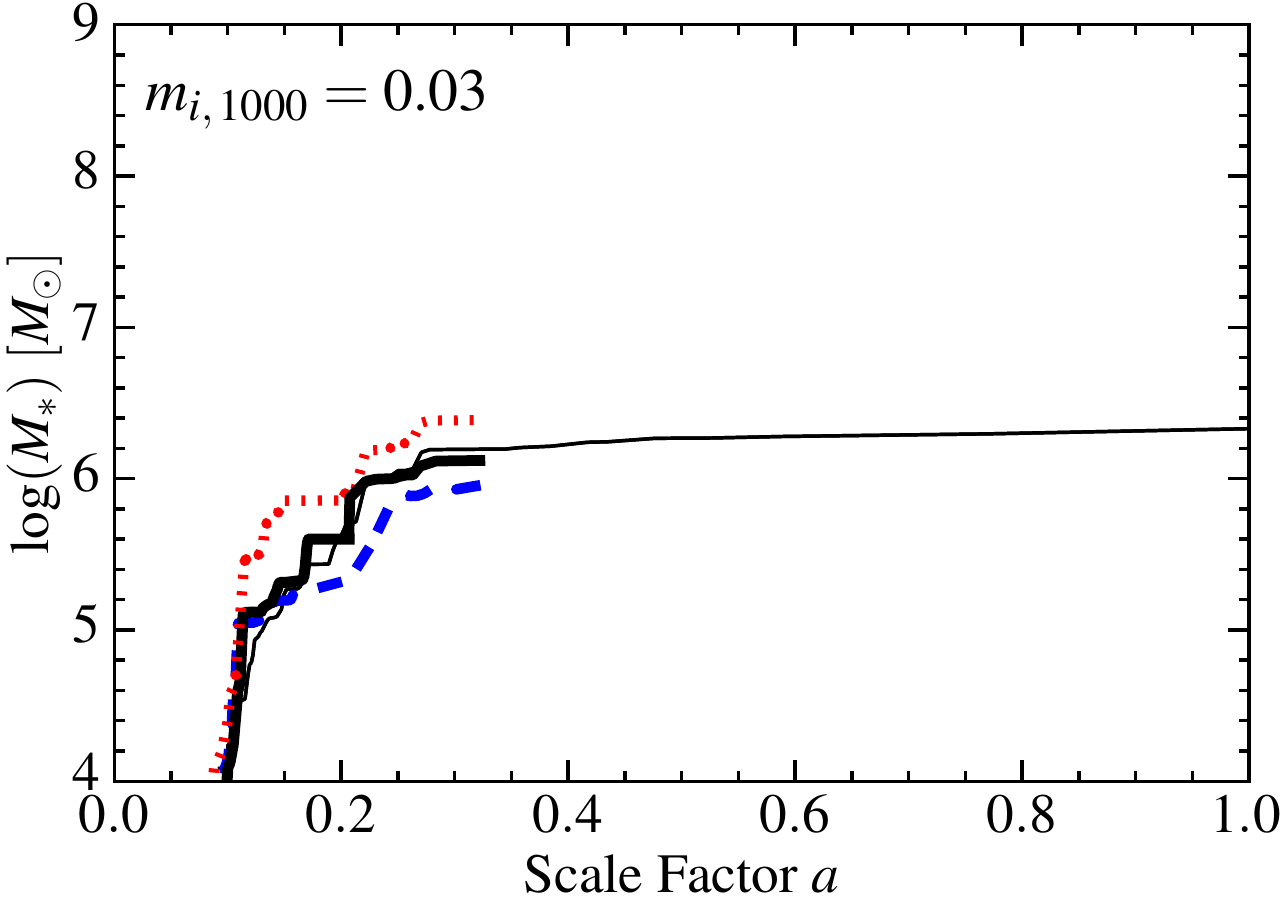} 
\end{tabular}
    \vspace{-0.25cm}
    \caption{Convergence properties of the sub-grid SNe treatments from Figs.~\ref{fig:sne.convergence.momentum}-\ref{fig:sf.history.sne.subgrid} in a cosmological simulation of a dwarf galaxy.
We show stellar mass, $M_{\ast}$, versus cosmic time (as Fig.~\ref{fig:sf.history.sne.subgrid}) for each treatment of unresolved phases of SNe explosions.
We did not run our highest-resolution simulations (with $\sim 30\,\msun$ resolution) to $z=0$ because of computational expense.
At low mass resolution, the ``Thermal (+Ejecta)'' assumption under-estimates the strength of feedback (producing {\em too-high} $M_{\ast}$), while the ``Fully-Kinetic'' assumption over-estimates the strength of feedback (producing {\em too-low} $M_{\ast}$). Our default FIRE sub-grid treatment resembles the high-resolution converged solution even at low mass resolution. Once we reach mass resolution $\ll 1000\,\msun$ ({\em bottom}), the cooling radius of individual SN blastwaves starts to become explicitly resolved, and the different treatments converge towards one another -- as predicted from our idealized tests of single explosions in Fig.~\ref{fig:sne.convergence.momentum}. Critically, the ``Thermal (+Ejecta)'' (similar to ``Fully-Thermal'') and ``Fully-Kinetic'' treatments {\em both} converge to our ``default FIRE sub-grid'' treatment at higher resolution.
   \vspace{-0.4cm}
    \label{fig:sf.fb.sne.subgrid.convergence}}
\end{figure*}

\vspace{-0.5cm}
\section{Numerical Tests: Subgrid Physics and the Need to Account For Thermal and Kinetic Energy}
\label{sec:feedback:mechanical:tests}

Having tested the algorithmic aspect of SNe coupling above, we now consider tests of the physical scalings in the feedback coupling, specifically how it assigns momentum versus thermal energy as described in \S~\ref{sec:feedback:mechanical:sedov}.

\vspace{-0.5cm}
\subsection{Validation: Ensuring ``Subgrid'' Scalings Reproduce High-Resolution Simulations in Resolution-Independent Fashion}
\label{sec:feedback:mechanical:tests:validation}

In Figs.~\ref{fig:sne.convergence.momentum}-\ref{fig:sne.convergence.energy}, we consider an idealized test problem that validates the sub-grid SNe treatment used in FIRE. We initialize a periodic box of arbitrarily large size with uniform density $n= 1\,{\rm cm^{-3}}$ and metallicity $Z=Z_{\sun}$, with constant gas particle mass $m_{i}$ (so the inter-particle separation is given by $\rho = m_{i}/h_{i}^{3}$, i.e.\ $h_{i}\sim 16\,{\rm pc}\,(m_{i}/100\,\msun)^{1/3}$), and with our full FIRE-2 cooling physics (with the $z=0$ meta-galactic background) and hydrodynamics, but no self-gravity. We then detonate a single SN explosion at the center of the box, using {\em exactly} our default FIRE-2 algorithm (same SN energy $=10^{51}\,{\rm erg}$, ejecta mass $=10.4\,\msun$, metal content, ejecta momentum, and algorithmic coupling scheme from Fig.~\ref{fig:mechanical.fb.cartoon} and \S~\ref{sec:feedback:mechanical:neighbor.finding}-\ref{sec:feedback:mechanical:assignment}). We also test several additional schemes for how to deal with the thermal versus kinetic (energy/momentum) component of the SN.

\begin{enumerate}

\item{{\em FIRE Sub-grid:} This is our default FIRE-2 treatment from \S~\ref{sec:feedback:mechanical:sedov} (Eq.~\ref{eqn:dp.subgrid.sub1}), where we account for the $PdV$ work done by the expanding blastwave out to the minimum of either the coupling radius or cooling radius (where the resulting momentum reaches the terminal momentum $p_{t}$ in Eq.~\ref{eqn:terminal.p}, and we assume any remaining thermal energy is dissipated outside the cooling radius). The coupled momentum ranges, therefore, between $p_{\rm ejecta} \le p_{\rm coupled} \le p_{\rm terminal}$ and {\em total} (kinetic+thermal) energy coupled ranges from $0 < E_{\rm coupled} \le E_{\rm ejecta} = 10^{51}\,{\rm erg}$, according to the total mass enclosed within the single gas particle (the smallest possible ``coupling radius''). Recall, at small particle mass, this becomes identical to coupling exactly the SN ejecta energy and momentum. At large particle mass, this reduces to coupling the terminal momentum and radiating (instantly) all residual (post-shock) thermal energy.}

\item{{\em Thermal (+Ejecta):} This couples only the ejecta momentum ($p_{\rm coupled} = p_{\rm ejecta} < p_{\rm terminal}$): any additional energy is coupled as thermal energy (not radiated away in the coupling step; $E_{\rm coupled} = E_{\rm ejecta}$). This ignores any accounting for whether the coupling is inside/outside the cooling radius, or any $PdV$ work done by the un-resolved blastwave expansion. It is equivalent to dropping the terms from \S~\ref{sec:feedback:mechanical:sedov} completely. A method like this was used in some previous work with non-cosmological simulations \citep{hopkins:fb.ism.prop}.}

\item{{\em Fully-Kinetic:} We assume that $100\%$ of the ejecta energy is converted into kinetic energy, i.e.\ coupled in ``pure momentum'' form ($p_{\rm coupled} = \sqrt{2\,E_{\rm ejecta}\,m_{\rm coupled}} \ge p_{\rm ejecta}$, $E_{\rm coupled} = E_{\rm kinetic} = E_{\rm ejecta}$). This ignores any un-resolved cooling. This is similar (algorithmically) to many common implementations in e.g.\ \citet{aguirre:2001.igm.metal.evol.sims,springel:multiphase,cen:2005.kinetic.sne.fb,dalla-vecchia:2008.iso.gal.wwo.freestream.winds,vogelsberger:2013.illustris.model} (although most of these authors alter the fraction of energy coupled).}

\item{{\em Fully-Thermal:} We assume $100\%$ of the ejecta energy is converted into thermal energy, with zero momentum (i.e.\ $E_{\rm coupled} = E_{\rm thermal} = E_{\rm ejecta}$, $p_{\rm coupled} = 0$). This also ignores any un-resolved cooling. This is a common implementation used in e.g.\ \citet{katz:1992.overcooling.problem.standard.sne,ceverino:cosmo.sne.fb,ceverino:2013.rad.fb,kim:agora.isolated.disk.test}.}

\end{enumerate}

We evolve the explosion until well after it reaches an asymptotic terminal momentum: when the momentum changes by $<1\%$ over a factor of $>2$ increase in the shock radius, or -- if this occurs before the shock reaches $>10$ inter-particle spacings -- when the shock radius moves by $<1\%$ over a factor of $2$ increase in time.

In Fig.~\ref{fig:sne.convergence.momentum} we plot the terminal momentum in each simulation and compare to our analytic scaling from Eq.~\ref{eqn:terminal.p}. In Fig.~\ref{fig:sne.convergence.energy}, we plot the radial profile of the shock properties as the shock radius expands: the total radial momentum, kinetic, and thermal energy (these depend on the time since explosion, so we plot each resolution at different times). We consider particle masses ranging from $m_{i}<0.1\,\msun$, sufficient to resolve even the free-expansion phase of explosion, (let alone the cooling radius), to $m_{i}> 10^{6}\,\msun$.

At sufficiently high resolution, all of the schemes above give identical, well converged solutions -- as they should, since in all cases (at high enough resolution) they generate a shock with the same initial energy, which undergoes an energy-conserving Sedov-Taylor type expansion (in which case the asymptotic solution is fully-determined by the ambient density and total blastwave energy). In this limit, the shock formation, Sedov-Taylor phase, conversion of energy into momentum, cooling radius, snowplow phase, and ultimate effective conversion of energy into momentum via $PdV$ work are explicitly resolved, so it does not matter how we initially input the energy. Reassuringly, Eq.~\ref{eqn:terminal.p} agrees well with the {\em predicted} terminal momentum in the highest-resolution simulations -- in other words, given the cooling physics in FIRE-2, we are using the correct $p_{t}$. 

At poor resolution, the different treatments diverge, as predicted in \S~\ref{sec:feedback:mechanical:sedov:resolution}. For ``Thermal (+Ejecta)'' and ``Fully-Thermal'' couplings, when the particle mass $m_{i} \gtrsim 100\,\msun$, the predicted momentum and kinetic energy drop rapidly compared to the converged, exact solutions. Physically, the cooling radius -- which is roughly the radius enclosing a fixed {\em mass} $m_{\rm cool}\sim 1000\,\msun$ (see \S~\ref{sec:feedback:mechanical:sedov:resolution}) -- becomes unresolved. Spreading only thermal energy among this large a gas mass leads to post-shock temperatures below the peak in the cooling curve, so the energy is immediately radiated before much work can be done to accelerate gas (increase the momentum). The terminal momentum and kinetic energy are under-estimated by constant factors of $\sim 60$ and $\sim 3600$, respectively. With the ``Thermal (+Ejecta)'' case, the same problem occurs, but the initial ejecta momentum remains present, so the terminal momentum and kinetic energy are under-estimated by factors of $\sim 14$ and $\sim 200$. 

The ``Fully-Kinetic'' coupling errs in the opposite direction at poor resolution: assuming perfect conversion of energy to momentum and ignoring cooling losses gives $p_{\rm coupled} = \sqrt{2\,E_{\rm ejecta}\,m_{\rm coupled}}$, so $p_{\rm coupled}/p_{\rm terminal} \propto m_{\rm coupled}^{1/2}$, and the terminal momentum is over-estimated by a factor $\sim 20\,(m_{i}/10^{4}\,\msun)^{1/2}$. The kinetic energy is over-estimated by a corresponding factor $\sim 400\,(m_{i}/10^{4}\,\msun)$. 

In contrast, the FIRE sub-grid model reproduces the high-resolution exact solutions correctly, {\em independent} of resolution (within $<10\%$ in momentum, kinetic and thermal energy, even at $m_{i}\sim 10^{6}\,\msun$). This is the desired behavior in a ``good'' sub-grid model. Of course, at poor resolution, the cooling radius is un-resolved, so the simulation cannot capture the early phases where gas is shock-heated to large temperatures. However the sub-grid treatment captures the correct behavior of the high-resolution blastwave once it has expanded to a mass or spatial resolution scale which {\em is} resolved in the low-resolution run. (For similar experiments which reach the same conclusions, see e.g.\ Fig.~6 in \citealt{kim:tigress.ism.model.sims}).

\vspace{-0.5cm}
\subsection{Effects In FIRE Simulations: Correctly Dealing With Energy \&\ Momentum Matters}
\label{sec:feedback:mechanical:tests:effects}

Having seen in \S~\ref{sec:feedback:mechanical:tests:validation} that correctly accounting for unresolved $PdV$ work in expanding SNe is critical to resolution-independent solutions, we now apply the different treatments therein to cosmological simulations. 

Fig.~\ref{fig:sf.history.sne.subgrid}-\ref{fig:sf.z0.sne.subgrid} show the results, for both dwarf- and MW-mass galaxies in cosmological simulations as well as controlled re-starts of the same  MW-mass galaxy at $z\sim0$ (to ensure identical late-time ICs).

At the mass resolution scales in Figs.~\ref{fig:sf.history.sne.subgrid}-\ref{fig:sf.z0.sne.subgrid}, our default FIRE-2 coupling scheme reproduces accurately the SN momentum, kinetic and thermal energies from much higher-resolution idealized simulations. In contrast, the ``Thermal (+Ejecta)'' and ``Fully-Kinetic'' models severely under and over-estimate, respectively, the kinetic energy imparted by SNe (relative to high resolution simulations and/or analytic solutions). Not surprisingly, then, this is immediately evident in the galaxy evolution. ``Fully-Thermal'' and ``Thermal (+Ejecta)'' cases resemble a ``no SNe'' case -- because the cooling radii are unresolved, the energy is radiated away immediately, and the terminal momentum that {\em should have been resolved} is not properly accounted for -- so SNe do far less work than they should and far more stars form. The Fully-Kinetic case, on the other hand, wildly over-estimates the conversion of thermal energy to kinetic (and ignores cooling losses), so star formation is radically suppressed.

Given this strong dependence, one might wonder whether the exact details of our FIRE treatment might change the results. However these are not so important. In Appendix~\ref{sec:appendix:implicit.cooling}, we consider a ``no implicit cooling'' model: here we take our standard FIRE-2 coupling (the coupled momentum, mass, and metals are unchanged), but even if the cooling radius is un-resolved, we still couple the full ejecta energy (i.e.\ we do not assume, implicitly, that the ejecta thermal energy has radiated away if we do not resolve the cooling radius, so couple a total thermal plus kinetic energy $=E_{\rm ejecta}$). This produces no detectable difference from our default model, which is completely expected. If the cooling radius is resolved, our default model does not radiate the energy away; if it is unresolved, ``keeping'' the thermal energy in the SNe coupling step simply leads to its being radiated away explicitly in the simulation cooling step on the subsequent timestep. 

Fig.~\ref{fig:sf.z0.sne.subgrid} considers the effects of changing the analytic terminal momentum $p_{t}$ in Eq.~\ref{eqn:terminal.p}, by a factor $\sim 4$. As discussed in \S~\ref{sec:feedback:mechanical:sedov}, while there are physical uncertainties in this scaling owing to uncertain microphysics of blastwave expansion, they are generally smaller. But in any case, the effect on our galaxy-scale simulations is relatively small, even at low resolution. As expected, smaller $p_{t}$ leads to higher SFRs, because the momentum coupled per SN is smaller, so more stellar mass is needed to self-regulate. In a simple picture where momentum input self-regulates SF and wind generation \citep[see e.g.][]{ostriker.shetty:2011.turb.disk.selfreg.ks,cafg:sf.fb.reg.kslaw,hayward.2015:stellar.feedback.analytic.model.winds}, we would expect the SFR to be inversely proportional to $p_{t}$ at low resolution. However, because of non-linear effects, and the fact that even at low resolution the simulations resolve massive super-bubbles (where $p_{t}$ does not matter because the cooling radius for overlapping explosions is resolved), the actual dependence is sub-linear, $\dot{M}_{\ast} \propto p_{t}^{-0.3}$. So given the (small) physical uncertainties, this is not a dominant source of error.\footnote{To be clear, in Fig.~\ref{fig:sf.z0.sne.subgrid} we alter {\em only} the terminal momentum, so e.g.\ if the cooling radius of super-bubbles is resolved the change has no effect whatsoever, and other feedback mechanisms (e.g.\ radiative feedback) are also un-altered. In contrast, in \citet{orr:ks.law} (Appendix~A) we show the results of multiplying/dividing {\em all} feedback mechanisms and strengths (total energy and momentum) by a uniform factor $=3$. Not surprisingly this produces a stronger effect closer to the expected inverse-linear dependence; however non-linear effects still reduce the dependence to somewhat sub-linear.}

Recently, \citet{rosdahl:2016.sne.method.isolated.gal.sims} performed a similar experiment, exploring different SNe implementations in the AMR code {\small RAMSES}. They used a different treatment of cooling and star formation, non-cosmological simulations, and no other feedback. However, their conclusions are similar, regarding the relative efficiencies of the ``Fully-Thermal,'' ``FIRE-sub grid'' (in their paper, the ``mechanical'' model), and ``Fully-Kinetic'' treatments of SNe. Our conclusions appear to be robust across a wide range of conditions and detailed numerical treatments.

Again, we have repeated these tests in other halos to ensure our conclusions are not unique to a single galaxy. Specifically we have compared a ``Fully-Thermal'' and ``Fully-Kinetic'' run in halo {\bf m10v} and {\bf m12f} from \paperone, and compared re-starts from $z=0.07$ of an {\bf m12f} run with $m_{i,\,1000}=56$ using the same set of parameter variations as Fig.~\ref{fig:sf.z0.sne.subgrid}. The results are nearly identical to our studies with {\bf m10q} and {\bf m12i}.

\vspace{-0.5cm}
\subsection{Convergence: Incorrect Sub-Grid Treatments Converge to the Resolution-Independent FIRE Scaling}
\label{sec:feedback:mechanical:tests:convergence}

In Fig.~\ref{fig:sf.fb.sne.subgrid.convergence}, we consider another convergence test of the SNe coupling scheme, but this time in cosmological simulations. We re-run our {\bf m10q} simulation with standard FIRE-2 physics, considering our default SNe treatment as well as the ``Thermal (+Ejecta)'' and ``Fully-Kinetic'' models, with mass resolution varied from $ 30 - 1.3\times10^{5}\,\msun$.

Not only does our default FIRE treatment of SNe produce excellent convergence in the star formation history across this entire resolution range, but {\em both} the ``Thermal (+Ejecta)'' model (which suffers from over-cooling, hence excessive SF, at low resolution because the SNe energy is almost all coupled thermally) and the ``Fully-Kinetic'' model (which over-estimates the kinetic energy of SNe, hence over-suppresses SF, at low resolution) converge {\em to our FIRE solution} at higher resolution, especially at $m_{i}\lesssim 100\,\msun$. Of course, even at our highest resolution, details of SNe shells and venting can differ in the early stages of ejecta expansion, so convergence is not perfect -- but the trends clearly approach the ``default'' model.

\vspace{-0.5cm}
\subsection{On ``Delayed-Cooling'' and ``Target-Temperature'' Models}
\label{sec:delayed.cooling.discussion}

Given the failure of ``Fully-Thermal'' models at low resolution, a popular ``fix'' in the galaxy formation literature is to artificially suppress gas cooling at large scales, either explicitly or implicitly. This is done via (a) ``delayed cooling'' prescriptions, for which energy injected by SNe is not allowed to cool for some large timescale $\Delta t_{\rm delay}\gtrsim t_{\rm dynamical} \sim 10^{7-8}$\,yr \citep[as in][]{thackercouchman00,thackercouchman01,stinson:2006.sne.fb.recipe,stinson:2013.new.early.stellar.fb.models,dubois:delayed.cooling.sne.models}, or (b) ``target temperature'' prescriptions, where SNe energy is ``stored'' until sufficient energy is accumulated to heat (in a single ``event'') a large resolved gas mass to some high temperature $T_{\rm target} \gg 10^{7}\,$K \citep[as in][]{gerritsen:target.temperature.models,mori:1997.target.temperature.sne.models,dalla.vecchia:target.temperature.sne.delayed.cooling.feedback,crain:eagle.sims}. 

Although these approximations may be useful in low-resolution simulations with $m_{i}\gtrsim 10^{6}\,\msun$ (typical of large-volume cosmological simulations), where ISM structure and the clustering of star formation cannot be resolved, they are fundamentally ill-posed for simulations with resolved ISM structure, for at least three reasons. 
{\bf (1)} Most importantly, they are {\em non-convergent} (at least as defined here). This is easy to show rigorously, but simply consider a case with arbitrarily good resolution: then either (a) turning off cooling for longer than the actual shock-cooling time, or (b) enforcing a ``target temperature'' that does not exactly match the initial reverse-shock temperature will produce un-physical results. Strictly speaking there is no define-able convergence criterion for these models: they do not interpolate to the correct solution as resolution increases, but to some other (non-physical) system. 
{\bf (2)} They do not represent the converged solution in Fig.~\ref{fig:sne.convergence.energy} at any low-resolution radius/mass. Once a SN has swept through, say, $\sim 10^{6}\,\msun$ of gas, it should, correctly, be a cold shell, not a hot bubble. Thus we are not reproducing the higher-resolution solutions correctly, at some finite practical resolution. 
{\bf (3)} They introduce an additional set of parameters: $\Delta t_{\rm delay}$ or $T_{\rm target}$, and the ``size'' (or mass) of the region that is influenced. Both of these strongly influence the results. For example, by increasing the region size, one does not simply ``spread'' the same energy among neighbors differently, but rather, because the models are binary, one either (a) increases the mass that cannot cool or (b) must change the number of SNe ``stored up'' (hence the implicit cooling-delay-time) to reach $T_{\rm target}$.

In Appendix~\ref{sec:delayed.cooling} we consider some implementations of these models, at the resolutions studied here. As expected, we show that they do not converge as we approach resolution $\sim 100\,\msun$, and that certain galaxy properties (metallicities, star formation histories) exhibit biases that are clear artifacts of the un-physical nature of these coupling schemes at high resolution. We therefore do not focus on them further.

\vspace{-0.5cm}
\section{Discussion \&\ Conclusions}
\label{sec:discussion}

We have presented an extensive study of both numerical and physical aspects of the coupling of mechanical feedback in galaxy formation simulations (most importantly, SNe, but the methods are relevant to stellar mass-loss and black hole feedback). We explored this in both idealized calculations of individual SN remnants and in the FIRE-2 cosmological simulations at both dwarf and MW mass scales. We conclude that there are two critical components to an optimal algorithm, summarized below.

\vspace{-0.5cm}
\subsection{Ensuring Conservation \&\ Statistical Isotropy} 
\label{sec:discussion:conservation} 

It is important to design an algorithm that is statistically isotropic (i.e.\ does not numerically bias the feedback to prefer certain directions), and manifestly conserves mass, metals, momentum, and energy. This is particularly non-trivial in mesh-free numerical methods. In particular, naively distributing ejecta with a simple kernel or area weight to ``neighbor'' cells or particles -- as is common practice in most numerical treatments -- can easily produce violations of linear momentum conservation and bias the ejecta so that in, for example, a thin disk, feedback preferentially acts (incorrectly) in the disk plane instead of venting out. This is especially important for {\em any} numerical method for which the gas resolution elements might be irregularly distributed around a star (e.g.\ moving-mesh codes, SPH, or AMR if the star is not at the exact cell center). If these constraints are not met, we show that spurious numerical torques or outflow geometries can artificially remove disk angular momentum and bias predicted morphologies. Worse yet, the momentum conservation errors may not converge and can become more important at high resolution. 

In fact, as discussed in detail in \S~\ref{sec:feedback:mechanical:ideal.tests:firesims}, our older published ``FIRE-1'' simulations suffered from some of these errors, but (owing to lower resolution) they were relatively small. Higher-resolution tests, however, demonstrated their importance, motivating the development of the new FIRE-2 algorithm.

In \S~\ref{sec:feedback:mechanical} we present a general algorithm (used in FIRE-2) that resolves all of these issues (as well as accounting for relative star-gas motions), and can trivially be applied in any numerical galaxy formation code (regardless of hydrodynamic method), for any mechanical feedback mechanism.

\vspace{-0.5cm}
\subsection{Accounting for Energy \&\ Momentum from Un-Resolved ``PdV Work''}
\label{sec:discussion:pdv} 

At the mass ($m_{i}$) or spatial resolution ($h_{i}$) of current cosmological simulations, it is {\em physically incorrect} to couple SNe to the gas either as entirely thermal energy (heating-only) or entirely kinetic energy (momentum transfer only), or the initial ejecta mix of momentum and energy. Because the SN blastwave has implicitly propagated through a region containing mass $\sim m_{i}$, it {\em must} have either (a) done some mechanical (``$PdV$'') work, increasing the momentum of the blastwave, and/or (b) radiated its energy away. In \citet{hopkins:2013.fire} we proposed a simple way to account for this in simulations, which we provide in detail in \S~\ref{sec:feedback:mechanical:sedov}. This method is used in all FIRE simulations, was further tested in idealized simulations by \citet{martizzi:sne.momentum.sims}, and similar methods have been developed and used in galaxy formation simulations by e.g.\ \citet{kimm.cen:escape.fraction,rosdahl:2016.sne.method.isolated.gal.sims}. Essentially, we account for the $PdV$ work by imposing energy conservation up to a terminal momentum (Eq.~\ref{eqn:terminal.p}), beyond which the energy is radiated, with the transition occurring at the cooling radius of the blastwave.

In this paper, we use high-resolution (reaching $<0.1\,\msun$) simulations of individual SN to show that this implementation, independent of the resolution at which it is applied, reproduces the exact, converged high-resolution simulation of a single SN blastwave, given the {\em same} physics. In other words, taking a high-resolution simulation of a SN in a homogenous medium and smoothing it at the resolved coupling radius produces the same result as what is directly applied to the large-scale simulations. Perhaps most importantly, we show that this method of partitioning thermal and kinetic energy leads to relatively rapid convergence in predicted stellar masses and star formation histories in galaxy-formation simulations. 

In contrast, coupling only thermal or kinetic energy (or the initial ejecta partitioning of the two) will over or under-predict the coupled momentum by orders of magnitude, in a strongly resolution-dependent fashion (Fig.~\ref{fig:sne.convergence.momentum}). Briefly, at poor resolution, coupling $\sim 10^{51}$\,erg as thermal energy (e.g.\ including no momentum or only the initial ejecta momentum) spreads the energy over an artificially-large mass, so the gas is barely heated and efficiently radiates the energy away without resolving the $PdV$ work. But simply converting all (or any resolution-independent fraction) of this energy into kinetic energy, on the other hand, ignores the cooling that should have occurred and will always, at sufficiently poor resolution, over-estimate the correct momentum generated in a resolution-dependent manner (since for fixed kinetic energy input, the {\em momentum} generated is a function of the mass resolution). This in turn leads to strongly resolution-dependent predictions for galaxy masses (Fig.~\ref{fig:sf.fb.sne.subgrid.convergence}). In principle, one could compensate for this by introducing explicitly resolution-dependent ``efficiency factors'' that are re-tuned at each resolution level to produce some ``desired'' result, but this severely limits the predictive power of the simulations and will still fail to produce the correct mix of phases in the ISM and outflows (because the correct thermal-kinetic energy mix is not present). Using cosmological simulations reaching $\sim 30\,\msun$ resolution, we show that {\em all} of these studied coupling methods do converge to the same solution when applied at sufficiently high resolution. The difference is, the proposed method in \S~\ref{sec:feedback:mechanical:sedov} from the FIRE simulations converges much more quickly (at a factor $\sim1000$ lower-resolution), while the unphysical ``Fully-Thermal'' or ``Fully-Kinetic'' approaches require mass resolution $\ll 100\,\msun$.

\vspace{-0.5cm}
\subsection{Caveats and Future Work}
\label{sec:discussion:future} 

While the SNe coupling algorithm studied here reproduces the converged, high-resolution solution at any practical resolution, it is of course possible that the actual conditions under which the SNe explode (the local resolved density, let alone density sub-structure) continue to change as simulation resolution increases. The small-scale density structure of the ISM might in turn depend on other physics (e.g.\ HII regions, radiation pressure), which could have different convergence properties from the SNe alone.

We stress that our conclusions are relevant for simulations of the ISM or galaxies with mass resolution in the range $10\,\msun \lesssim m_{i} \lesssim 10^{6}\,\msun$. Below $\ll 100\,\msun$, simulations directly resolve early stages of SNe remnant evolution, and it is less important that the coupling is done accurately because the relevant dynamics will be explicitly resolved. Above $\gg 10^{6}\,\msun$, it quickly becomes impossible to resolve even the largest scales of fragmentation and multi-phase structure in the ISM. Such star formation cannot cluster and SNe are not individually time-resolved (i.e.\ a resolution element has many SNe per timestep), so there is no possibility of explicitly resolving overlap of many SNe into super-bubbles, regardless of how the SNe are treated. In that limit, it is necessary to implement a {\em galaxy-wide} sub-grid model for SNe feedback (e.g.\ a model that directly implements a mass-loading of galactic winds as presented in e.g.\ \citealt{vogelsberger:2013.illustris.model,dave:mufasa.followup.gas.metal.sfr.props.vs.time,dave.2016:mufasa.fire.inspired.cosmo.boxes}). 

The scalings above for un-resolved ``PdV work'' are well-studied for SNe, but much less well-constrained for quasi-continuous processes such as stellar mass-loss (OB \&\ AGB winds) and AGN accretion-disk winds. In both cases, the problem is complicated by the fact that the structure and time-variability (e.g.\ ``burstiness'') of the mass-loss processes themselves is poorly understood. Especially for energetic AGN-driven winds, more work is needed to better understand these regimes.

Finally, new physics not included here could alter our conclusions. For example, magnetic fields, or anisotropic thermal conduction, or plasma instabilities altering fluid mixing, or cosmic rays, could all influence the SNe cooling and expansion. Different stellar evolution models could change the predicted SNe rates and/or energetics. It is not our intent to say that the solution here includes all possible physics. However, {\em independent} of these physics, the two key points (\S~\ref{sec:discussion:conservation}-\ref{sec:discussion:pdv}) must still hold! And the goal of any ``sub-grid'' representation of SNe should be to represent the converged solution {\em given the same physics} as the large-scale simulation -- otherwise convergence cannot even be defined in any meaningful sense. So in future work it would be valuable to repeat the exercises in this paper for modified physical assumptions. However, the extensive literature studying the effect of different physical conditions on SNe remnant evolution (see references in \S~\ref{sec:feedback:mechanical:sedov:background}) has shown that the terminal momentum is weakly sensitive to these additional physics. 

\vspace{-0.7cm}
\acknowledgments 
We thank our referee, Joakim Rosdahl, for a number of insightful comments. 
Support for PFH and co-authors was provided by an Alfred P. Sloan Research Fellowship, NASA ATP Grant NNX14AH35G, and NSF Collaborative Research Grant \#1411920 and CAREER grant \#1455342. 
AW was supported by a Caltech-Carnegie Fellowship, in part through the Moore Center for Theoretical Cosmology and Physics at Caltech, and by NASA through grant HST-GO-14734 from STScI.
CAFG was supported by NSF through grants AST-1412836 and AST-1517491, and by NASA through grant NNX15AB22G. 
DK was supported by NSF Grant AST1412153 and a Cottrell Scholar Award from the Research Corporation for Science Advancement. 
The Flatiron Institute is supported by the Simons Foundation. 
Numerical calculations were run on the Caltech compute cluster ``Wheeler,'' allocations TG-AST120025, TG-AST130039 \&\ TG-AST150080 granted by the Extreme Science and Engineering Discovery Environment (XSEDE) supported by the NSF, and the NASA HEC Program through the NAS Division at Ames Research Center and the NCCS at Goddard Space Flight Center. \\

\vspace{-0.2cm}
\bibliography{/Users/phopkins/Dropbox/Public/ms}

\begin{appendix}

\vspace{-0.5cm}
\section{Additional Resolution Tests}
\label{sec:resolution}

\begin{figure*}
\plotsidesize{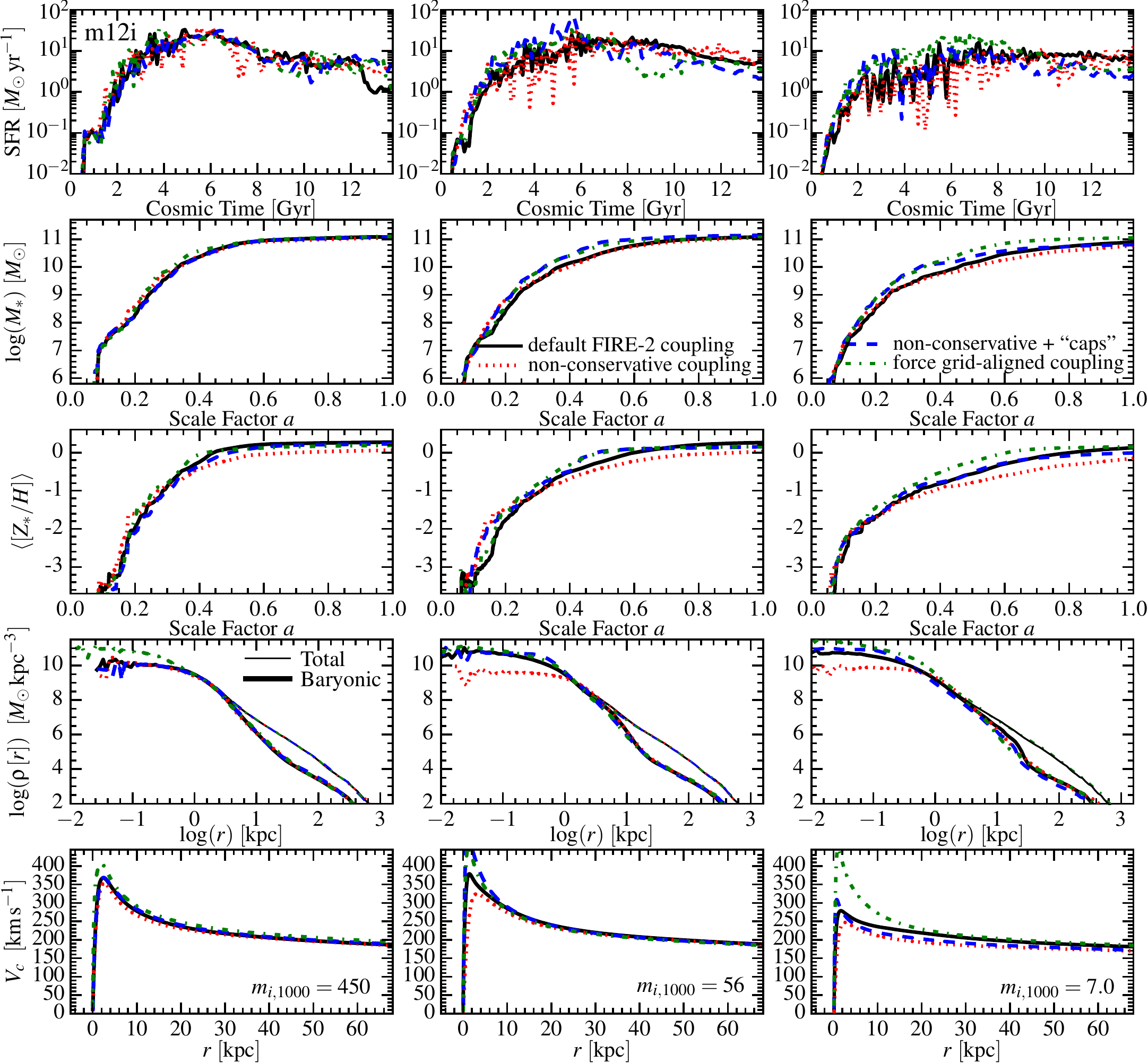}{0.9}
    \vspace{-0.25cm}
    \caption{Additional mechanical FB coupling tests, as Fig.~\ref{fig:sf.history.sne.algorithm}. We focus on MW-mass galaxies, where differences are maximized. Here we compare the ``default FIRE-2 coupling'' and the ``non-conservative coupling'' algorithms from the text, at three resolution levels (labeled). We also compare: 
    {\em Non-conservative+``caps'':} a modified version of the non-conservative algorithm which adds simple ``caps'' to the feedback to prevent conservation errors from getting too large, at the expense of not always coupling the feedback that should be present (\S~\ref{sec:non.con.extreme}). The artificial suppression slightly weakens feedback but resolves the larger differences between ``default'' and ``non-conservative'' algorithms. 
    {\em Force grid-aligned coupling:} a scheme where we artificially inject all feedback in the Cartesian $x/y/z$ grid directions (regardless of the physical gas particle/cell shapes and distribution around the star), mimicking simple coupling in a fixed grid (\S~\ref{sec:grid}). This artificially forces alignment of the disk \&\ winds with coordinate axes, generating spurious torques that remove angular momentum and make the disk more compact. 
    Our conclusions from the text do not change with resolution. 
    \vspace{-0.4cm}
    \label{fig:sf.sne.algorithm.hires.models}}
\end{figure*}

Extensive resolution tests of our ``default'' algorithm, at both dwarf and MW mass scales (and considering both mass and spatial resolution, and additional halos) are presented in \paperone. The main text here also directly compares the different sub-grid treatments of un-resolved cooling (``Fully-Thermal,'' ``Fully-Kinetic,'' etc.\ models) as a function of resolution. Here we simply note that we have re-run tests of the different purely numerical SNe coupling schemes from Figs.~\ref{fig:sf.history.sne.algorithm}-\ref{fig:images.resolution.nonsymmetric}, at both dwarf and MW mass scales, at several resolution levels. In both cases we find our conclusions from the main text are not sensitive to resolution. In the dwarf case this is unsurprising, since there was no significant effect from the coupling algorithm. For MW-mass halos, we demonstrate this explicitly in Fig.~\ref{fig:sf.sne.algorithm.hires.models}. 

\vspace{-0.5cm}
\section{Confirmation that the Errors in the Non-Conservative Algorithm Are Dominated By Extreme Events}
\label{sec:non.con.extreme}

\begin{figure}
\begin{tabular}{cc}
\vspace{-0.1cm}
\hspace{-0.20cm}
\includegraphics[width=0.49\columnwidth]{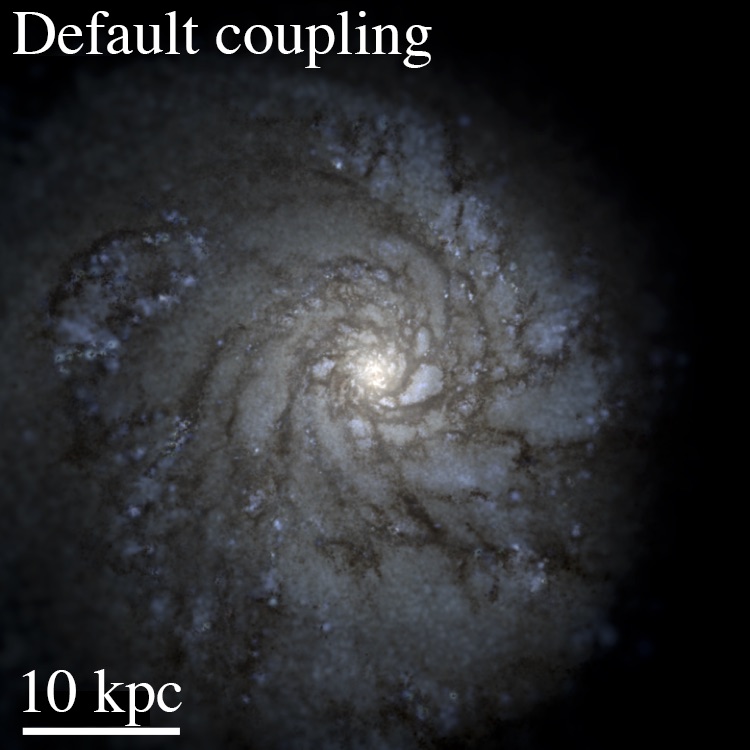} &
\hspace{-0.50cm}
\includegraphics[width=0.49\columnwidth]{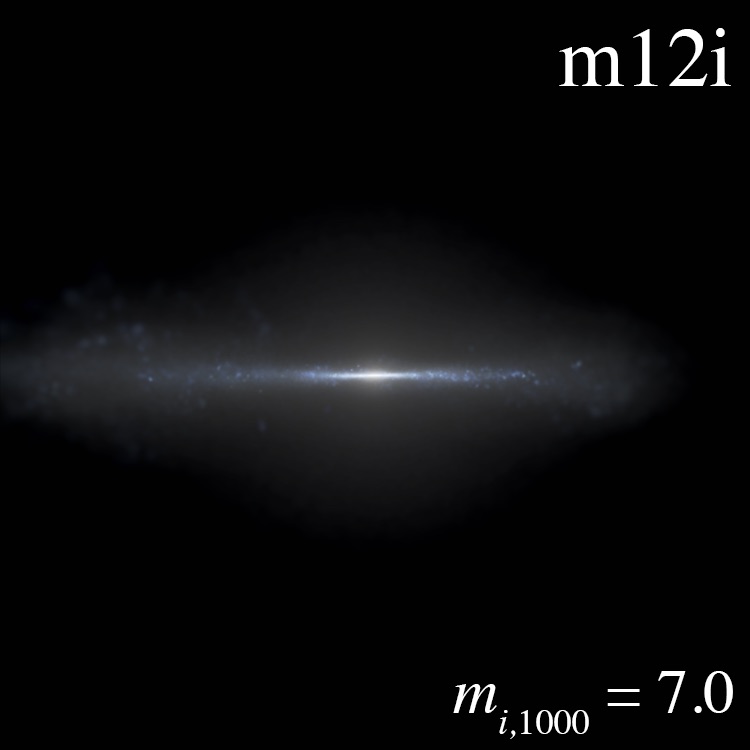} \\
\vspace{-0.09cm}
\hspace{-0.20cm}
\includegraphics[width=0.49\columnwidth]{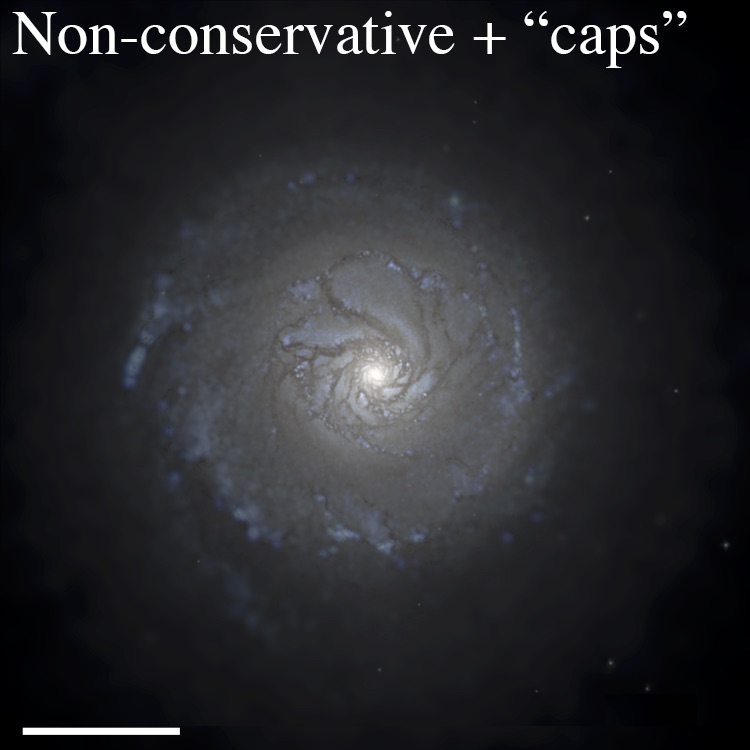} &
\hspace{-0.50cm}
\includegraphics[width=0.49\columnwidth]{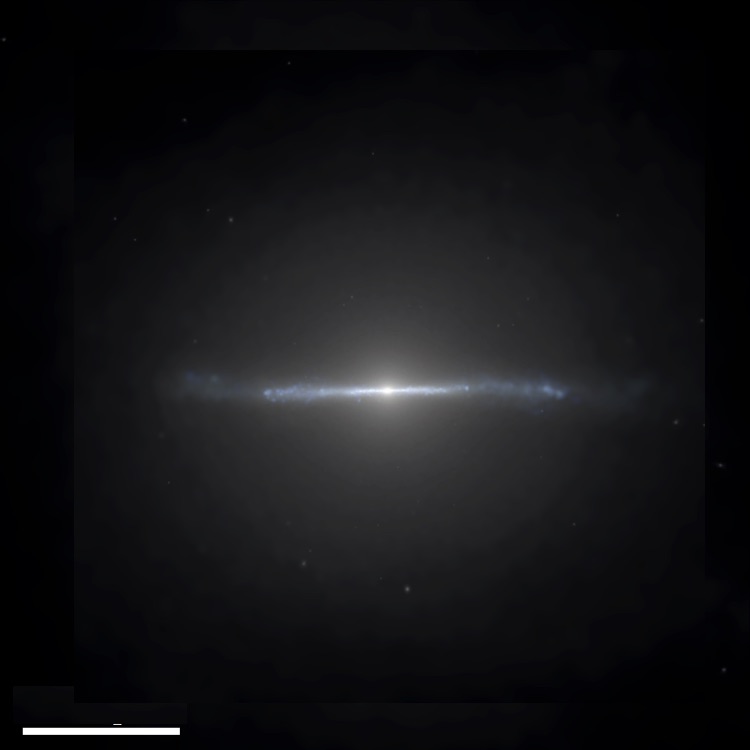} \\
\vspace{-0.04cm}
\hspace{-0.20cm}
\includegraphics[width=0.49\columnwidth]{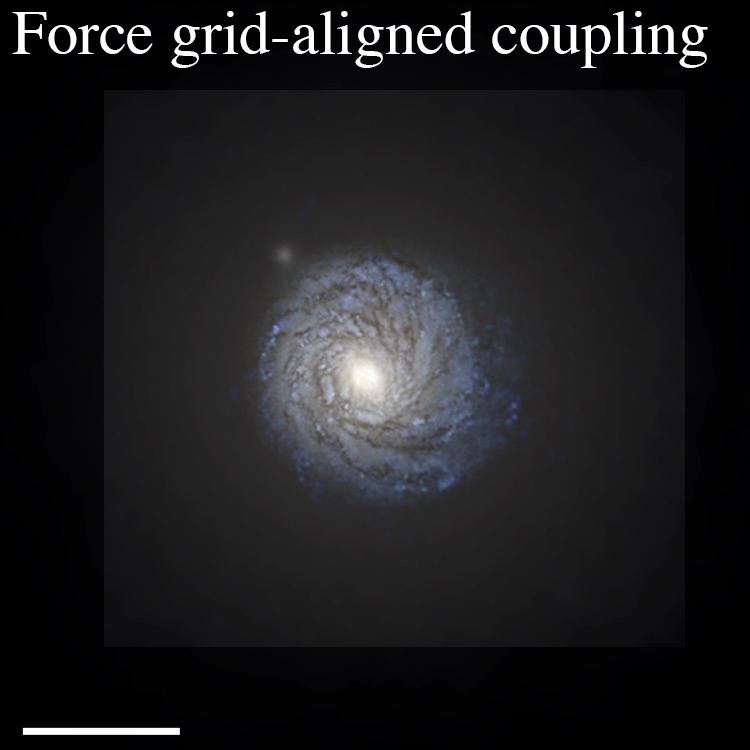} &
\hspace{-0.50cm}
\includegraphics[width=0.49\columnwidth]{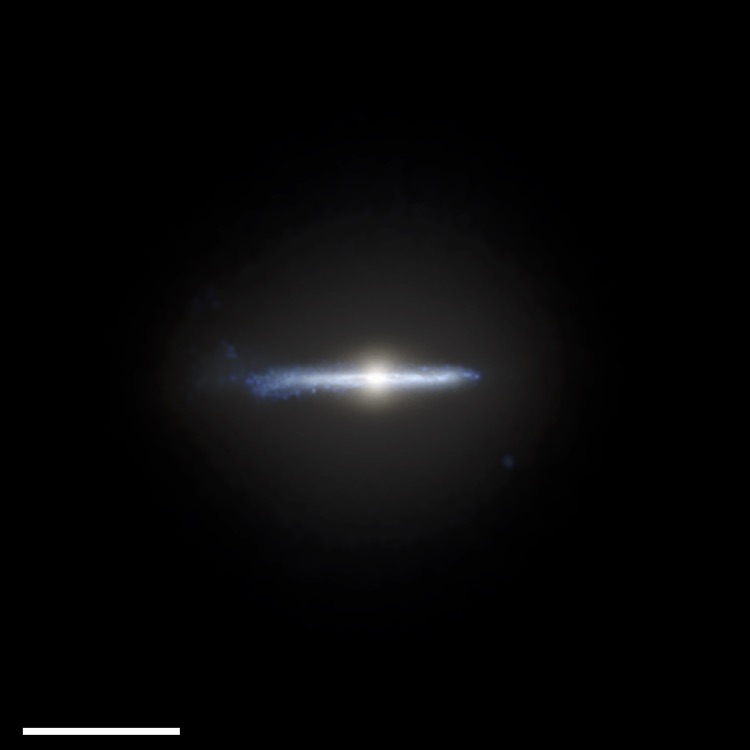} \\
\end{tabular}
    \vspace{-0.25cm}
    \caption{Mock images in HST bands of our {\bf m12i} run at $z=0$ at our highest resolution ($m_{i,\,1000}=7.0$), for the alternative SNe coupling tests in Fig.~\ref{fig:sf.sne.algorithm.hires.models}. With the ``caps'' added to the non-conservative method, the catastrophic errors in Fig.~\ref{fig:images.resolution.nonsymmetric} are suppressed and the morphology agrees with our ``default'' run reasonably well. In the ``force grid-aligned coupling'' run, the spurious torques from numerically forcing the winds along the coordinate axes (incorrectly) drive the disk into alignment with these axes, removing angular momentum from recycling material and producing a more compact disk.
    \vspace{-0.4cm}
    \label{fig:images.resolution.diff.coupling}}
\end{figure}

In \S~\ref{sec:feedback:mechanical:ideal.tests:firesims}, we showed that at sufficiently high resolution, the non-conservative algorithm can produce momentum errors which destroy the thin-disk morphology of a simulated MW-mass galaxy. Here we confirm that the dramatic effects seen there are dominated by the smaller number of ``extreme'' or ``worst case'' events, rather than smaller errors that occur more ubiquitously in a non-conservative algorithm.

Specifically, in Figs.~\ref{fig:sf.sne.algorithm.hires.models} \&\ \ref{fig:images.resolution.diff.coupling}, we conduct the same tests as Figs.~\ref{fig:sf.history.sne.algorithm}-\ref{fig:images.resolution.nonsymmetric}, but with a modified non-conservative algorithm (``non-conservative + caps''). Here we take the non-conservative formulation from \S~\ref{sec:feedback:mechanical:ideal.tests:firesims} and -- for testing purposes only -- limit the most serious errors by enforcing an upper limit to the fraction of SNe momentum coupled to any one particle ($=h_{b}^{2}/4\,[{\rm MAX}(h_{a},\,h_{b})^{2} + |{\bf x}_{ba}|^{2}]$; where $h_{a}^{3}\equiv m_{a}/\rho_{a}$) and an upper limit to the maximum velocity change of $\sim 50\,{\rm km\,s^{-1}}$ (per ``event'').

Figs.~\ref{fig:sf.sne.algorithm.hires.models} \&\ \ref{fig:images.resolution.diff.coupling} clearly demonstrate that it is only the most severe, pathological local coupling cases in the ``non-conservative'' algorithm which generate the ``disk destruction'' (as opposed to an integrated sum of small errors). Running this ``capped'' model at the same resolution, we see a reasonable, clearly thin-disk morphology emerge, in good agreement with our default run. So long as we control (or better yet, eliminate) these errors at a reasonable level, they do not corrupt our solutions. This is why at lower resolution (where the ``worst case'' kick magnitude was much smaller, $< 10\,{\rm km\,s^{-1}}$ for a single gas element), as we studied in FIRE-1, we do not see problematic behavior.

\vspace{-0.5cm}
\section{Problems with Explicitly-Grid-Aligned Feedback Coupling}
\label{sec:grid}

In \S~\ref{sec:feedback:mechanical:ideal.tests:firesims}, we discuss the effects of the purely numerical mechanical feedback coupling algorithm. We discussed the importance of algorithms which respect statistical isotropy. Here we compare another algorithm which is not statistically isotropic, for a different reason. 

In Fig.~\ref{fig:sf.sne.algorithm.hires.models}, we conduct the same tests as Figs.~\ref{fig:sf.history.sne.algorithm}-\ref{fig:images.resolution.nonsymmetric}, but we consider a ``Force grid-aligned coupling'' model. The coupling follows our default algorithm, except we treat the particles around the SN {\em as if} they were distributed in a perfect Cartesian lattice with the SN at the center (as if the SN exploded at the exact center of a cell in a Cartesian grid code), and so enforce the exact same coupling in the $\pm x$, $\pm y$, $\pm z$ coordinate directions. This trivially ensures momentum conservation but is not the correct solution given the actual non-grid distribution of particles. Moreover it imprints the coordinate axes of the simulation directly onto the galaxy -- it is a fundamentally non-statistically-isotropic algorithm. But this is useful for comparison, because such ``preferred directions'' are generic to Cartesian grid-based simulations (e.g.\ AMR) and their SNe coupling schemes. 

The ``grid-aligned'' implementation shows a higher central $V_{c}$, especially at our highest resolution ($m_{i,\,1000}=7$), owing to a more compact disk. This is evident in Fig.~\ref{fig:images.resolution.diff.coupling}, where we compare the $z=0$ visual morphologies of the MW-mass simulations run at our highest resolution. In the ``grid-aligned'' implementation (uniquely), the disk is nearly perfectly-aligned with the simulation coordinate axes -- not surprising given that feedback is forcibly aligned in this case. This artificial alignment generates strong torques on outflowing/recycling material, as well as material within the disk (it must be torqued from its ``natural'' orientation); as winds recycle and the disk first forms, this in turn produces a significant loss of angular momentum. As a result, the late-time inflowing/recycled material (which forms the disk) has  lower angular momentum in this run, and produces a more compact disk, with a much higher central $V_{c}$. Note the error is essentially independent of resolution (whereas the central $V_{c}$ decreases with resolution, in all other algorithms tested), because the grid alignment is resolution-independent. 

We show this to emphasize that this can be a serious worry for fixed-grid or adaptive mesh refinement (AMR) codes, where grid-alignment of disks is a ubiquitous and well-known problem \citep{davis:1984.rotating.upwind.eulerian.scheme}, even at extremely high resolution and independent of feedback \citep[because the hydrodynamics themselves are grid-aligned; see e.g.][]{de-val-borro:2006.disk.planet.interaction.comparison,byerly:2014.hybrid.cartesian.scheme.for.ang.mom,hopkins:gizmo}, especially in simulations of cosmological disk formation \citep[see][]{hahn:2010.disk.gal.orientations.ramses}. This may bias these simulations to smaller, more compact galaxies.

Given the highly-irregular dSph morphology of {\bf m10q}, there is not an obvious difference in that galaxy with this algorithm (there is no thin disk to torque); we therefore do not show a detailed comparison. 

We have re-run halos {\bf m09} and {\bf m10v} (both dwarfs), and {\bf m12f} and {\bf m12m} (MW-mass) from \paperone\ with this algorithm to confirm the results are robust across halos at both dwarf and MW mass scales.

\vspace{-0.5cm}
\section{Delayed-Cooling and Target-Temperature Models: Tests}
\label{sec:delayed.cooling}

\begin{figure}
\hspace{-0.4cm}\plotonesize{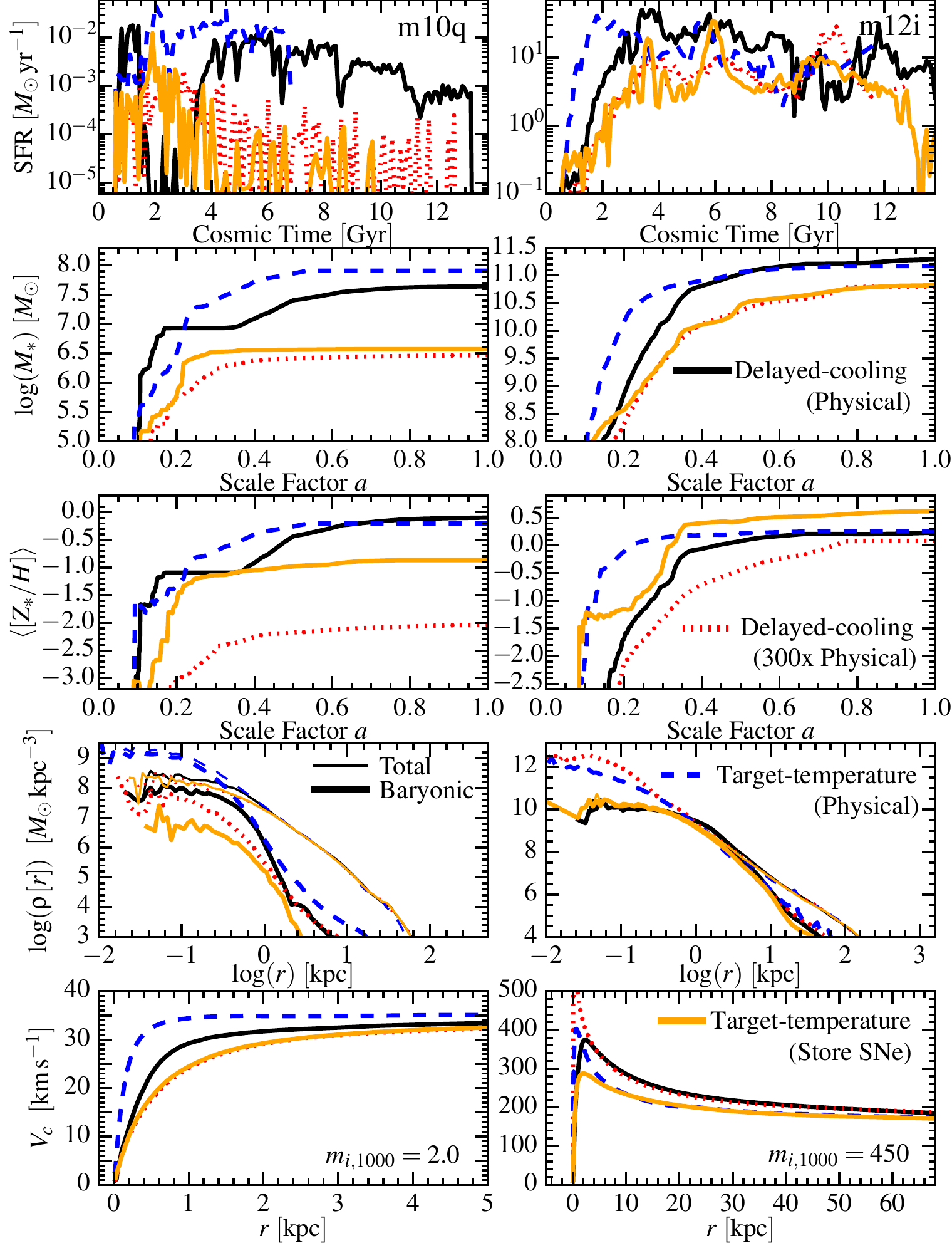}{1.04}
    \vspace{-0.5cm}
    \caption{Tests of ``Delayed-cooling'' or ``Target-temperature'' models (\S~\ref{sec:delayed.cooling}), as Fig.~\ref{fig:sf.history.sne.algorithm}. 
     In these models cooling is artificially ``turned off'' either explicitly (not allowing SN-heated gas to cool for some $\Delta t_{\rm delay}$) or implicitly (by ``storing'' SNe energy for an arbitrarily long time, until sufficient energy has accumulated to heat gas particles to some target temperature $T_{\rm target}$). We compare: {\em Delayed-cooling (Physical):} Cooling is turned off only for the $\Delta t_{\rm delay}$ corresponding to the actual blastwave cooling time (time when the blast reaches the cooling radius). If the cooling radius is un-resolved, this timescale is necessarily much shorter than the dynamical time. {\em Delayed-cooling (300xPhysical):} $\Delta t_{\rm delay}$ is increased by an arbitrary factor of $\sim 300$ to force it to $\gtrsim 10^{7}$\,yr, comparable to the galaxy dynamical time. {\em Target-temperature (Physical):} SNe energy is deposited in such a way (distributed among neighbors) to bring ``coupled'' neighbors as close as possible to a target temperature $T_{\rm target}=10^{7.5}$\,K, but {\em without} artificially turning off cooling or ``saving'' SNe after they should explode. {\em Target-temperature (Store SNe):} SNe are ``stored'' until the gas neighbors can all be raised to exactly $T_{\rm target}=10^{7.5}$\,K (roughly equivalent to a cooling-delay time of $\Delta t_{\rm delay} \sim 30\,$Myr; or forcing $10^{5}$\,SNe to explode simultaneously at a single place and time in the {\bf m12i} run). With a physical cooling time or post-SNe temperature, these models are very similar to the ``Fully-Thermal'' model in the text, and produce severe over-cooling at low resolution (as expected). By making the effective cooling-delay or ``storage'' time very large, cooling becomes inefficient and galactic outflows are driven; however this requires un-physical values that do not resemble the solution for resolved explosions. As a result the star formation histories, rotation curves, metal abundance distributions, and gas phase structure do not resemble the converged solutions in the main text. 
    \vspace{-0.4cm}
    \label{fig:delaycool.physics}}
\end{figure}

\begin{figure}
\hspace{-0.4cm}\plotonesize{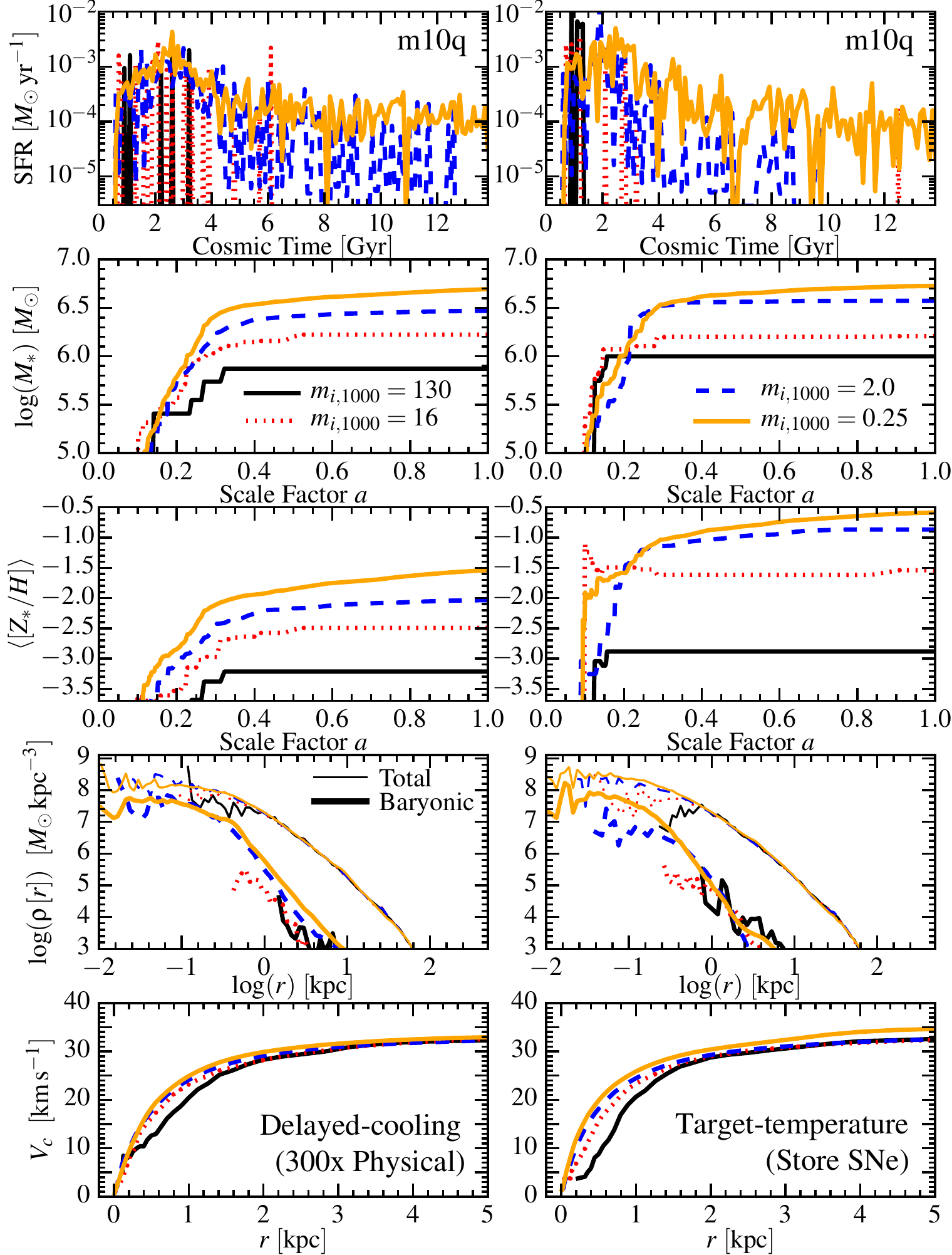}{1.04}
    \vspace{-0.5cm}
    \vspace{-0.25cm}
    \caption{Comparison of the ``Delayed-cooling (300xPhysical)'' ({\em left}) and ``Target-temperature (Store SNe)'' ({\em right}) models from Fig.~\ref{fig:delaycool.physics}, as a function of resolution (in {\bf m10q}). Neither exhibits convergence down to mass resolution $m_{i}=250\,\msun$: stellar masses, metallicities, central galaxy (and dark matter) densities, rotation curves in the central $\sim$\,kpc, and late-time star formation rates all increase systematically with resolution. Because the models do not physically correspond to any stage of a physical SNe explosion, they in fact {\em should not} converge to the correct solution even at infinite resolution (for example, even if the cooling radius is resolved, the ``Delayed-cooling (300xPhysical)'' model will necessarily prevent it from cooling for $\sim 300$ times longer than it should). For this reason neither model is well-suited to the convergence studies in the main text, and they should not be used in simulations which can resolve super-bubble overlap and physical cooling radii. 
    \vspace{-0.4cm}
    \label{fig:delaycool.resolution}}
\end{figure}

We briefly discussed ``delayed-cooling'' and ``target-temperature'' models in the text in \S~\ref{sec:delayed.cooling.discussion}. There we emphasized that such models are fundamentally ill-posed at high resolution. Here we demonstrate this explicitly in zoom-in cosmological simulations at dwarf and MW mass scales. 

We compare four simple models, which resemble common implementations in the literature. 
\begin{enumerate}

\item{\em Delayed-Cooling (Physical):} Here we take the ``Fully-Thermal'' model from the text (injecting the full $10^{51}\,{\rm erg}$ per SNe as thermal energy), but particles which are heated by the SNe are not allowed to cool for a time $\Delta t_{\rm delay}$. Physically, the cooling time of an explosion is (by definition) the time it takes to reach $R_{\rm cool}$: since it is in an energy-conserving Sedov-phase before this, the time is $t^{\rm shock}_{\rm cool} = (2/5)\,R_{\rm cool}/v(R_{\rm cool})$ where $v(R_{\rm cool}) \equiv p_{\rm t}/M_{\rm cool}$ is the velocity at this stage. Given the expression for $p_{\rm t}$ in the text (Eq.~\ref{eqn:terminal.p}), this is $t^{\rm shock}_{\rm cool} = 5\times10^{4}\,{\rm yr}\,(n/{\rm cm^{-3}})^{-4/7}\,(Z/Z_{\sun})^{-5/14}$ (this is, as it should be, approximately the physical cooling time for metal-enriched gas with the post-shock temperature appropriate for a shock velocity $v(R_{\rm cool})\sim 200\,{\rm km\,s^{-1}}$). So we adopt $\Delta t_{\rm delay} = t^{\rm shock}_{\rm cool}$. Note, though, that $t^{\rm shock}_{\rm cool}$ is $\sim 1000$ times shorter than the gas dynamical time ($1/\sqrt{G\,\rho}$) -- so unless the cooling radius (stage of blastwave expansion where the expansion time is shorter than $t^{\rm shock}_{\rm cool}$) is resolved, this will do little work.\footnote{In our delayed-cooling experiments, we have considered both turning off {\em all} cooling for a particle, and tracking a separate reservoir of SNe-injected energy, which is not allowed to cool (while other energy can cool). Both give similar results for our comparison here. We also ``reset'' the delay time $\Delta t_{\rm delay}$ whenever a new SNe injects energy into a gas element.}

\item{\em Delayed-Cooling (300xPhysical):} We take the ``Delayed-Cooling (Physical)'' model, but arbitrarily multiply the delay timescale by a factor of $300$. This brings it to $\gtrsim 10^{7}\,$yr, comparable to the galaxy dynamical time. 

\item{\em Target-Temperature (Physical):} We take the ``Fully-Thermal'' model from the text but wish to heat the targeted gas particles as close as possible to some desired target temperature, $T_{\rm target}\sim 10^{7.5}\,$K, without artificially changing the physics. Therefore we adjust the number of neighbor particles on-the-fly as needed to get as close as possible to this goal. However, even putting $10^{51}\,{\rm erg}$ into a single particle can only increase the temperature by $\Delta T \approx 2.4\times10^{6}\,\,(m_{i}/1000\,\msun)^{-1}\,$K. So typically this amounts to putting $100\%$ of the energy of each SN into the neighbor particle which is closest to (but still below) $T_{\rm target}$. 

\item{\em Target-Temperature (Store SNe):} We {\em require} that all gas particles heated by a SN receive sufficient energy such that their temperature rises by $T_{\rm target}=10^{7.5}\,$K. We follow \citet{dalla.vecchia:target.temperature.sne.delayed.cooling.feedback}\footnote{The \citet{dalla.vecchia:target.temperature.sne.delayed.cooling.feedback} ``target-temperature'' implementation released the SNe energy stochastically rather than deterministically after a fixed time -- we have implemented this as well and the results are identical to the ``target-temperature (store SNe)'' implementation shown.} and achieve this by implicitly turning off cooling -- we ``store'' SNe until a sufficient number have accumulated in order to heat a target gas mass by the desired $T_{\rm target}$. Then all the SNe energy is deposited ``at once'' in that gas in a thermal-energy dump. To minimize the number of SNe which must be ``stored,'' we set a target gas mass (for each ``heating event'') of just $10$ gas particles. Given this, the number of SNe which must be ``stored'' and then injected simultaneously is $N_{\rm SNe} \sim 10^{5}\,(m_{i}/10^{6}\,\msun)$; this is physically similar to delaying cooling for $\sim 30\,$Myr (while the SNe accumulate) for a gas particle surrounded by $\sim 10$ star particles.

\end{enumerate}

Figs.~\ref{fig:delaycool.physics}-\ref{fig:delaycool.resolution} repeat the experiments from \S~\ref{sec:delayed.cooling.discussion} in the main text, for these models. Not surprisingly, at the resolution shown ($m_{i}\gg 100\,\msun$), the ``physically-motivated'' models (either delayed-cooling or target-temperature) resemble the ``Fully-Thermal'' model from the text, which itself resembled the ``no SNe'' result. Turning off cooling only for the real cooling time, or heating gas only to the correct physical temperature, ignoring momentum, leads to over-cooling at low resolution. 

Of course, in this class of models we can simply adjust the model parameters until a reasonable stellar mass is obtained. The ``Delayed-cooling (300xPhysical)'' and ``Target-temperature (Store SNe)'' models manage to produce order-of-magnitude similar galaxy masses to our converged default model at low resolution. However there are serious issues. 
\begin{enumerate}

\item The actual explicit or implicit ``cooling turnoff times'' are wildly unphysical ($\gtrsim 10\,$Myr) -- many orders of magnitude larger than physical in both cases \citep[see][]{martizzi:sne.momentum.sims,agertz:sf.feedback.multiple.mechanisms}. Thus the solutions we ``insert'' on large scales do {\em not} in any way resemble a ``down-sampled'' high-resolution simulation; nor can the relevant parameters be predicted {\em a priori} from higher-resolution simulations. Note that such unphysically-long delayed cooling times are what are actually used in most simulations with these ``delayed cooling'' models \citep[e.g.][]{stinson:2006.sne.fb.recipe,shen:2014.seven.dwarfs,crain:eagle.sims}. 

It has been suggested that these models, while obviously unphysical for a single SN explosion, could represent the result of SNe which are strongly clustered in both space and time. However, all the simulations here, by allowing resolved cooling into GMCs, explicitly resolve stellar clustering (and if anything, we show in \paperone\ that low resolution tends to over-estimate clustering, owing to discrete star-particle sampling). Therefore if such clustering were to occur, one would not need to artificially turn off cooling or store SNe (one could simply allow the explosions to occur rapidly and create a super-bubble, as occurs in our default models). In contrast, these models {\em impose}, rather than predict, a strong and {\em explicitly resolution-dependent} assumption about clustering: for e.g.\ the target temperature model it is that SNe explode in ``units'' in both time and space of $\sim 10^{5}\,{\rm SNe}\,(m_{i}/10^{6}\,\msun)$. 

\citet{walch.naab:sne.momentum,martizzi:sne.momentum.sims,kim:tigress.ism.model.sims} and most explicitly \citet{kim:superbubble.mass.loading} have demonstrated this in greater detail, in studies of idealized single-SN explosions or clustered SNe in a sub-volume of the ISM. There, these authors demonstrate more explicitly that delayed cooling or target-temperature models are not a good approximation to the ``down-sampled'' results of high-resolution simulations.

\item As also noted by \citet{agertz:sf.feedback.multiple.mechanisms}, this un-physical feedback coupling produces several artifacts in the galaxy properties. 
{\bf (1)} Shapes of the star formation histories are biased: in dwarfs the star formation in both cases is much more concentrated at early times, compared to our converged solutions in Fig.~\ref{fig:sf.fb.sne.subgrid.convergence}. 
{\bf (2)} In massive galaxies, the ``delayed cooling'' model accumulates a massive reservoir of gas (with its cooling turned off by successive generations of SNe) at the galaxy center, which finally (because of the dependence of $t_{\rm delay}$ on density and metallicity) achieves a short cooling time even with the imposed ``delay,'' then forms a strong starburst (at time $\sim 10\,$Gyr) and leaves an extremely compact bulge (the $\sim 500\,{\rm km\,s^{-1}}$ rotation-curve peak). 
{\bf (3)} The metal abundances are highly sensitive to the ``delayed cooling'' and ``target temperature'' model implementations, and vary by several orders of magnitude in the variants explored here. The metallicities for dwarfs are extremely high in the ``target temperature'' models, because the ``stored'' SNe inject a huge metal mass simultaneously,\footnote{If we ``store up'' SNe each with $\sim 10^{51}\,{\rm erg}$ until we can heat a discrete mass $\Delta m$ to a temperature $\sim 10^{7.5}\,$K, then if each SN deposits $\sim 2\,\msun$ worth of metals, the mass $\Delta m$ will be immediately enriched to metallicity $Z\approx 2\,Z_{\sun}$!} which is ejected from the galaxy but is so metal-rich that it re-cools and preferentially forms the next generation of stars. We have verified this feature remains regardless of whether we include or exclude explicit ``turbulent metal mixing'' (numerical metal diffusion) terms as described in \paperone. 
{\bf (4)} The gas phase structure is quite different from our converged solutions in the text. Since these models rely only on hot gas, there is little or no cool ($\sim 10^{4}-10^{5}\,$K) or cold ($\lesssim 10^{4}\,$K) gas in the outflows here, unlike our default simulations \citep[see][]{muratov:2015.fire.winds,muratov:2016.fire.metal.outflow.loading,faucher-giguere:2014.fire.neutral.hydrogen.absorption,faucher.2016:high.mass.qso.halo.covering.fraction.neutral.gas.fire,angles.alcazar:particle.tracking.fire.baryon.cycle.intergalactic.transfer}, although \citet{rosdahl:2016.sne.method.isolated.gal.sims} show that alternative ``delayed cooling'' implementations err in the opposite manner and produce far too much cold, dense gas in outflows.

\item The solutions are non-convergent. Fig.~\ref{fig:delaycool.resolution} shows this explicitly, re-running {\bf m10q} at resolution from $250-10^{5}\,\msun$, with the ``Delayed-cooling (300xPhysical)'' and ``Target-temperature (Store SNe)'' models. In both, the stellar masses, metallicities, central galaxy (and dark matter) densities, rotation curves in the central $\sim$\,kpc, and late-time star formation rates all increase systematically as the resolution increases. 

Of course, this owes to the explicit resolution-dependence of the assumed clustering and blastwave structure of SNe. In target-temperature models, the SNe cluster and are synchronized in time and space in an explicitly particle-mass dependent manner. In delayed-cooling models, the ``cooling mass'' $M_{\rm cool}$ is essentially defined to be the mass of the kernel over which the SNe are distributed (some multiple of the particle mass): since the terminal momentum for an energy-conserving blast (which this is forced to be, by not allowing cooling) is $p_{t} \sim \sqrt{E_{\rm SNe}\,M_{\rm cool}}$, the momentum injected increases $\propto M_{\rm cool}^{1/2} \propto m_{i}^{1/2}$, so feedback becomes more efficient at lower resolution (analogous to the ``Fully-Kinetic'' models discussed in the text). 

Interestingly, while the lack of convergence for delayed-cooling models is ``smooth,'' the ``target temperature'' models exhibit false convergence in some properties (such as stellar mass) at low resolution, then ``jump'' in the predicted values once a critical mass resolution (here $\sim 2000\,\msun$) is reached. That is of course the mass resolution where the {\em physical} cooling radii of SNe begin to be resolved: so the fundamental meaning and behavior of the sub-grid model changes. At even higher resolution, the ``target temperature'' of $\sim 10^{7.5}$\,K would actually become {\em lower} than the correct, resolved blastwave temperatures: this would lead one to ``store'' $<1$ SN at a time. Clearly, in this limit the ``delayed cooling'' and ``target temperature'' models simply become ill-defined.

\end{enumerate}

\vspace{-0.5cm}
\section{Energy-Conserving Solutions Accounting for Arbitrary Star-Gas Motions}
\label{sec:energy.cons.w.motion}

\begin{figure}
\hspace{-0.4cm}\plotonesize{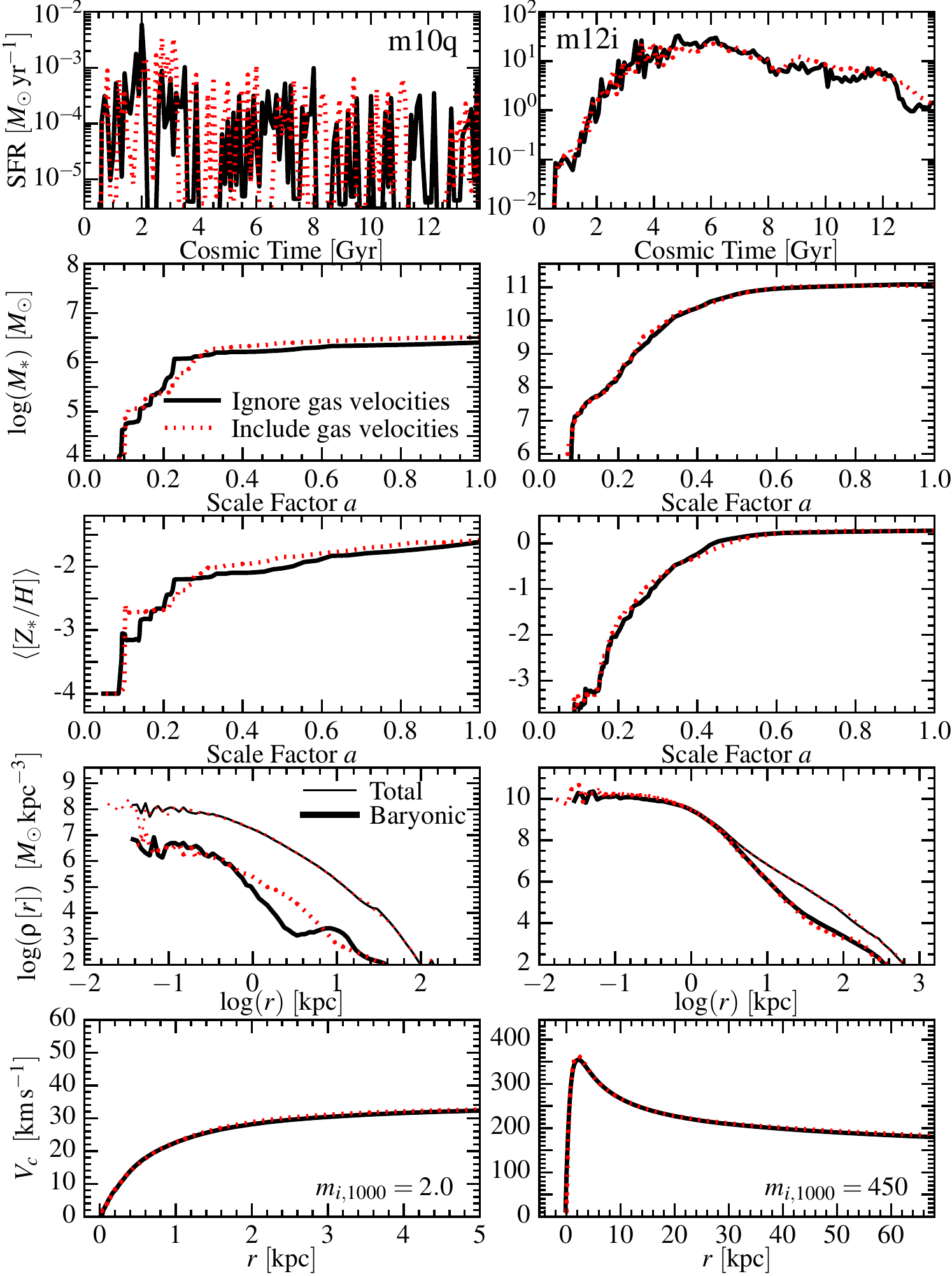}{1.04}
    \vspace{-0.5cm}
    \caption{Comparison of our ``default'' feedback implementation from the text (as Fig.~\ref{fig:sf.history.sne.algorithm}), with and without accounting for the relative gas-star motions as described in \S~\ref{sec:energy.cons.w.motion}. The ``Ignore gas velocities'' implementation treats the momentum injection by SNe as if the ambient medium is at rest (i.e.\ $\psi=\phi=1$, in \S~\ref{sec:energy.cons.w.motion}); so the kinetic energy input is variable. The ``Include gas velocities'' implementation treats the energy injection as fixed and re-scales the injected momentum if there are large gas-star velocities. The latter can have significant effects on the coupled momentum at sufficiently low resolution, if the gas is uniformly approaching or receding from SN locations at high velocities ($\gg 200\,{\rm km\,s^{-1}}$). Both implementations become identical at sufficiently high resolution. Even at low resolution (shown here), the differences in galaxy properties are extremely small.
    \vspace{-0.4cm}
    \label{fig:sneenergy.method.tests}}
\end{figure}

In the text, we noted that, for a spherically symmetric blastwave propagating into a medium initially at rest, converting an energy $E_{\rm ej} = (1/2)\,m_{\rm ej}\,{v}_{\rm ej}^{2}$ into kinetic energy (pure radial momentum), after coupling to a total mass $m_{b}$, simply implies a final kinetic energy $p_{\rm final}^{2}/(2\,(m_{b} + m_{\rm ej})) = E_{\rm ej}$, giving $p_{\rm final} = (1+m_{b}/m_{\rm ej})^{1/2}\,m_{\rm ej}\,(2\,E_{\rm ej}/m_{\rm ej})^{1/2} = (1+m_{b}/m_{\rm ej})^{1/2}\,p_{\rm ej}$, where $p_{\rm ej}=m_{\rm ej}\,v_{\rm ej}$ is the initial ejecta momentum. The situation is more complex if we allow for arbitrary initial gas and stellar velocities. 

First recall the mass conservation condition $\sum \Delta m_{b} = m_{\rm ej}$ is un-altered by star or gas motion. The momentum condition is, in the rest frame of the star, $\sum \Delta {\bf p}_{b} = \mathbf{0}$, which in the lab/simulation frame becomes $\sum \Delta {\bf p}^{\prime}_{b} = m_{\rm ej}\,{\bf v}_{a}$, trivially satisfied by the boost $\Delta {\bf p}_{b}^{\prime} = \Delta {\bf p}_{b} + \Delta m_{b}\,{\bf v}_{a}$. In these two cases, no net mass or linear momentum is created/destroyed. For energy, we must account for the energy injected. Consider a hypothetical instant ``just after'' explosion, but ``before'' coupling. Then the mass of the star particle is $m_{a}-m_{\rm ej}$, moving at ${\bf v}_{a}$. Gas neighbors $b$ have their ``unperturbed'' mass and velocities, etc. In the rest-frame of the star, the ejecta contain the energy $E_{\rm ej}=(1/2)\,m_{\rm ej}\,{v}_{\rm ej}^{2}$. Assume the ejecta have negligible initial internal energy, then $v_{\rm ej}$ is the real radial velocity. If the ejecta are isotropic in the rest frame, each parcel in some solid angle $d\Omega$ carries mass $dm = m_{\rm ej}/(4\pi)\,d\Omega$, with velocity ${\bf v}_{\rm ej} = v_{\rm ej}\,\hat{r}$ (where $\hat{r}$ points from the star radially outward). If the star is moving initially at velocity ${\bf v}_{a}$, the whole system is boosted, and ${\bf v}^{\prime}_{\rm ej} = v_{\rm ej}\,\hat{r} + {\bf v}_{a}$. To calculate $E_{\rm ej}^{\prime} = (1/2)\,\int |{\bf v}^{\prime}_{\rm ej}|^{2}\,dm$ in the lab frame, note $|{\bf v}^{\prime}_{\rm ej}|^{2} = v_{\rm ej}^{2} + 2\,{\bf v}_{\rm ej}\cdot {\bf v}_{a} + v_{a}^{2} = v_{\rm ej}^{2} + v_{a}^{2} + 2\,v_{\rm ej}\,v_{a}\,\cos{\theta}_{ea}$ (where we can define standard spherical coordinates such that $\hat{r}\cdot \hat{\bf v}_{a} \equiv \cos{\theta}_{ea}$). Using $dm = m_{\rm ej}/(4\pi)\,d\Omega = m_{\rm ej}/(4\pi)\,d\phi\,d\cos{\theta}_{ea}$, we trivially obtain that the ``cross-term'' ${\bf v}_{\rm ej}\cdot {\bf v}_{a}$ vanishes (integrating over all ejecta), so we have $E_{\rm ej}^{\prime} = (m_{\rm ej}/2)\,(v_{\rm ej}^{2} + v_{a}^{2})$.

Now assume we couple some energy and momentum to all the neighbors $b$ -- the exact {\em discrete} energy conservation condition must be satisfied, summed over all elements which receive some ejects mass/energy/momentum. In the simulation/lab frame, the energy conservation condition can be written: 
\begin{align}
 E_{\rm initial}& + E_{\rm ej}^{\prime} = \\
\nonumber & \frac{(m_{a}-m_{\rm ej})}{2}\,{\bf v}_{a}^{2} + \sum_{b}\,\frac{{\bf p}_{b}^{2}}{2\,m_{b}}+
\frac{m_{\rm ej}}{2}\,({\bf v}_{a}^{2} + v_{\rm ej}^{2}) + U_{0} \\ 
\nonumber &= \frac{m_{a}}{2}\,{\bf v}_{a}^{2} +  \sum_{b}\,\frac{{\bf p}_{b}^{2}}{2\,m_{b}} + \frac{m_{ej}}{2}\,v_{\rm ej}^{2} + U_{0} \\ 
\nonumber &= E_{\rm final} = \frac{(m_{a}-m_{\rm ej})}{2}\,{\bf v}_{a}^{2} + \sum_{b}\,\frac{|{\bf p}_{b} + \Delta {\bf p}^{\prime}_{b}|^{2}}{2\,(m_{b}+\Delta m_{b})} + U_{f}
\end{align}
where (as in the main text) we use $x_{b}$ to denote the pre-coupling value of $x$. Here $U_{0}$ and $U_{f} \equiv U_{0}+\Delta U$ collect the non-kinetic energy terms (e.g.\ thermal energy, discussed below). 

Now using the fact that $\Delta {\bf p}^{\prime}_{b} \equiv \Delta {\bf p}_{b} + \Delta m_{b}\,{\bf v}_{a}$, we can write this in terms of the coupled momenta $\Delta {\bf p}_{b}$. Use the identities ${\bf v}_{b} = {\bf p}_{b}/m_{b}$, $\sum \Delta m_{b} = m_{\rm ej}$, $\sum \Delta {\bf p}_{b} = \mathbf{0}$, $\mu_{b} \equiv \Delta m_{b}/m_{b}$, $1/(1+x) = 1-x/(1+x)$, and following through with some tedious algebra, we can re-arrange terms and write the energy conservation condition as:
\begin{align}
\label{eqn:egycon.reduced} \epsilon &= \sum_{b}\,\left[ \frac{|\Delta {\bf p}_{b}|^{2}}{2\,m_{b}\,(1+\mu_{b})}
+ \frac{{\bf v}_{ba} \cdot \Delta {\bf p}_{b}}{(1+\mu_{b})} 
\right] 
\end{align}
where ${\bf v}_{ba} \equiv {\bf v}_{b}-{\bf v}_{a}$ and the energy $\epsilon$ is defined by: 
\begin{align}
\epsilon  &\equiv \frac{1}{2}\,m_{\rm ej}\,\left( v_{\rm ej}^{2} + \sum_{b} w^{\prime}_{b}\,|{\bf v}_{ba}|^{2} \right) - \Delta U \\ 
w^{\prime}_{b} &\equiv \frac{1}{1+\mu_{b}}\,\left(\frac{\Delta m_{b}}{m_{\rm ej}}\right)\ .
\end{align}
This makes it clear that the dynamics depend only on the {\em relative} velocity ${\bf v}_{ba}$ of gas relative to the star (i.e.\ a uniform boost will not change the dynamics, as it should not). In $\epsilon$, the term in ${\bf v}_{ba}^{2}$ reflects the additional energy generated by relative gas-star motion -- since $\sum w^{\prime}_{b}\approx1$, this is negligible for SNe where ${\bf v}_{ba}^{2} \ll v_{\rm ej}^{2}$, but potentially important for slow winds. 

Now without loss of generality, define the coupled momentum $\Delta {\bf p}_{b}$ as the value we used in the text (for the case where the gas is not moving relative to the star) multiplied by an arbitrary function $\psi_{b}$: 
\begin{align}
\label{eqn:delta.p.defn}\Delta {\bf p}_{b} &\equiv \psi_{b}\,\Delta m_{b}\,\left(1 + \frac{m_{b}}{\Delta m_{b}} \right)^{1/2}\,\left(\frac{2\,\epsilon}{m_{\rm ej}} \right)^{1/2} \Delta\hat{\bf p}_{b}
\end{align}
where $\Delta\hat{\bf p}_{b}\equiv \Delta{\bf p}_{b}/|\Delta{\bf p}_{b}|$. Inserting this into the energy conservation condition in Eq.~\ref{eqn:egycon.reduced}, we obtain the constraint equation in terms of $\psi$: 
\begin{align}
\label{eqn:psi.simplified.b} 1 &= \sum_{b}\,\psi_{b}^{2}\,\frac{\Delta m_{b}}{m_{\rm ej}} + 2\,\sum_{b}\,\psi_{b}\,\cos{\theta}_{ba}\,\left({\frac{w^{\prime}_{b}\,m_{b}\,|{\bf v}_{ba}|^{2}}{2\,\epsilon}}\right)^{1/2} \\ 
&\cos{\theta}_{ba} \equiv \hat{\bf v}_{ba} \cdot \Delta\hat{\bf p}_{b} = \frac{{\bf v}_{ba}\cdot \Delta {\bf p}_{b}}{|{\bf v}_{ba}|\,|\Delta {\bf p}_{b}|}
\end{align}
If ${\bf v}_{ba}=0$ (no initial gas-star motion), then the term in $\cos{\theta}_{ba}$ vanishes, and this is trivially solved for $\psi_{b}=1$, and Eq.~\ref{eqn:delta.p.defn} reduces to our solution for a spherically symmetric explosion in a stationary medium (as it should). More generally, any $\psi_{b}$ and $\Delta\hat{\bf p}_{b}$ must still produce $\sum \Delta{\bf p}_{b} = {\bf 0}$. If we have defined a set of vector weights, as in the main text, such that this is true for the ${\bf v}_{ba}=0$ (stationary) case, then the simplest choice which guarantees $\sum \Delta{\bf p}_{b} = {\bf 0}$ is preserved is to take $\psi_{b}=\psi$, such that $\sum \Delta {\bf p}_{b} \rightarrow \psi\,\sum \Delta {\bf p}_{b}^{\rm stationary} = \mathbf{0}$. Of course in detail the true solution for a blastwave in an inhomogeneous medium with locally varying velocities could feature variable $\psi_{b}$ or changes in the direction $\Delta\hat{\bf p}_{b}$ (i.e.\ work being done in different directions from the initial ejecta expansion); but by definition this sub-structure is un-resolved at the coupling radius so we should think of $\psi$ as an average over the un-resolved structure. The solution to Eq.~\ref{eqn:psi.simplified.b} is then simply:
\begin{align}
\psi_{b} &= \psi \equiv \sqrt{1+\beta_{\psi}^{2}} - \beta_{\psi} \\ 
\beta_{\psi} &\equiv \sum_{b}\,\cos{\theta}_{ba}\,\left({\frac{w^{\prime}_{b}\,m_{b}\,|{\bf v}_{ba}|^{2}}{2\,\epsilon}}\right)^{1/2} 
\end{align}

This gives us the desired expression in the strictly energy-conserving limit. But at low resolution (large particle masses) the solution is not energy-conserving (the blastwave has reached the terminal momentum and radiated energy away). Following the text, return to Eq.~\ref{eqn:egycon.reduced} and insert the terminal momentum $\Delta {\bf p}_{b} = \Delta {\bf p}_{b}^{\rm terminal} = \phi_{b}\,\Delta {\bf p}_{b}^{\rm terminal}({\bf v}_{ba}=0) = \phi_{b}\,(p_{\rm t}/p_{\rm ej})\,\Delta {\bf p}_{b}^{\rm initial} = \phi_{b}\,(p_{\rm t}/p_{\rm ej})\,\Delta m_{b}\,(2\,\epsilon/m_{\rm ej})^{1/2}\,\Delta \hat{\bf p}_{b}$, where $\phi_{b}$ is an arbitrary constant analogous to $\psi_{b}$. Following the solution above we will take $\phi_{b}\rightarrow\phi$. Since some energy has been radiated, the right hand side of Eq.~\ref{eqn:egycon.reduced} must be $< \epsilon$ (the constraint is an inequality). This is solved by:
\begin{align}
\phi &= {\rm MIN}\left[1,\,{\alpha_{\phi}^{-1}}\,\left(\sqrt{\beta_{\phi}^{2}+{\alpha_{\phi}}} -\beta_{\phi} \right) \right] \\ 
\alpha_{\phi} &\equiv \sum_{b}\,w_{b}^{\prime}\,\frac{\Delta m_{b}}{m_{\rm ej}}\,\frac{p_{t}}{m_{b}\,v_{t}} \\ 
\beta_{\phi} &\equiv \sum_{b}\,w_{b}^{\prime}\,\cos{\theta}_{ba}\,\frac{|{\bf v}_{ba}|}{v_{t}} 
\end{align}
where $v_{t} \equiv 2\,\epsilon/p_{t}$ (approximately the velocity at which the blastwave becomes radiative). In the limit where the terminal momentum is reached, $p_{t} \ll m_{b}\,v_{t}$ by definition, so $\alpha_{\phi}$ is vanishingly small and $\phi \approx {\rm MIN}[1,\,1/2\beta_{\phi}]$ (for $\beta_{\phi}>0$). This has a simple interpretation then: $\beta_{\phi}$ is just the (kernel-averaged) ratio of the net outward gas velocity from the SN to the velocity where the blastwave becomes radiative. If the recession velocity exceeds $v_{t}\sim 200\,{\rm km\,s^{-1}}$ (the velocity at which the terminal momentum is reached, for a stationary surrounding medium), the SN must reach terminal momentum earlier (at a higher velocity therefore lower terminal momentum) before the ambient medium ``outruns'' the blastwave: mathematically $\beta_{\phi} \gtrsim 1$ and $\phi \lesssim 1$, accordingly. 

Having computed $\psi$ and $\phi$, we can then decide which limit (the energy-conserving or terminal-momentum limit) a gas element should be in, as in Eqs.~\ref{eqn:dp.subgrid.sub1}-\ref{eqn:dp.subgrid}, by comparing the corrected $\Delta {\bf p}_{b}^{\rm energy-conserving}$ (Eq.~\ref{eqn:delta.p.defn}) and corrected $\Delta {\bf p}_{b}^{\rm terminal}$ above. If $|\Delta {\bf p}_{b}^{\rm energy-conserving}| > |\Delta {\bf p}_{b}^{\rm terminal}|$ (the momentum implied by the energy-conserving limit exceeds the terminal momentum), {or} $m_{b} > m^{b}_{\rm cool} \equiv |\Delta {\bf p}_{b}^{\rm terminal}| / v_{t}$ (the mass of particle $b$ exceeds the swept-up-mass at which the energy-conserving solution would de-celerate to below $v_{t} = 2\,\epsilon/p_{t}$), then the terminal solution is applied (otherwise the energy-conserving solution is applied). If the gas-particle motion is negligible ($\psi\approx\phi\approx1$), these two conditions are {\em exactly} equivalent; more generally we need to check both.

Physically, note that for ${\bf v}_{ba}=0$ (non-moving cases or uniformly boosting the whole simulation), $\beta_{\psi}=\beta_{\phi}=0$ so $\psi=\phi=1$ and we obtain the stationary case as expected. For ${\bf v}_{ba}=$\,constant (assuming $\Delta m_{b} \ll m_{b}$), the condition $\sum \Delta{\bf p}_{b}={\bf 0}$ becomes mathematically identical to $\beta_{\psi}=\beta_{\phi}=0$, so there is no change in the coupled momentum or energy relative to what would occur in the stationary case (this is easiest to see by returning to Eq.~\ref{eqn:egycon.reduced} and simply taking the ${\bf v}_{ba}$ term outside the sum). In a medium moving with uniform velocity relative to the star, the blastwave produce a stronger, slower-moving shock in the ``upwind'' direction and weaker, faster-moving shock in the ``downwind'' direction, but it is easy to verify that the difference in the energies produced in both directions cancel one another (and of course, the momentum imparted in both directions must, by conservation, be equal). In a turbulent medium, different velocities will tend to cancel, so $\beta_{\psi,\,\phi}$ will both be small. 

However, when there is a large net inflow/outflow motion around the star, $\beta_{\psi,\,\phi}$ can be non-negligible. Consider a spherically symmetric case with ${\bf v}_{ba}=v_{r}\,\hat{r}$ so $\cos{\theta}_{ba}=1$ if $v_{r}>0$, or $\cos{\theta}_{ba}=-1$ if $v_{r}<0$, and assume the total mass to which the ejecta is coupled is distributed in a shell with mass $M_{\rm coupled}$. Then $|\beta_{\psi}|\sim (M_{\rm coupled}\,v_{r}^{2}/2\,\epsilon)^{1/2} \sim (KE_{\rm initial}/KE_{\rm ejecta})^{1/2}$, i.e.\ $\beta_{\psi}^{2}$ scales with the ratio of the initial (pre-coupling) kinetic energy of the surrounding gas elements (across which the ejecta are distributed) to the ejecta energy. Although $|{\bf v}_{ba}| = |v_{r}| \ll v_{\rm ej}$ for SNe, the kinetic energy is weighted by the particle mass, so it is {\em not} necessarily negligible: $|\beta_{\psi}|\gtrsim 1$ if the typical $ |v_{r}| \gtrsim (m_{\rm ejecta}/M_{\rm coupled})^{1/2}\,v_{\rm ej} \sim 350\,{\rm km\,s^{-1}}\,(m_{i}/100\,\msun)^{-1/2}$ for typical core-collapse SNe. At sufficiently high resolution, then, this term becomes negligible (the gas velocities are never so coherently large), but at low resolution it can be important. Of course, at low resolution, the cooling radius is un-resolved and we should use the terminal momentum expression, where $\beta_{\phi} \gtrsim 1$ requires $|v_{r}|\gtrsim v_{t} \sim 200\,{\rm km\,s^{-1}}$ -- the expression becomes resolution-independent in this limit (once $m_{i}$ exceeds a few hundred solar masses). In either case, if such large $v_{r}$ is reached, $\psi\approx 1/2\beta_{\psi}$ (or $\phi\approx1/2\beta_{\phi}$) becomes $<1$. This comes from the ${\bf p}_{b}\cdot \Delta {\bf p}_{b}$ term, which dominates over $\Delta {\bf p}_{b}^{2}$ in this limit -- physically, it requires more energy to accelerate a shell which is already moving rapidly away from the origin. Conversely, when $|v_{r}|$ is large and $\beta_{\psi}<0$, $\psi\approx2\,|\beta_{\psi}|\gg1$, i.e.\ this implies a larger momentum injection, with the energy for the additional $PdV$ work coming from the shocked external medium falling onto the shock. 

If we choose to keep our simple momentum scaling from the main text (setting $\psi=\phi=1$ always) -- i.e.\ assume that the momentum scaling of SNe is robust across variations in the surrounding gas velocity field -- then this necessarily means the kinetic energy coupled by SNe varies, with larger kinetic energy coupled in cases with a net ``outflow'' motion around the star, and smaller kinetic energy change in cases with net ``inflow'' motion around the star. This is not necessarily unphysical -- it depends, to some extent, on whether the more robust property of SNe blastwaves in a non-uniform flow is their kinetic energy or their momentum. Similarly, it is quite possible that the general scaling for the terminal momentum $p_{t}$ from the text could have a complicated dependence on the detailed structure of the velocity field, although simulations in turbulent media discussed in \S~\ref{sec:feedback:mechanical:sedov:background} suggest that, on average, broadly similar results are obtained as in simulations where the background is stationary. Clearly, future work is warranted to explore these conditions in more detail. 

In practice we find that whether we include this more detailed correction, or set $\psi=\phi=1$, almost always has a small effect on galaxy properties at all resolution levels: some examples are shown in Fig.~\ref{fig:sneenergy.method.tests}. Galaxy masses, star formation histories, mass profiles, visual morphologies, metal abundance distribution functions, rotation curves, CGM gas content, and mean outflow rates are essentially unchanged (with at most a systematic $\sim 0.1$\,dex shift in the masses of very low-mass dwarfs, and smaller effects in higher-mass galaxies). We have specifically tested this in the {\bf m10q} and {\bf m12i} galaxies in this paper as well as galaxies {\bf m10v}, {\bf m11q}, {\bf m12f} and {\bf m12m} from \paperone; we have compared all properties discussed in this paper and in \paperone. The fact that these corrections produce such small effects owes to the fact that coherent, large inflow/outflow velocities around star particles are rare and, even when they occur, tend to average out over time and space. Even in the worst-possible-case (maximal $\beta_{\psi}$) scenario, namely violent post-starburst outflow episodes around dwarf galaxies at low resolution, where most of the ISM of the galaxy is evacuated, the net change in kinetic energy of the gas setting $\psi=\phi=1$ only differs from the kinetic energy coupled with the exact formulation here by a factor $\sim 2$. And, critically, the difference between methods vanishes ($\beta_{\psi},\,\beta_{\phi}\rightarrow0$) at sufficiently high resolution.

\vspace{-0.5cm}
\section{Details of Unresolved Cooling Do Not Influence Predictions of Our Default Model}
\label{sec:appendix:implicit.cooling}

\begin{figure}
\hspace{-0.4cm}\plotonesize{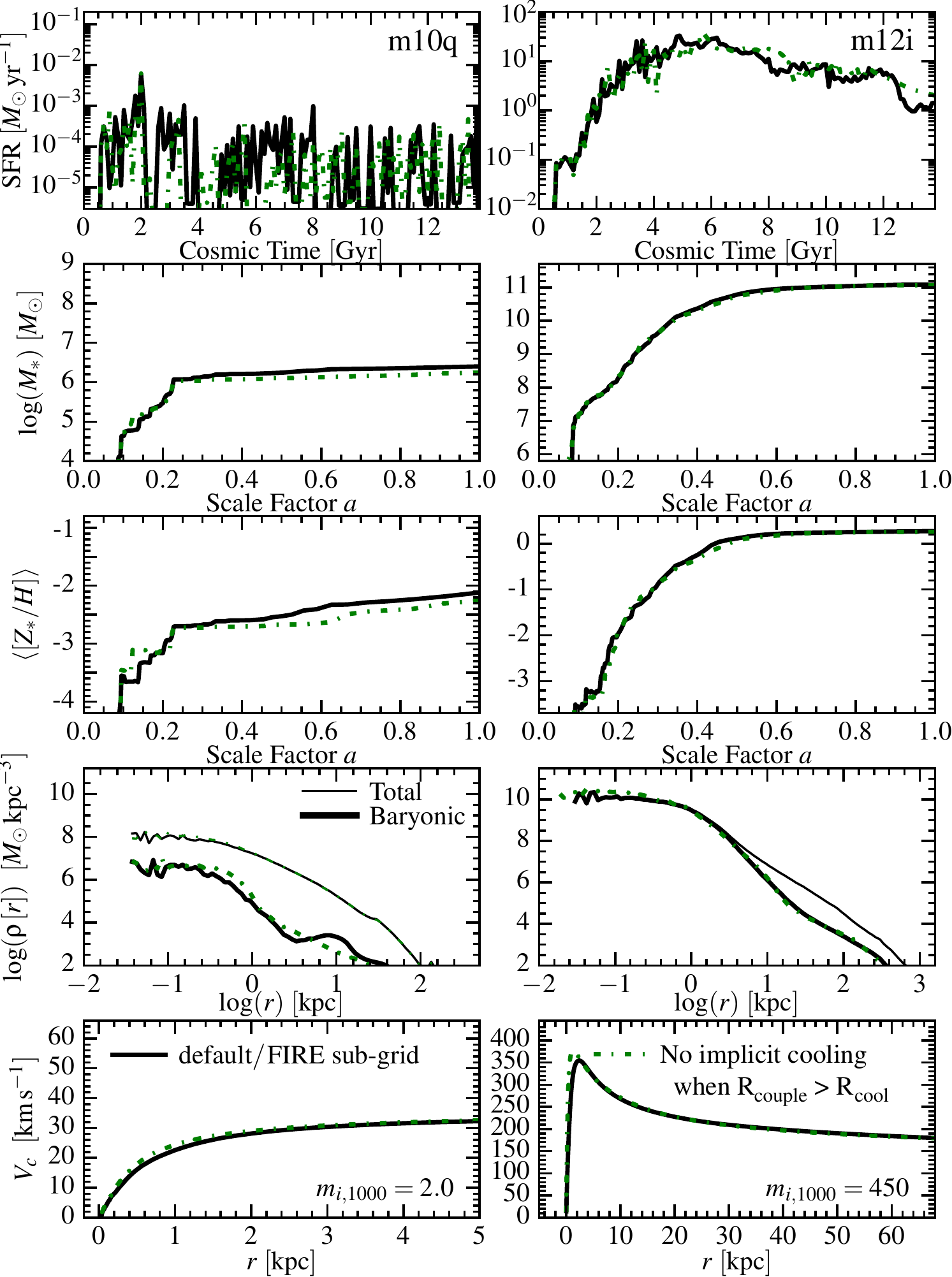}{1.04}
    \vspace{-0.5cm}
    \caption{Comparison of the default ``FIRE sub-grid'' model (as Fig.~\ref{fig:sf.history.sne.subgrid}) to a variant implementation in which we do not assume the residual thermal energy from the initial energy-conserving phase has been radiated away if the cooling radius is not resolved (i.e.\ coupling radius $R_{\rm couple} > R_{\rm cool}$). Instead we simply couple it explicitly, using the same kinetic-thermal solution as the default model, and rely on the code cooling physics to radiate it explicitly. This has no effect because, by definition, in this regime the code will radiate the thermal energy away rapidly in either case.
   \vspace{-0.4cm}
    \label{fig:implicit.cooling}}
\end{figure}

As noted in \S~\ref{sec:feedback:mechanical:tests:effects}, we have verified in a number of tests that, within the context of our default FIRE sub-grid model, the details of how we treat the ``unresolved cooling phase'' when the simulation does not resolve the local cooling radius are secondary, so long as the correct momentum is coupled to the gas. Fig.~\ref{fig:implicit.cooling} shows this explicitly for both {\bf m10q} and {\bf m12i} simulations. In this figure we compare a model where we take our standard sub-grid coupling (the momentum, mass, and metals are unchanged) but always couple the ``full'' total energy -- we do not assume (as in our default model) that the residual thermal energy has been radiated away when the cooling radius is unresolved. As expected, this produces nearly identical results to our default model -- in this limit, {\em by definition}, the cooling time is shorter than the dynamical time at the radius where the energy is deposited. So the code simply radiates away the energy in the next few timesteps, without doing significant work. This is a non-trivial statement, however, in that it clearly shows that in this regime, the {\em momentum} coupled, not the thermal energy, is the important physical ingredient.

\end{appendix}

\end{document}